\documentclass[fleqn,usenatbib]{mnras}

\usepackage{newtxtext,newtxmath}
\usepackage{longtable}
\usepackage[T1]{fontenc}

\DeclareRobustCommand{\VAN}[3]{#2}
\let\VANthebibliography\thebibliography
\def\thebibliography{\DeclareRobustCommand{\VAN}[3]{##3}\VANthebibliography}

\usepackage{graphicx}	% Including figure files
\usepackage{amsmath}	% Advanced maths commands
\usepackage{rotating}
\usepackage{graphicx}
\usepackage{amsmath}
\usepackage{threeparttable}
\usepackage{caption}
\usepackage{bibunits}
\usepackage{color}
\usepackage{adjustbox}
\usepackage{url}
\usepackage{soul}
\newcommand{\cii}{[C\,{\sc ii}]}
\newcommand{\ci}{[C\,{\sc i}]}
\newcommand{\oi}{[O\,{\sc i}]}
\newcommand{\nii}{[N\,{\sc ii}]}
\newcommand{\niii}{[N\,{\sc iii}]}

\newcommand{\oiii}{[O\,{\sc iii}]}

\newcommand{\oiiil}{[O\,{\sc iii}] 88\,$\mu{\rm m}$}
\newcommand{\ciil}{[C\,{\sc ii}] 158\,$\mu{\rm m}$}
\newcommand{\niil}{[N\,{\sc ii}] 205\,$\mu{\rm m}$}
\newcommand{\oil}{[O\,{\sc i}] 145\,$\mu{\rm m}$}

\definecolor{referee}{RGB}{0,0,0}
\definecolor{referee2}{RGB}{0,0,0}
\definecolor{referee3}{RGB}{0,0,0}
\usepackage{ragged2e}

\title[Efficient high-resolution line survey]{A Novel high-$z$ submm Galaxy Efficient Line Survey in ALMA bands 3 through 8 - An ANGELS Pilot}

\author[Bakx et al.]{T. J. L. C. Bakx$^{1,2,3}$\thanks{E-mail: tom.bakx@chalmers.se},
A. Amvrosiadis$^{4}$,
G. J. Bendo$^{5}$,
H. S. B. Algera$^{6,3}$,
S. Serjeant$^{7}$,
L. Bonavera$^{8,9}$,\newauthor
E. Borsato$^{10}$,
X. Chen$^{3}$,
P. Cox$^{11}$,
J. González-Nuevo$^{8,9}$,
M. Hagimoto$^{3}$,
K. C. Harrington$^{12}$,\newauthor
R. J. Ivison$^{13,14,15,16}$,
P. Kamieneski$^{17}$
L. Marchetti$^{18,19}$,
D. A. Riechers$^{20}$,
T. Tsukui$^{21,16}$,\newauthor
P. P. van der Werf$^{22}$,
C. Yang$^{1}$,
J. A. Zavala$^{3}$,
P. Andreani$^{23}$,
S. Berta$^{24}$,
A. R. Cooray$^{25}$,
G. De Zotti$^{10}$,\newauthor
S. Eales$^{26}$,
R. Ikeda$^{27,28}$,
K. K. Knudsen$^{1}$,
I. Mitsuhashi$^{3}$,
M. Negrello$^{26}$,
R. Neri$^{24}$,\newauthor
A. Omont$^{11}$,
D. Scott$^{29}$,
Y. Tamura$^{3}$,
P. Temi$^{30}$, and
S. A. Urquhart$^{7}$.
\\
Affiliations can be found after the references
}

\date{Accepted 2024 October 21. Received 2024 October 21; in original form 2024 June 28}

\pubyear{2024}

\begin{document}
\label{firstpage}
\pagerange{\pageref{firstpage}--\pageref{lastpage}}
\maketitle

\begin{abstract}
We use the Atacama Large sub/Millimetre Array (ALMA) to efficiently observe spectral lines across Bands~3, 4, 5, 6, 7, and 8 at high-resolution (0\farcs{}5 -- 0\farcs{}1) for 16 bright southern \textit{Herschel} sources at $1.5 < z < 4.2$. With only six and a half hours of observations, we reveal 66 spectral lines in {17} galaxies.
These observations detect emission from CO (3--2) to CO(18--17), as well as atomic (\ci{}(1-0), (2-1), \oi{}~145~$\mu$m and \nii{}~205~$\mu$m) lines. Additional molecular lines are seen in emission (${\rm H_2O}$ and ${\rm H_2O^+}$) and absorption (OH$^+$ and CH$^+$). 
The morphologies based on dust continuum ranges from extended sources to strong lensed galaxies with magnifications between 2 and 30.
CO line transitions indicate a diverse set of excitation conditions with a fraction of the sources ($\sim 35$\%) showcasing dense, warm gas. 
The resolved gas to star-formation surface densities vary strongly per source, and suggest that the observed diversity of dusty star-forming galaxies could be a combination of lensed, compact dusty starbursts and extended, potentially-merging galaxies. The predicted gas depletion timescales are consistent with 100~Myr to 1~Gyr, but require efficient fueling from the extended gas reservoirs onto the more central starbursts, {\color{referee} in line with the Doppler-shifted absorption lines that indicate inflowing gas for two out of six sources.} 
This pilot paper explores a successful new method of observing spectral lines in large samples of galaxies, supports future studies of larger samples, and finds that the efficiency of this new observational method will be further improved with the planned ALMA Wideband Sensitivity Upgrade. 
\end{abstract}


%%%%%%%%%%%%%%%%%%%%%%%%%%%%%%%%%%%%%%%%%%%%%%%%%%

%%%%%%%%%%%%%%%%% BODY OF PAPER %%%%%%%%%%%%%%%%%%

\section{Introduction}
\begin{figure*}
    \centering
    \includegraphics[width=\textwidth]{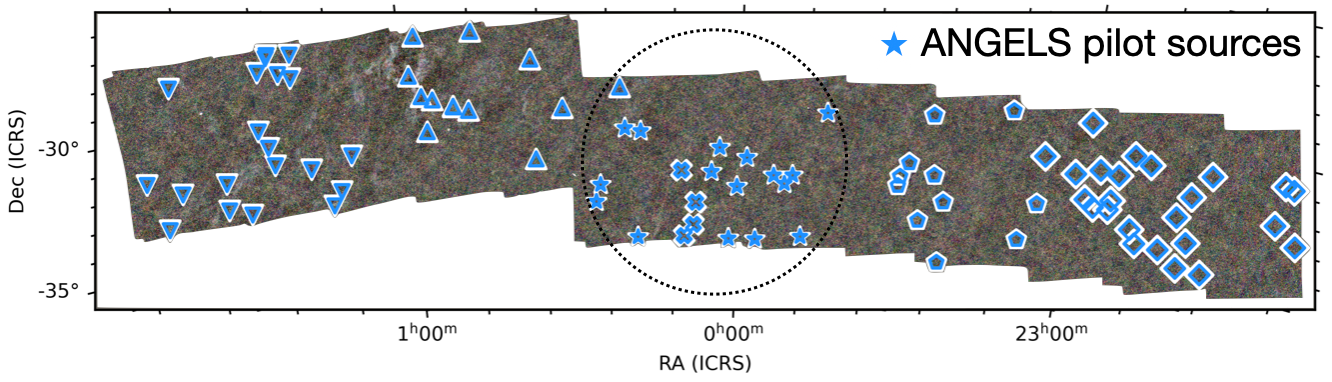}
    \caption{The 88 \textit{Herschel} Bright Sources \citep{bakx18,Bakx2020Erratum} in the 285~square degree South Galactic Pole field (85 observed in the BEARS survey; \citealt{Urquhart2022}). Observations within ten degrees on the sky can be done in rapid succession, without the need for additional calibration. Here we show the five different regions across the SGP field indicated with different symbols, with the central field used for Pilot observations (\textit{star-symbols}), while the four remaining fields (\textit{upward- and downward triangles, hexagons and diamonds}) are left for subsequent follow-up. Note that four sources were excluded at the time of observation (\textit{crosses}). The circle with a radius of $\sim 7$~degrees indicates the typical size for sources that the ALMA Observing Tool predicts can be observed with only a single calibration step for ALMA configuration schedules C43-6 and smaller.}
    \label{fig:ANGELSFIELDS}
\end{figure*}
The discovery of a population of dusty galaxies -- called either sub-mm galaxies (SMGs; \citealt{Smail1997,Hughes1998, blain2002}) or dusty star-forming galaxies (DSFGs; \citealt{Casey2014}) -- showed that more than half of the star-formation in the early Universe was shrouded in dust. With total infrared luminosities exceeding several times 10$^{12}$ L$_{\odot}$, the star-formation rates in these SMGs and DSFGs (hereafter, we use DSFGs for simplicity) approach the stability limit of 1000\,M$_{\odot}$ year$^{-1}$ and may even exceed this limit \citep{Andrews2011,fudamoto2017,RowanRobinson2018}. These intense star-formation rates -- enough to build a large galaxy in only 0.1\,Gyr -- suggest that these may be the ancestors of present-day early-type galaxies \citep{Eales1999,Lilly1999}. Evidence of this evolutionary pathway of DSFGs to quenched elliptical galaxies is mounting, given their high stellar masses \citep{Hainline2011,Aravena2016,Long2023,Smail2023}, high specific star-formation rates \citep{Straatman2014,Spilker2016,Glazebrook2017,Schreiber2018, Merlin2019} and location in overdense regions \citep{Blain2004,weiss2009,Hickox2012}. However, the DSFG population appears diverse \citep[e.g.,][]{Hagimoto2023}, and because the types of galaxies studied in this paper are very rare (a few / deg$^2$ for $>10^{12.5}\,\rm L_{\odot}$), hydrodynamical models struggle to include a large enough volume to simulate these galaxies accurately to test the evolutionary pathways of these DSFGs \citep[e.g.,][]{narayanan2015}. Observations of large numbers of DSFGs are thus the only direct way of understanding this population.

The {\it Herschel} Space Observatory increased the number of known DSFGs from hundreds to over half a million sources through several large-area surveys. In these surveys, the SPIRE instrument \citep{griffin2010} observed these large areas at the peak of the redshifted DSFG dust spectra without the hindering effects of the atmosphere. The total mapping area of {\it Herschel} surveys is over one thousand square degrees, with 660 deg$^2$ {\it Herschel} ATLAS (H-ATLAS; \citealt{eales2010}) and the 380 deg$^2$ {\it Herschel} Multi-tiered Extragalactic Survey (HerMES; \citealt{oliver2012}) and the \textit{Herschel} Stripe 82 Survey (HerS; \citealt{Viero2014}) accounting for the bulk of the area increase \citep{shirley2021}.  
The vast increase in survey area (at the time, ground-based surveys only covered $\sim$1 deg$^2$ in total) also meant that {\it Herschel} discovered many more of the most extreme DSFGs. This has been complemented by observations in the mm regime, such as the South Pole Telescope survey, which revealed around a hundred sources that are also at high redshift from over 2500\,deg$^2$ \citep{reuter20}. 

To examine the nature of these extreme high-redshift sources, it is crucial to perform detailed studies characterizing the conditions of the Inter-Stellar Medium (ISM) conditions seen through resolved spectral line observations \citep[e.g.,][]{Hodge2019}. Recent high-resolution follow-up studies of H-ATLAS sources, to use one survey as an example, have revealed 14 sources with $4 < z < 6.5$ with star-formation rates up to 8,000 M$_{\odot}$ year$^{-1}$, well over the \textit{maximal starburst} limit \citep{fudamoto2017,Zavala2018,Montana2021}, and the discovery of a source at $z = 4.002$ which when followed up with Atacama Large sub/Millimetre Array (ALMA) turned out to be the core of a proto-cluster \citep{Oteo2018}, with its large scale environment forming stars at a rate of several tens of thousands of solar masses per year.

Given the large beam size of single-dish telescopes, high-redshift sub-mm astronomy was often jokingly referred to as \textit{blobology} prior to the advent of interferometer facilities such as ALMA, CARMA, NOEMA (Northern Extended Millimeter Array; previously PdBI) and the SMA. These interferometers were able to reveal large numbers of spectral lines of individual sources with NOEMA \citep[e.g.,][]{Yang2023}, as well as stunning lensing arcs and Einstein rings \citep[e.g.,][]{dye2015,Dye2018,Spilker2016} while targeting spectral lines at high resolution with ALMA \citep[e.g.,][]{Rizzo2020}. Particularly since the advent of large samples of DSFGs with robust redshifts \citep{reuter20,Urquhart2022,Cox2023}, spectroscopic follow-up studies are the key next step in a full characterisation of this extreme population.
Unfortunately, while interferometers are powerful instruments, the modest bandwidths currently available at ALMA mean that spectral line observations require dedicated tunings for every source, resulting in either runaway observation times or statistically-insignificant samples. Consequently, a comprehensive high-resolution spectroscopic study of complete bright {\it Herschel} samples, or for that matter any survey from large-area mapping studies, is beyond the scope of current instrumentation. The Wideband Sensitivity Upgrade (WSU; \citealt{Carpenter2023}) will provide some relief, but as we will show, even its most optimistic goals can benefit from the methods detailed in this exploratory paper.

To break the current paradigm -- i.e., long observations or small samples -- this paper attempts short snapshot-style observations with ALMA \citep[similar to][]{Dye2018,Amvrosiadis2018} while targeting as many spectral lines as possible: \textbf{A} \textbf{N}ovel high-$z$ submm \textbf{G}alaxy \textbf{E}fficient \textbf{L}ine \textbf{S}urvey (ANGELS; Appendix Figure~\ref{fig:RGBAngels}). We describe this  technique in Section~\ref{sec:Observations}, together with the definition of the sample of galaxies, and the ALMA data reduction steps.
We describe the results of the observations, both for continuum and line measurements, in Section~\ref{sec:Results}. We find lensing morphology for around half of the sample, and we report the lensing and ancillary data properties in Section~\ref{sec:Lensing}. We place the line properties into context of the literature in Section~\ref{sec:SpectralLineProperties}. The resolved studies of ANGELS sources are discussed in Section~\ref{sec:ResolvedStudies}.
We collate an overview of the nature of the ANGELS sources, as became apparent by these observations, in Section~\ref{sec:NatureOfANGELSSources}.
We evaluate this efficient line survey in Section~\ref{sec:ANGELSasAsurvey}, where we also detail the inevitable caveats and look into the future of surveys including the benefits of the WSU \citep{Carpenter2023}.  Conclusions are summarized in Section~\ref{sec:conclusions}.
Throughout this paper, we assume a flat $\Lambda$-CDM cosmology with the best-fit parameters derived from the \textit{Planck} results \citep{Planck2020}, which are $\Omega_\mathrm{m} = 0.315$, $\Omega_\mathrm{\Lambda} = 0.685$ and $h = 0.674$.

\section{Observations}
\label{sec:Observations}
We target a group of sources within 10 degrees on the sky using rapid two minute snapshots across Bands 3 through 8 to test the viability of a rapid line survey with ALMA across the available bands with the best atmospheric transmission. Prior high-redshift observations with ALMA were limited to observations that have to retune the local oscillator and subsequently repeat the calibration observations. This approach results in large overheads ($\sim 20 - 30$~min. per target) and requires a Large Program before one hundred lines are reached. Our method instead asks: \textit{what specific tuning would allow us to target as many spectral lines as possible?} The purpose of this paper is to test the usefulness of a snapshot-style line survey with bandwidths optimised to cover as many lines as possible. In Section~\ref{sec:targetSelection}, we describe the selected sources, and detail the ANGELS method in Section~\ref{sec:ANGELSMETHODSec2}, before describing the data reduction steps in Section~\ref{sec:DataReduction}.

\subsection{Target selection}
\label{sec:targetSelection}
The group of sources is selected from the 285~square degree South Galactic Pole (SGP) field from the H-ATLAS survey \citep{eales2010}. The field is centered on RA, DEC = ~23:24:46, $-$33:00:00, and is shown in Figure~\ref{fig:ANGELSFIELDS} \citep{valiante2016,Ward2022}. In total, this field contains 88 bright high-redshift {\it Herschel} sources ($5 > z_{\rm phot} > 2$ and $350 >$~S$_{500} > 80$~mJy; \citealt{bakx18,Bakx2020Erratum}), of which 65 sources have one or more ALMA-resolved sources with spectroscopic redshifts (which increased to 68 sources by including the spectroscopic redshifts identified by this program). {\color{referee} These sources represent a larger sample of roughly the 300 brightest high-redshift objects \citep{nayyeri2016,negrello2017,bakx18,Bakx2020Erratum} selected from a thousand square degrees of {\it Herschel} surveys, and typically have slightly lower observed luminosities than the majority-lensed SPT- (\citealt[{\color{referee2} $\overline{\mu L_{\rm IR}} = 4.4 \times{}10^{13} L_{\odot}$}]{reuter20}) and {\it Planck}-selected galaxies (\citealt[{\color{referee2} $\overline{ \mu L_{\rm IR}} = 1.2 \times{}10^{14} L_{\odot}$}]{Berman2022}) from $\sim 2500$~square degrees and an all-sky survey, respectively. Meanwhile, these {\it Herschel} sources have higher apparent luminosities than the more abundant, typically-unlensed SMGs from ground-based sub-mm observations (e.g., \citealt{Garratt2023}), that are selected from surveys of up to $\sim 10$~square degrees. }
The Bright Extragalactic ALMA Redshift Survey (BEARS) provided the bulk of the spectroscopic redshifts across 62 {\it Herschel} sources \citep{Urquhart2022}, for a total of 71 ALMA-resolved galaxies {\color{referee2} with average lensed infrared luminosities of $ \overline{\mu L_{\rm IR}} = 4.0 \times{} 10^{13} L_{\odot}$}. This survey efficiently used Band~3 and 4 to derive robust spectroscopic redshifts using multiple Carbon-monoxide (CO) and atomic carbon (\ci{}) lines \citep{Bakx2022}; and in addition, the data enabled subsequent marginally-resolved studies of the dust continuum \citep{Bendo2023} and molecular gas \citep{Hagimoto2023}.

The observations in this pilot program aimed to target a group of sources close on the sky ($< 10$~deg) with spectroscopic redshifts. We identified this group of sources using the ALMA Observing Tool, which split the 88 sources across six groups on the sky that can each be observed within a single calibration, i.e., all sources that are closer together than 10 degrees on the sky. The group that was selected ($RA, DEC =$~00:00:00, $-$33:29:00) contains most fields with spectroscopic redshifts; thirteen out of sixteen fields had at least one source with a spectroscopic redshift, ranging between $z = 1.536$ and $4.182$ with a median value of $\bar{z} = 2.843$, see Table~\ref{tab:sources}. {\color{referee2} The average infrared luminosities of these sixteen fields are on average $\overline{\mu L_{\rm IR}} = 4.2 \times{} 10^{13} L_{\odot}$ based on the luminosities documented in \citep{bakx18,Bakx2020Erratum}\footnote{This value is slightly higher than the luminosities from a re-fit using the spectroscopic redshifts with the Eyelash template \citep[e.g.][]{ivison16} of $2.9 \times{} 10^{13} L_{\odot}$, which still comparable to the average luminosity of the BEARS sample. }, and thus these fields are representative of the larger BEARS and HerBS samples. } In total, 26 resolved galaxies exist within these sixteen fields. An additional four fields were not targeted in these initial ANGELS observations, although they were within the range of this field (shown as crosses in Figure~\ref{fig:ANGELSFIELDS}).
Only one line was detected in the spectra of the three remaining sources (HerBS-87, HerBS-104, and HerBS-170), and it was not possible to determine robust redshifts for these targets.
The additional observations from ANGELS allowed us to cut through this redshift degeneracy and measure redshifts for these three sources (see Section~\ref{sec:angelsasaredshiftmachine}).

% The sources with their redshifts
\begin{table}
    \centering
    \caption{Sources targeted by ANGELS Pilot}
    \label{tab:sources}
    \begin{tabular}{lcccc}
    \hline \hline
Source & $z$ & RA & Dec & $\Delta$V [km/s] \\ \hline
HerBS-21AB & 3.323 & 23:44:18.112 	&	$-$30:39:36.58  & $550 \pm 110$\\
HerBS-22A & 3.050 & 00:26:25.000 	&	$-$34:17:38.22  & $680 \pm 150$  \\
HerBS-22B$^{\ddagger}$  & --  & 00:26:25.560 &  $-$34:17:23.30 & -- \\
HerBS-25 & 2.912 & 23:58:27.505 	&	$-$32:32:45.05  & $210	\pm 40$\\
HerBS-36 & 3.095 & 23:56:23.079 	&	$-$35:41:19.64  & $550	\pm 80$\\
HerBS-41A & 4.098 & 00:01:24.796 	&	$-$35:42:11.15  & $680 \pm 100$\\
HerBS-41B$^{\ddagger}$  & -- & 00:01:23.240 & $-$35:42:10.80 & -- \\
HerBS-41C$^{\ddagger}$  & -- & 00:01:25.820 & $-$35:42:18.00 & -- \\
HerBS-42A & 3.307 & 00:00:07.458 	&	$-$33:41:03.07  & $640 \pm 70$\\
HerBS-42B & 3.314 & 00:00:07.428 	&	$-$33:40:55.82  & $490 \pm 90$\\
HerBS-42C$^{\ddagger}$   & -- & 00:00:07.050 & $-$33:41:03.40  & -- \\
HerBS-81A & 3.160 & 00:20:54.739 	&	$-$31:27:50.96  & $670 \pm 200$ \\
HerBS-81B & 2.588 & 00:20:54.201 	&	$-$31:27:57.49  & $650 \pm 190$ \\
HerBS-86 & 2.564 & 23:53:24.569 	&	$-$33:11:11.93  & $460 \pm 100$\\
HerBS-87$^{\dagger}$ & 2.059 & 00:25:33.683 	&	$-$33:38:26.21  & $370 \pm 110$ \\
HerBS-93 & 2.402 & 23:47:50.443 	&	$-$35:29:30.13 & $640 \pm 160$ \\
HerBS-104A$^{\ddagger}$  & -- & 00:18:39.470 & $-$35:41:48.00 & -- \\
HerBS-104B$^{\dagger}$ & 1.536 & 00:18:38.851 	&	$-$35:41:33.09  &  $430 \pm 130$ \\
HerBS-106A & 2.369 & 00:18:02.467 	&	$-$31:35:05.17  & $500 \pm 170$ \\
HerBS-106B$^{\ddagger}$  & --  & 00:18:01.104 & $-$31:35:08.00 & -- \\
HerBS-155A & 3.077 & 00:03:30.644 	&	$-$32:11:35.00  & $500 \pm 150$\\
HerBS-155B$^{\ddagger}$  & -- &  00:03:30.060 & $-$32:11:39.30 & -- \\
HerBS-159A & 2.236 & 23:51:21.750 	&	$-$33:29:00.53  &  $280 \pm 80$\\
HerBS-159B & 2.236 & 23:51:22.363 	&	$-$33:29:08.14  &  $330 \pm 90$\\
HerBS-170$^{\dagger}$ & 4.182 & 00:04:55.445 	&	$-$33:08:12.85  & $680 \pm 180$ \\
HerBS-184 & 2.507 & 23:49:55.661 	&	$-$33:08:34.48  & $320 \pm 40$\\ \hline
    \end{tabular}
    \raggedright \justify \vspace{-0.2cm}
\textbf{Notes:} 
Col. 1: Source name. In the case of source multiplicity, we denote the brightest component with the letter A, and subsequent letters mark the source sensitivity.
Col. 2: Spectroscopic redshift from sub-mm observations.
Col. 3: Right Ascension in [hms] units.
Col. 4: Declination in [dms] units.
Col. 5: Velocity-width in [km/s] units from unresolved ALMA observations \citep{Urquhart2022}.
$^{\dagger}$ The redshifts of these sources are confirmed using the ANGELS observations, and are based on the lines included in this paper, see Section~\ref{sec:angelsasaredshiftmachine}. 
$^{\ddagger}$ These sources do not have spectroscopic redshifts and are not completely covered by the primary beam across Bands~3 through 8. If the field only has one galaxy with a spectroscopic redshift, we drop the sub-index (i.e., A, B etc.) throughout the paper for legibility reasons. These sources are shown in Appendix~\ref{sec:offsetSources}.
\end{table}

\subsection{The ANGELS method}
\label{sec:ANGELSMETHODSec2}
\begin{figure*}
    \centering
    \includegraphics[width=0.75\paperheight,angle=270]{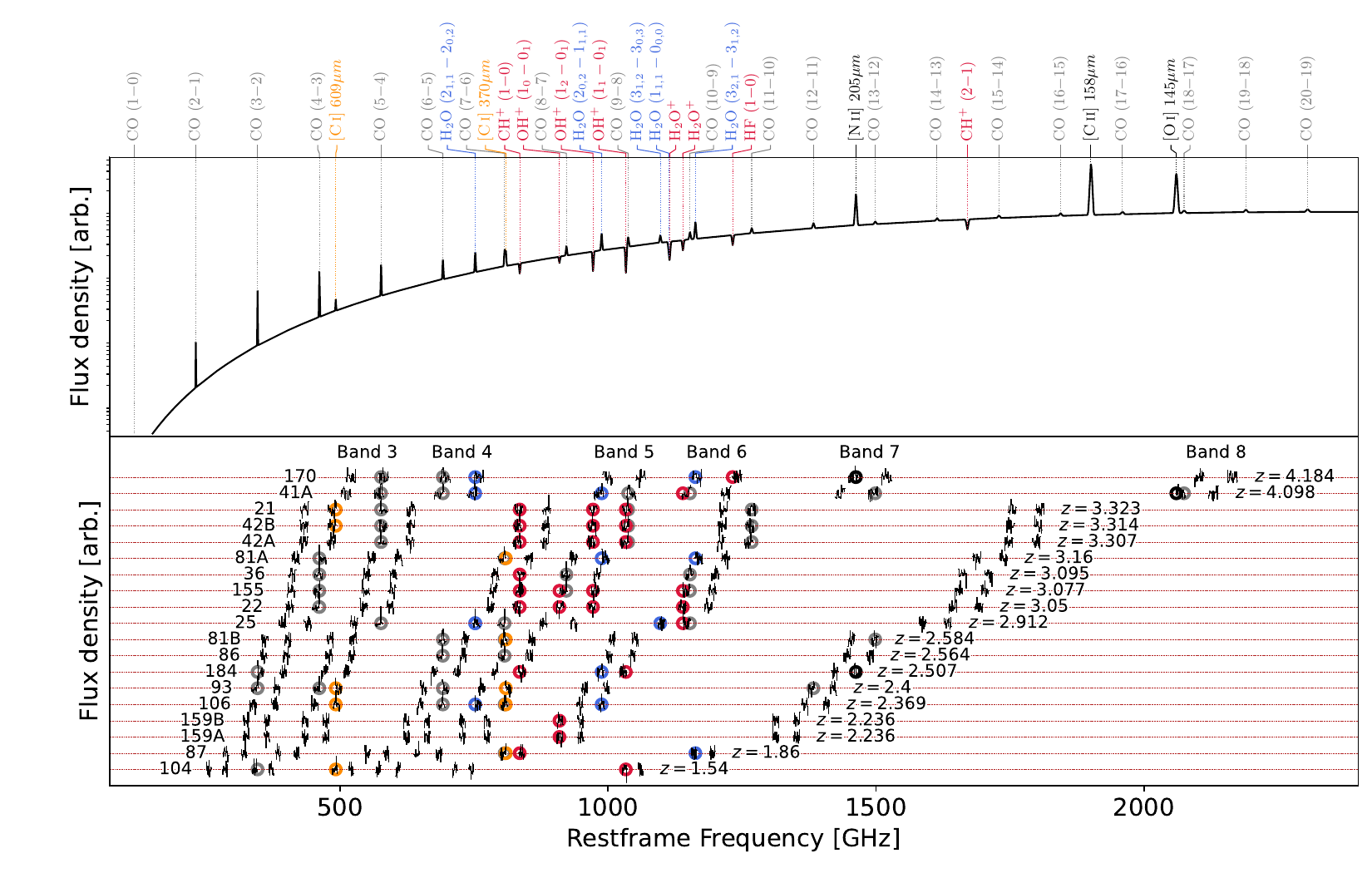}
    \caption{\textit{The top panel} shows the rest-frame spectra of a typical dusty galaxy, assuming a two-body dust spectrum based on {\it Herschel} observations \citep{pearson13} in conjunction with the spectral lines of a local ULIRG, in this case based on the line ratios of Arp220 by \citet{Rangwala2011}. Most lines are seen in emission, with carbon-monoxide (CO; grey), atomic carbon (\ci{}, orange) and water (blue, of course) lines at low frequencies, and atomic lines at high frequencies (shown in black). A {\it forest} of absorption lines is seen at the $\sim 1000$~GHz frequency range, consisting of CH$^+$, OH$^+$ and (ionized) water lines, shown in red. 
    \textit{The bottom panel} shows the restframe frequencies of the optimized tunings in Bands 3, 4, 5, 6, 7 and 8 for each of the sources in ANGELS. The tunings (shown as the aperture-extracted spectra in black lines) were optimized to target as many lines as possible, and wherever they are expected to observe a line, we mark the frequency with a circle. The sources are sorted by redshift, with the lowest-redshift source at the bottom, and the highest-redshift source at the top. 
    }
    \label{fig:tunings}
\end{figure*}

% The goal of these observations is to observe as many spectral lines as possible in a relatively short amount of time.
The {\color{referee}number density} of sources allowed the use of short snapshots ($\approx 2$~min) with only a single $\sim 20-30$~min. calibration step to reduce the overheads typically associated with line surveys ($\sim$one calibration per targeted line). As a result, each observation required roughly one hour per band. This style of snapshot observation is not unique (e.g., \citealt{Amvrosiadis2018} showed short snapshots can reveal the morphology of 15 lensed galaxies with only 2~min. of on-source time), however this ANGELS pilot project explicitly focuses to include as many spectral lines as possible, while covering a much larger range of frequencies between Bands 3 through 8. Since we knew the redshifts of all of our targets (except for HerBS-87, -104 and -170, at the time of observing), we could use the cosmological frequency shift of the Universe as our spectrometer and evaluate the coverage of the spectral windows at the rest-frame frequencies of all of the {\it Herschel} galaxies (see Figure~\ref{fig:tunings}). The sources are shown by their spectral windows shifted to their restframe frequency, and any lines within the spectral windows are marked as circles in the bottom panel. Typically, Bands 3 and 4 are able to target carbon-monoxide (CO) and atomic Carbon (\ci{} at 492.161~GHz) emission lines. Bands 5, 6 and 7 are sensitive to molecular lines such as water, and absorption lines including CH$^+$ and OH$^+$. The highest-frequency Band 8 is able to target atomic emission lines, such as \oi{} (2060~GHz) and \nii{} (1461~GHz).

This shared calibration step is only enabled for sources within 10~deg. on the sky for configurations below $\sim 3$~km (C43-6 or lower), while for longer baselines this condition is more stringent ($< 1$~deg.). This places a rough limit on the maximum resolution between 0\farcs{}3 and $\sim 0$\farcs{}07 for Bands~3 through 8, respectively. Given the spatial extent of DSFGs, any higher resolution could run into problems with maximum resolvable scales and the dilution of emission across the increasing number of beams.

The ALMA observations were carried out in Cycle 8 between November 2021 and July 2022 (Program ID: 2021.1.01628.S; P.I. T. Bakx). 
Table~\ref{tab:observationSetup} shows the exact tuning frequencies used in this ANGELS pilot, together with the observing depth across the entire image (i.e., the continuum depth), and across 35~km/s bins. A quasar was used for the complex gain calibration (J2359-3133), while another quasar was used for bandpass calibration (J2258-2758). 
We found this exact set-up of the spectral windows by parameterizing each of the spectral set-ups of the bands by their starting frequency. Appendix Figure~\ref{fig:optiTuneGraph} compares each potential tuning with ALMA to the total number of lines that can be targeted. The bottom part of the graph shows the atmospheric transmission, normalized between 0 and 1, to indicate the effect of tuning at this specific frequency. The variability in the observable number of lines with frequency reveals an important fact: {\it The simple act of optimizing the tuning towards the known spectroscopic redshifts can triple the number lines within any snapshot program.}

% The set-ups used in ANGELS
\begin{table*}
    \caption{ANGELS optimized observation set-up}
    \label{tab:observationSetup}
    \centering
    \begin{tabular}{lccccccccc}
\hline \hline
Tuning & Bandwidth & \multicolumn{2}{c}{Observing depth} & Resolution &   FoV & MRS & Pix. size & Cal. unc. \\
 &           & Cont. & 35~km/s & &  &
   &
    &
   \\
 & [GHz]     & \multicolumn{2}{c}{[mJy / beam]} & [" $\times{}$ "] & (arcsec) &
   (arcsec) &
  (arcsec) & \\
\hline
Band 3$^{@}$ & \ \ 98.3705 ... 102.0895 \& 110.3705 ... 114.0895 & 0.051 & 1.25 & 0.51 $\times{}$ 0.30 & 82 & 4.2 & 0.05 & 5 \%  \\
Band 4$^{\#}$  &    132.4405 ... 136.1605 \& 144.4405 ... 148.1605 & 0.049 & 1.04 & 0.38 $\times{}$ 0.34 & 62 & 4.1 & 0.05 & 5 \%  \\
Band 5$^{\dagger}$ &    190.8645 ... 194.5855 \& 202.8645 ... 206.5855 & 0.051 & 0.92 & 0.31 $\times{}$ 0.26 & 44 & 2.9 & 0.04 & 5 \%  \\
Band 6$^{\ddagger}$ &    222.9465 ... 226.6665 \& 237.4465 ... 241.1665 & 0.069 & 1.89 & 0.20 $\times{}$ 0.18 & 38 & 2.5 & 0.03 & 10 \% \\
Band 7$^{*}$ &    279.2225 ... 282.9425 \& 291.2225 ... 294.9425 & 0.102 & 1.52 & 0.17 $\times{}$ 0.12 & 30 & 2.1 & 0.02 & 10 \% \\
Band 8$^{**}$ &    403.9195 ... 407.6395 \& 415.9195 ... 419.6395 & 0.210 & 2.61 & 0.17 $\times{}$ 0.16 & 21 & 1.5 & 0.02 & 20 \% \\
\hline
    \end{tabular}    
{    \raggedright \justify \vspace{-0.2cm}
\textbf{Notes:} 
Col. 1: ALMA Band used for the tuning.
Col. 2: Frequency set-up, contiguous between the triple-dots, and with the lower- and upper sidebands separated by an ampersand.
Col. 3: Sensitivity of the observations across the complete $\approx 7.5$~GHz.
Col. 4: Average sensitivity of the observations across a 35~km/s bin.
Col. 5: Average beam size in units of arcsec by arcsec.
Col. 6: Field of View in units of arcsec, this corresponds to the diameter where the primary beam reaches 0.35 of its peak value.
Col. 7: Maximum recoverable scale in units of arcsec. Note that none of our sources extend beyond this scale.
Col. 8: Pixel size in units of arcsec.
Col. 9: Calibration uncertainty in flux density measurements. {\color{referee2}Following \cite{Bakx2024Receivers}, our observations are conducted using the Local Oscillator \citep{Bryerton2013} and the following receivers:}
$^{@}$ The Band 3 performance is documented in \cite{Claude2008,Kerr2014}
$^{\#}$ The Band 4 performance is documented in \cite{Asayama2014}
$^{\dagger}$ The Band 5 performance is documented in \citet{Belitsky2018}
$^{\ddagger}$ The Band 6 performance is documented in \cite{Ediss2004,Kerr2004,Kerr2014}
$^{*}$ The Band 7 receiver performance is reported in \citet{Mahieu2012}.
$^{**}$ The Band 8 receiver performance is reported in \citet{Sekimoto2008}.
}
\end{table*}

\subsection{Data reduction}
\label{sec:DataReduction}

All data were processed using the {\sc common astronomy software applications} ({\sc casa}) package \citep{mcmullin2007, casateam2022}.  The pipeline calibration of the visibility data were first restored using the standard scripts from the ALMA Science Archive and using {\sc casa} version 6.2.1.  After this, the data were imaged using {\sc tclean}.  The continuum images were created with {\sc casa} version 6.2.1, while the image cubes were created with {\sc casa} version 6.4.1.  

We initially created image cubes binning the channels by a factor of 16 to identify detectable spectral lines from each target within each spectral window {\color{referee}(corresponding to a channel width between 50 and 15~km/s between Bands 3 and 8, respectively)}.  After this, we subtracted the continuum from the visibility data for each field and spectral window containing a detectable spectral line.  We also had spectral windows that covered spectral lines that were expected to be present based on the redshifts of the individual targets from \citet{Urquhart2022} but that were not detected.  In these situations, we identified all channels that lied within a 0.4~GHz region centred on the expected frequency of the spectral line for each target as potentially containing line emission and used all channels not covering the potential line emission to measure and subtract the continuum from the visibility data.  These continuum-subtracted image cubes were then used for the final analysis.  The channels in the visibility data without the continuum subtraction that were identified as not containing emission were used to create the final continuum images.

Natural weighting was used to create the continuum images, but when creating the image cubes, we set the weighting to {\sc briggsbwtaper}, which is a variant of Briggs weighting that makes adjustments for variations in the beam size with frequency, and we set the robust factor to 2, which is equivalent to natural weighting and which is best for detecting faint emission in ALMA data.  To further optimize the images for the detection of faint line or continuum emission, we used the Hogbom deconvolver.  Additionally, a {\it uv} taper of 0.05~arcsec was applied to remove noise with very high spatial frequencies.  The channel width in the image cubes was adjusted to attempt to maximize the peak signal-to-noise of the beam while preserving as much spectral structure of the line as possible.  The final images have pixel scales that sample the full-width at half maxima (FWHM) of the beams by at least $5\times$; the specific pixel scales and reconstructed beam sizes are listed in Table~\ref{tab:observationSetup}.  This table also lists the fields of view in the ALMA images in each band, the maximum recoverable scales, and the calibration uncertainties, which are based on information from the ALMA Technical Handbook\footnote{Available from \url{https://almascience.eso.org/documents-and-tools/cycle10/alma-technical-handbook}.} \citep{Cortes2023}.

\section{Results}
\label{sec:Results}
The ANGELS observations resulted in a high-angular resolution spectral perspective on sixteen Herschel galaxies, across the frequency range from ~200 to 2000 GHz in the rest-frame. In this Section, we present the main results on the underlying dust continuum (Section~\ref{sec:ContSnapshotsResults}) and the line measurements (Section~\ref{sec:lineMeasurements}). 

\subsection{Continuum snapshots}
\label{sec:ContSnapshotsResults}
Figure~\ref{fig:cont1} displays the dust continuum images for all the sources that have complete coverage in Bands 3 through 8. 
% Figures~\ref{fig:cont1} show the dust continua of the pilot targets for all sources that have complete coverage in Bands 3 through 8. 
We further present sources separated beyond the primary beams of the higher-frequency bands in Appendix Figure~\ref{fig:offsetSources}. 
The sources have starkly varying morphologies, with lensing structures such as (partial) Einstein rings seen for HerBS-21, -22, -25, -36, -41A, -87, -104, and -155. HerBS-21 has a much more complicated gravitational lensing morphology when compared to the other sources, which are single galaxy-galaxy lensed systems (see Section~\ref{sec:lensingAmongHerBSSources}). Other sources are resolved into bright single components, such as HerBS-42A and -B, -81A, -86, -93, -170 and -184. HerBS-81B and -106 (and potentially -170) are resolved into multiple components on the scale of $< 1.5$~arcsec, and were seen as individual clumps in the BEARS observations (discussed further in Section~\ref{sec:hyperluminousGalaxies}). Since the maximum recoverable scale (MRS) is between 4 and 1.5~arcseconds for the ALMA observations (see Table~\ref{tab:observationSetup}), we expect this to play little effect in the final imaging quality.

\begin{figure*}
    \centering
    \includegraphics[width=0.9\linewidth]{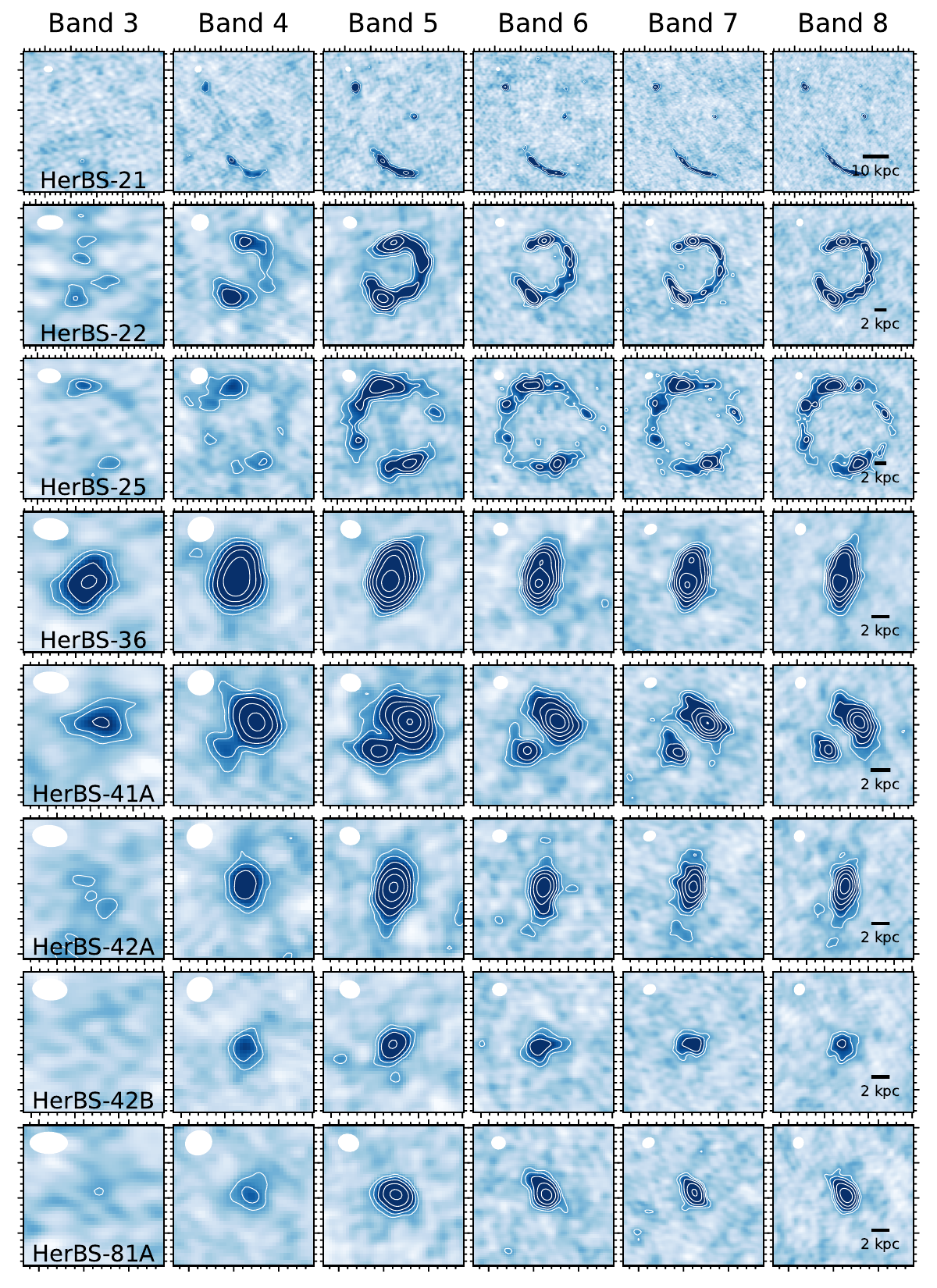}
    \caption{The dust continua of HerBS-21 through -81A are shown as white contours (3, 5, $8 \sigma$ and beyond, following the Fibonacci sequence, see Table~\ref{tab:observationSetup} for the $\sigma$-values) between Bands~3 through 8 from left to right. The beam size is shown as a white ellipsoid in the top-left, and the scale bar is shown in the bottom-right of the Band~8 image, using the proper distance at the redshift of the {\it Herschel} source. The images are scaled to contain the source, using a 7~arcsec square poststamp for HerBS-21 field, a 3~arcsec poststamp for HerBS-22 and -25, and a 2~arcsec poststamp for HerBS-36 through 81A. For fair comparison, we show the images before primary-beam corrections, as these effects could vary across the different bands.   }
    \label{fig:cont1}
\end{figure*}\addtocounter{figure}{-1}

\begin{figure*}
    \centering
    \includegraphics[width=0.9\linewidth]{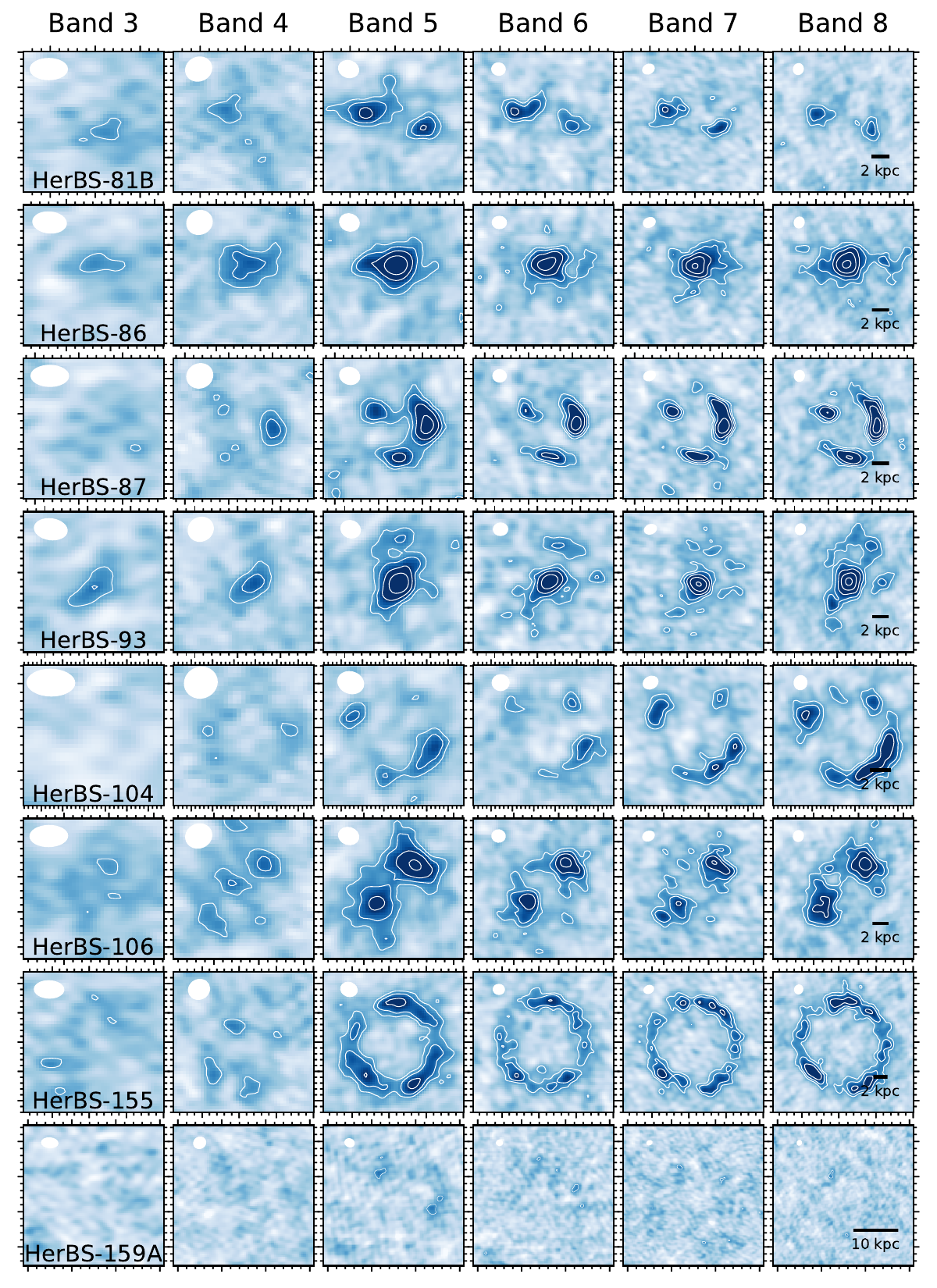}
    \caption{\textit{Continued.} The dust continua of HerBS-81B through -159A are shown as white contours (3, 5, $8 \sigma$ and beyond, following the Fibonacci sequence, see Table~\ref{tab:observationSetup} for the $\sigma$-values). The images are scaled to contain the source, using a 2~arcsec square poststamp for all fields, except a 1.6~arcsec square poststamp for HerBS-104, a 2.4~arcsec square poststamp for HerBS-155, and a 4~arcsec square poststamp for HerBS-159A. For fair comparison, we show the images before primary-beam corrections, as these effects could vary across the different bands. }
    \label{fig:cont2}
\end{figure*}

\begin{figure*}\addtocounter{figure}{-1}
    \centering
    \includegraphics[width=0.9\linewidth]{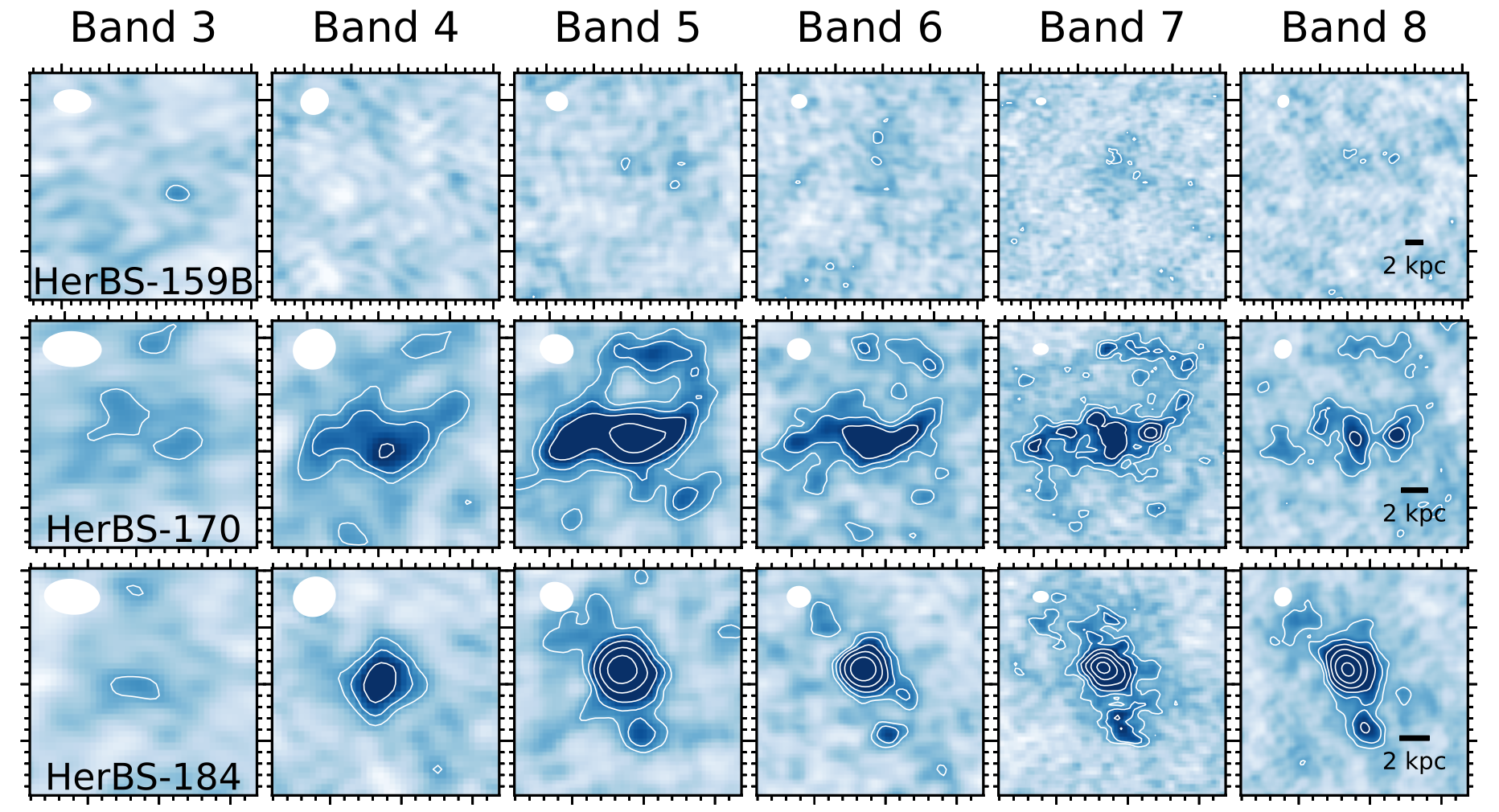}
    \caption{\textit{Continued.} The dust continua of HerBS-159B through -184 are shown as white contours (3, 5, $8 \sigma$ and beyond, following the Fibonacci sequence, see Table~\ref{tab:observationSetup} for the $\sigma$-values). The images are scaled to contain the source, using a 3~arcsec square poststamp for HerBS-159B, and a 2~arcsec square poststamp for HerBS-170 and -184. For fair comparison, we show the images before primary-beam corrections, as these effects could vary across the different bands. }
    \label{fig:cont3}
\end{figure*}

\begin{figure*}
    \centering
    \includegraphics[width=0.9\linewidth]{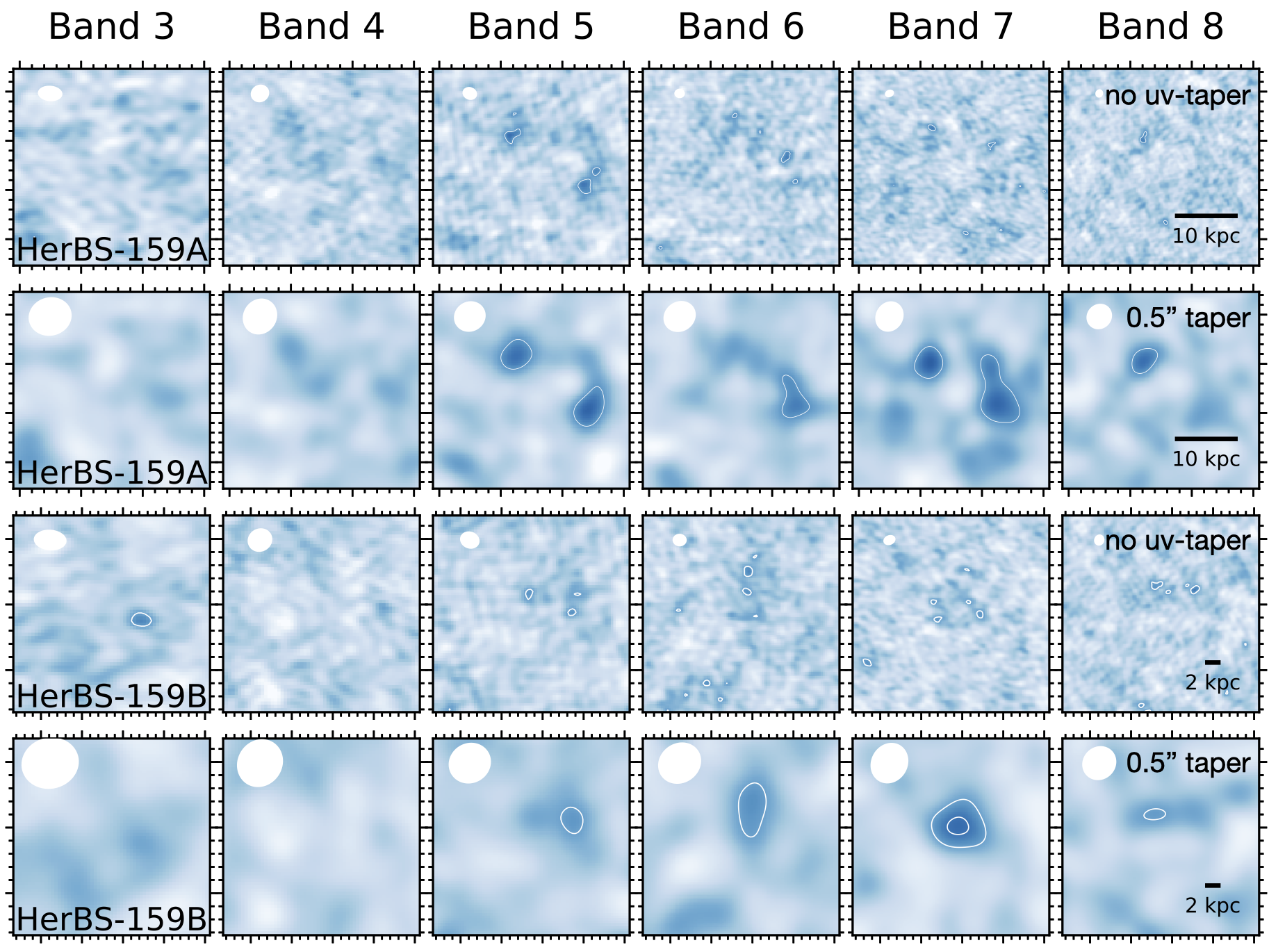}
    \caption{HerBS-159A and -B are undetected in the natural weighted maps. Instead, we create tapered images with an additional weighting term on the $uv$-data of 0.5~arcsec. These reveal the extended structure missed at higher resolutions. HerBS-159A appears to consist of multiple components separated by $\sim 1$~arcsec, while HerBS-159B appears to be a single bright component. Similar to Figures~\ref{fig:cont1}, the dust continua are shown as white contours (3, $5 \sigma$) between Bands~3 through 8 from left to right. The beam size is shown as a white ellipsoid in the top-left, and the scale length is shown in the bottom-right of the Band 8 image. For fair comparison, we show the images before primary-beam corrections, as these effects could vary across the different bands. }
    \label{fig:tapering}
\end{figure*}

Interestingly, HerBS-159A and -B are not seen in the imaging. For these sources, we create tapered images in Figure~\ref{fig:tapering} to see if any emission of these 500~\micron{}-bright sources is seen at these longer wavelengths. Tapering reveals the dust emission in Bands 5 and beyond. For HerBS-159A, the emission appears to be split across several sources separated at $\sim 1$~arcsec, while HerBS-159B appears to be a single bright component. We discuss the significance of these types of sources in Section~\ref{sec:spatiallyExtendedSources}.

The morphology of the ANGELS sources varies across the different bands, and provides a strong argument towards a multi-band interpretation of dusty galaxies, particularly if only lower frequencies are available. For example, even though the observations are at high enough resolution in Bands~3 and 4, the lensing nature of HerBS-41A remained unclear until the convincing arcs seen in Bands 7 and 8.

We provide the continuum spectral energy distribution fitting in a subsequent paper (Bendo et al. in prep.) that will investigate the nature of dust in these ANGELS sources.
In this paper, we employ two methods for calculating the far-infrared luminosity, which are broadly in agreement between one-another \citep{Bendo2023}. Firstly, where it is necessary to discuss the resolved luminosity (e.g., in the resolved star-formation law; Section~\ref{sec:resolvedKS}), we use a single-temperature modified black-body estimate of the far-infrared luminosity of these sources with 35~K and $\beta_{\rm dust} = 2$ \citep{Bendo2023} fit to the Band~7 flux densities of the sources, which typically have the highest signal-to-noise ratio. {\color{referee2} 
Secondly, in scenarios where the bulk luminosity is important (e.g., for the total intrinsic luminosity after lensing correction; Section~\ref{sec:eddingtonLimits}), we use the \textit{Herschel}-photometry to re-derive the luminosity by fixing the redshift to the $z_{\rm spec}$ \citep{bakx18,Bakx2020Erratum} assuming the spectrum of the famous Eyelash galaxy \citep[see e.g.][]{ivison16}. For sources resolved into multiple components, we distribute this luminosity based on the relative flux densities of the individual components as measured in Band~7. Note that this implicitly assumes these sources have the same colour temperature and are at the same redshift. In Appendix~\ref{sec:infraredLuminosities}, we provide the infrared luminosities derived using three methods of fitting, including the Eyelash template, the {\it Herschel}-derived template from \cite{pearson13,bakx18,Bakx2020Erratum}, and a single-temperature black-body. The latter also provides an estimate for the dust temperature and dust mass.
As a sanity check, when we investigate the luminosity of the sources based on the total Band~7 flux both assuming a constant dust temperature and a best-fit dust temperature, we find a comparable, albeit uncertain (a standard deviation of $\sim 50$~per cent) luminosity when compared to the second method, with a better fit of the constant dust temperature ($ = 35 K$) when compared to the best-fit dust temperature. A more thorough investigation of the ALMA and {\it Herschel} dust properties is planned for a subsequent paper (Bendo et al. in prep.).
}

\subsection{Line measurements}
\label{sec:lineMeasurements}
The main aim of this study is to perform efficient line surveys across a large sample of galaxies in an effort to out-perform the {\it point-tune-and-shoot} style observations that are the norm for extragalactic surveys with ALMA. We present the line results in Table~\ref{tab:spectralLineData} and show the spectral line profiles in Appendix Figures~\ref{fig:ANGELSSpectra}.

These sources are mostly resolved systems with line emission extending across multiple beams. Calculating the total line fluxes is therefore a non-trivial exercise, and we use the following procedure on the image-plane data to collect the properties of the emission lines across our spectrum.

In an effort to extract the total line fluxes, we tried several flux measurement techniques, including $2 \sigma$ moment-0 apertures (biased to including false positive 2$\sigma$ features and missing the $0 - 2 \sigma$ regions on the edges of the contours; \citealt{Stanley2023}) and dust continuum apertures (presuming dust and star-formation are coincident with line emission). Instead, we combine these methods into an algorithm that matches the highest-fidelity (band~7) observation to the cube resolution, and extending the $2 \sigma$ continuum contour by up to 3 beams of the cube. Effectively, we test the best aperture for each line to include as much flux as possible, favouring larger apertures to measure the total flux as accurately as possible at the potential cost of signal-to-noise ratio.

For sources where the apertures cover a large number of beams, an additional aperture selection criterion is used, based on the $2 \sigma$ contours of the moment-0 map before extending the aperture by up to 3 beams. This step guides the aperture to match the line emission, but also reduces the effect of $p$-hacking. Note that, in the case of absorption lines, typically no additional spatial smoothing is used, since the expected absorption is only seen at the location where continuum emission is seen. 

The subsequent spectrum is then fitted with a single Gaussian profile to provide a line estimate. We list all line properties in Table~\ref{tab:spectralLineData}, and provide the moment-0 and spectra of the detected lines in the Appendix Figures~\ref{fig:ANGELSSpectra}, where we further highlight the thresholds used to create the aperture (i.e., $\sigma_{\rm cont}, \sigma_{\rm mom-0}, N_{\rm beam}$).

\section{Lensing}
\label{sec:Lensing}
Nine sources show lensing features in their dust continuum. As gravitational lensing is a purely geometric process, it is achromatic, and we can use the $uv$-based lensing reconstruction of Band 7 -- which typically has the highest signal-to-noise detections for our sample -- to deduce the properties of our galaxies across the entire frequency range. 

\subsection{Lensing reconstruction}

In this section we give a general description of the methodology and models that we used to obtain source-plane reconstructions. We perform this lens modelling (lens mass model optimization and background source reconstructions) using the open-source software {\tt PyAutoLens}, which is described in \citet{2015MNRAS.452.2940N, 2018MNRAS.478.4738N, 2021JOSS....6.2825N} and builds on the methods from previous works in the literature \citep[e.g.][]{2003ApJ...590..673W, 2006MNRAS.371..983S}. 
We perform lens modelling in the visibility (i.e., $uv$) plane
\citep[e.g.][]{2018MNRAS.475.3467E, 2022MNRAS.512.2426M, 2021MNRAS.501..515P}.  

The analysis pipeline is broken down in two phases. In the first phase we fit a parametric model for the source (i.e. a Sersic profile) while in the second phase we switch to pixelated reconstructions. In both phases the mass model for the lensing galaxy is a Singular Isothermal Ellipsoid (SIE).

{\color{referee} The analysis pipeline was divided into two phases for both technical and observational reasons. Technically, the non-linear search involving a pixelated reconstruction, which is computationally intensive, benefits from a robust initial estimate of the lens's mass model to converge more efficiently. While a parametric model of the source may not fully capture the complexity of the background source, it provides a reliable initial mass model. This model, derived from the parametric phase, serves as a prior for the subsequent pixelated phase. Observationally, at higher resolutions, complex structures in some sources emerge that cannot be accurately modeled with a parametric approach. In such cases, relying on a parametric model could introduce biases in the lens parameters and, by extension, in the derived magnification factors, which are primarily what we aim to determine from the lensing analysis in this paper. In the sub-sections that follow we give more details about the various models that are used in our analysis.} 

\subsubsection{Lens}\label{sec:lens}

The convergence of the SIE model is defined as,
\begin{equation}
    \kappa(x, y) = \frac{1}{1 + q} \left( \frac{\theta_{\rm E}}{\sqrt{x^2 + y^2 / q^2}} \right)^{-1} \, ,
\end{equation}
where $q$ is the axis ratio (minor to major axis) and $\theta_{\rm E}$ is the Einstein radius. Our model also has additional free parameters that control the position of the centre ($x_c, \, y_c$) and its position angle, $\theta$, which is measured counterclockwise from the positive x-axis. In practise we parametrise the model's axis ratio, $q$, and position angle, $\theta$, in terms of two components of ellipticity,
\begin{equation}
    e_1 = \frac{1 - q}{1 + q} {\rm sin}\left(2\theta\right) \, , \,\,\,\,\,
    e_2 = \frac{1 - q}{1 + q} {\rm cos}\left(2\theta\right) \, ,
\end{equation}
which help to prevent periodic boundaries and discontinuities in the parameter space, which are associated with the position angle and axis ratio, respectively. Finally, our model also has an additional component to model the potential presence of external shear and/or any mass distributions with more complexity than a single SIE \citep{Etherington2023}. This is typically described by two parameters, its magnitude, $\gamma_{\rm ext}$, and position angle, $\theta_{\rm ext}$, but here we express them in terms of $\gamma_{\rm ext, 1}$ and $\gamma_{\rm ext, 2}$, with relations:
\begin{equation}
    \gamma_{\rm ext} = \sqrt{\gamma^2_{\rm ext, 1} + \gamma^2_{\rm ext, 2}} \, , \,\,\,\,\,
    {\rm tan}2\theta_{\rm ext} = \frac{\gamma_{\rm ext, 2}}{\gamma_{\rm ext, 1}} \, .
\end{equation}

\subsubsection{Source}\label{sec:source}

As mentioned before two different approaches are used to represent the background source. The first approach is to use a parametric profile, specifically a Sersic \citep{1963BAAA....6...41S}, which is the most common type of profile that is used to describe the light distribution of galaxies. This has a functional form that is given by,
\begin{equation}
    I(r) = I_{\rm e} \exp{} \left\{-b_{\rm n} \left[ \left(\frac{r}{r_{\rm e}}\right)^{1/n} 
 - 1\right] \right\}\, ,
\end{equation}
where $r_e$ is the effective radius, $I_{\rm e}$ the intensity at the effective radius, $n$ the Sersic index and $b_{\rm n}$ a constant parameter that only depends on $n$ \citep{2004AJ....127.1917T}. 

The second approach is to use a pixelated grid to reconstruct the source's surface brightness distribution, independent of the parametric method described above. For this we use an irregular Voronoi
mesh grid \citep[e.g.][]{2015MNRAS.452.2940N}. This irregular grid has the advantage that adapts to the magnification pattern in the source-plane, meaning that more source-plane pixels are dedicated to regions with higher magnifications (since the resolution will also be higher in these regions, roughly scaling as $\mu^{-1/2}$) compared to regions with lower magnification. In addition, it is preferred in cases where the morphology of the background source is complex and can not be described with simple parametric profiles (e.g. Sersic). When reconstructing a source on a pixelated grid it is also necessary to introduce some form of regularization in order to avoid over-fitting \citep[e.g.][]{2006MNRAS.371..983S}. In this work we use a constant regularization scheme, where the level of regularization is controlled by the regularization coefficient, $\lambda_{\rm reg}$. 

\begin{figure*}
    \centering
    \includegraphics[width=0.975\textwidth]{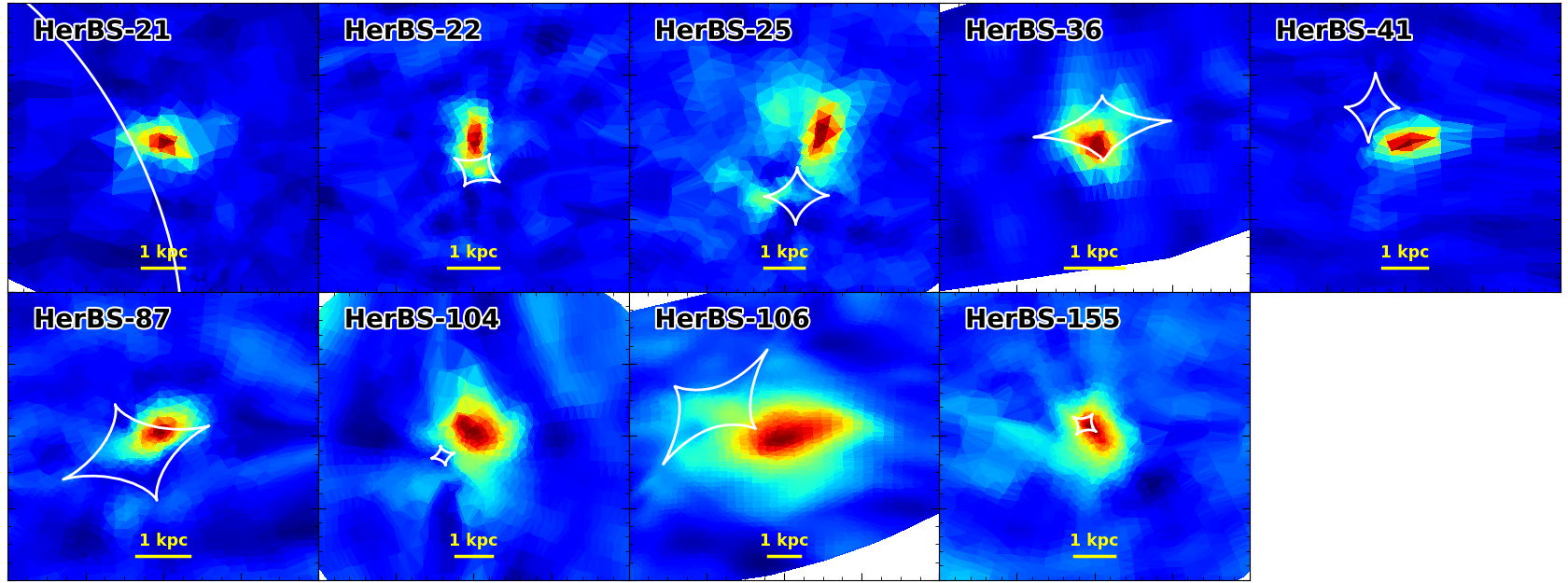}
    \caption{Source-plane reconstructions for the 9 confirmed strong lenses in our sample. The gray curve in each panel of the figure shows the caustic curve while the white scalebar at the bottom corresponds to a physical scale of 1 kpc.
    }
    \label{fig:lensingreconstruction}
\end{figure*}

As seen in Figure~\ref{fig:lensingreconstruction}, some sources display extended structures at faint flux levels, possibly indicating interactions. However, the majority of sources show simple morphologies, in particular at the brighter flux levels.
We can therefore use our best-fit parametric profiles to get an estimate of the sizes of the star-forming regions, which are presented in Table~\ref{tab:intrinsicProperties}. For the same reason we also use the parametric fits to estimate magnifications, also presented in Table~\ref{tab:intrinsicProperties} as the ratio of the image-plane to source-plane flux. 

\begin{table}
    \caption{Intrinsic properties of ANGELS sources}
    \label{tab:intrinsicProperties}
    \centering
    \begin{tabular}{lccc} \hline \hline
    Source & Magnification & \multicolumn{2}{c}{Intrinsic source radii}\\
    & $\mu \pm \sigma_{\mu}$ & [mas] & [kpc] \\ \hline
HerBS-21 		&  9.0  $\pm$ 1.8  & $121_{-10}^{+13}$  & $0.924_{-0.080}^{+0.100}$ \\
HerBS-22 		& 18.8  $\pm$ 3.7  & $38_{-1}^{+1}$  & $0.301_{-0.009}^{+0.009}$ \\
HerBS-25 		&  9.2  $\pm$ 1.8  & $279_{-27}^{+46}$ & $2.217_{-0.222}^{+0.368}$  \\
HerBS-36 		&  4.1 $\pm$ 0.8     & $86_{-3}^{+3}$ & $0.672_{-0.023}^{+0.027}$  \\
% HerBS-36$^*$ 	& {\it 1.0} & 371 $\pm$ 7 & 2.90 $\pm$ 0.05 \\
HerBS-41 		&  2.6  $\pm$ 0.5  & $90_{-2}^{+2}$  & $0.633_{-0.018}^{+0.018}$ \\
HerBS-42A$^*$   & {\it 1.0} & 265 $\pm$ 9 & 2.02 $\pm$ 0.07 \\
HerBS-42B$^*$   & {\it 1.0} & 287 $\pm$ 38 & 2.19 $\pm$ 0.29 \\
HerBS-81A$^*$   & {\it 1.0} & 301 $\pm$ 70 & 2.33 $\pm$ 0.54 \\
HerBS-81B$^*$   & {\it 1.0} & 322 $\pm$ 63 & 2.64 $\pm$ 0.52 \\
HerBS-86$^*$    & {\it 1.0} & 422 $\pm$ 28 & 3.47 $\pm$ 0.23 \\
HerBS-87 		&  8.7  $\pm$ 1.7  & $49_{-2}^{+3}$  & $0.417_{-0.021}^{+0.024}$ \\
HerBS-93$^*$    & {\it 1.0} & 284 $\pm$ 24 & 2.36 $\pm$ 0.20 \\
HerBS-104 		& 6.3  $\pm$ 1.2  & $83_{-2}^{+5}$  & $0.721_{-0.024}^{+0.046}$ \\
HerBS-106 		& 2.3  $\pm$ 0.4  & $239_{-15}^{+16}$ & $2.001_{-0.129}^{+0.131}$  \\
HerBS-155 		& 28.3 $\pm$ 5.6  & $71_{-3}^{4}$ & $0.553_{-0.027}^{+0.030}$  \\
HerBS-159A$^*$  & {\it 1.0} & 1415 $\pm$ 530 & 11.9 $\pm$ 4.48 \\
HerBS-159B$^*$  & {\it 1.0} & 433 $\pm$ 189 & 3.65 $\pm$ 1.60 \\
HerBS-170$^*$   & {\it 1.0} & 952 $\pm$ 67 & 6.64 $\pm$ 0.47 \\
HerBS-184$^*$   & {\it 1.0} & 246 $\pm$ 14 & 2.03 $\pm$ 0.12 \\ \hline
    \end{tabular}
    \raggedright \justify \vspace{-0.2cm}
\textbf{Notes:} 
Col. 1: Source name.
Col. 2: Magnification and uncertainties. Numbers in italics assume no lensing magnification as is apparent in the $\sim 0.1$~arcsec Band~7 observations. 
Col. 3: The source-plane size estimates in units of milli-arcseconds.
Col. 4: The source-plane size estimates in units of kpc.
The source sizes were calculated for both Sersic profiles of the lensed sources, which is left as a free parameter. The {\sc IMFIT} routine is used for the non-lensed sources, which are indicated by a star ($^*$). 
\end{table}

\subsection{Multi-wavelength perspective}
\begin{figure*}
    \centering
    \includegraphics[width=\linewidth]{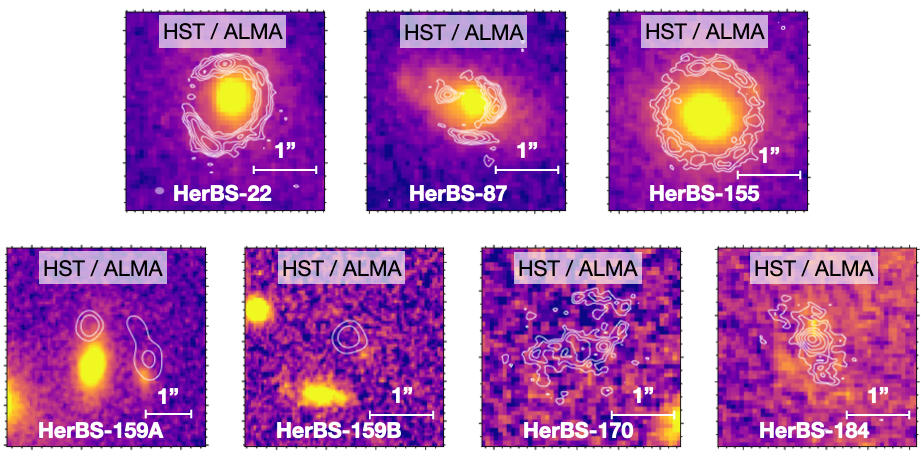}    
    \caption{\textit{Hubble} JH-band snapshot data is available for seven of our target galaxies, which provide a high-resolution imaging of several lensed sources, particularly on the top row. The contours of Band~7 continuum emission are drawn at levels of 3, 5, $8 \sigma$ and beyond, following the Fibonacci sequence (see Table~\ref{tab:observationSetup} for the $\sigma$-values). HerBS-159A, a source with extended emission requiring tapering to image, is shown to lie in front of a foreground galaxy which is likely causing a gravitational lensing scenario. Gravitational lensing could be a likely explanation for the extended emission seen in this source. }
    \label{fig:HSTcomparison}
\end{figure*}

In total, six ANGELS fields have \textit{Hubble} imaging from surveys reported in \cite{Borsato2023}, shown in Figure~\ref{fig:HSTcomparison}. These \textit{Hubble} observations use JH-band snapshot-style observations between 7 and 11~minutes to reveal the foreground structure of these galaxies. The near-complete Einstein rings of HerBS-22, -87 and -155 are shown to lie directly behind foreground lensing galaxies. In fact, HerBS-155 was already identified as a gravitational lens through a foreground removal study \citep{Borsato2023}, where the lensing features were already apparent in the pre-subtracted images. They find a comparable lensing magnification ($\mu \approx 20 - 21$) to the ALMA-based lensing magnification of $\mu = 28.3 \pm 5.6$, although they are derived independently. The astrometry of the \textit{HST} images is corrected using \textit{GAIA} stars and the imaging is thus accurate to $< 0.1$~arcsec. Given the high astrometric accuracy of ALMA\footnote{Estimated to lie below $0.05$~arcsec for $> 5 \sigma$ observations in Band~7 using eq. 10.7 in \url{https://almascience.eso.org/documents-and-tools/cycle10/alma-technical-handbook}.}, the relative astrometric accuracy between \textit{HST} and ALMA imaging thus lies below 0.1~arcsec.
The sources with extended emission (HerBS-159A and -B) show a complex foreground system. This suggests that extended emission, particularly in HerBS-159A, could be caused by gravitational lensing extending across multiple arcseconds. Unfortunately, the signal-to-noise of HerBS-159A does not allow for a robust lensing model to test this hypothesis.

\begin{figure*}
    \centering
    \includegraphics[width=\linewidth]{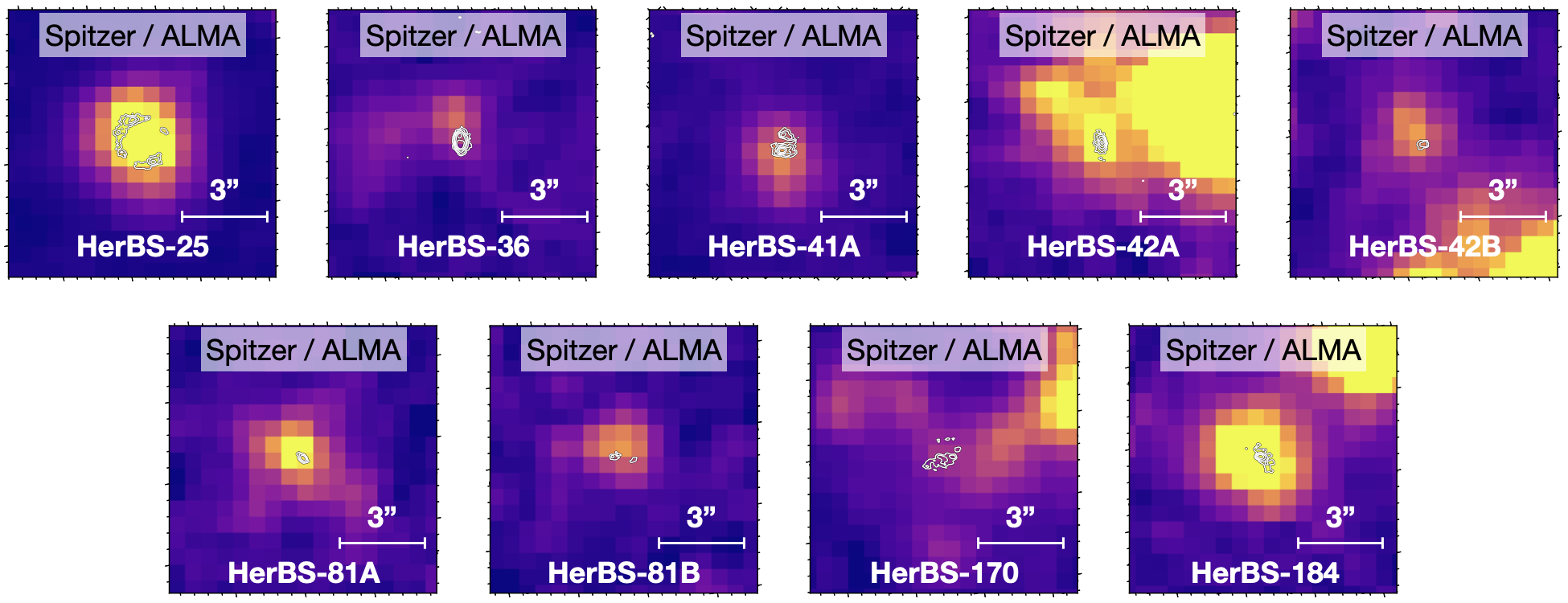}
    \caption{\textit{Spitzer} 3.6~$\mu$m data is available for seven of our ANGELS fields (nine sources in total), which are shown with the \textit{Spitzer} imaging in the background, and the ALMA Band~7 contours in the foreground (at 3, 5, $8 \sigma$ and beyond according to Fibonacci's sequence). The majority of sources lie close to, or on top of a bright \textit{Spitzer} galaxy. The nature of sources with obvious lensing features (e.g., HerBS-25) is not in doubt, but it is unclear whether sources such as HerBS-170, or -184 are lensed from the current ANGELS and multi-wavelength imaging. }
    \label{fig:SpitzerComparison}
\end{figure*}

On top of the \textit{HST} fields, seven {\it Herschel} fields (nine ALMA sources including the multiplicity) have \textit{Spitzer} imaging. Note that two of these sources (HerBS-170 and -184) also have \textit{HST} imaging, shown in Figure~\ref{fig:SpitzerComparison}. The image PSFs are wider, complicating the astrometric correction ($\sim 1$"), and as a result, provide a less clear picture of the lensing nature of several galaxies. HerBS-25 and -41A, together with the \textit{HST}-identified Einstein rings, are shown to lie directly on top of a foreground galaxy. 
HerBS-36 is likely lensed with a very small Einstein radius (predicted down to $\sim 0.2$~arcsec, \citealt{Amvrosiadis2018}). The continuum emission is a barely-resolved clump with a double-peaked internal structure of $\approx 3 \times{} 7$~kpc, and the southern clump appears to arc (see Appendix Figure~\ref{fig:RGBAngels}).
The lensing nature of the HerBS-42, -81, and -184 systems remain unclear. The emission of HerBS-42A and -B coincide with \textit{Spitzer}-bright emission, although the emission does not appear to resemble an obvious lensing feature. 
The double emission peaks of HerBS-81B are close to a \textit{Spitzer}-imaged source, although the geometry is not reminiscent of a lensing system. 
While the \textit{HST} image of HerBS-184 does not suggest a bright foreground system, \textit{Spitzer} shows a bright source adjacent to the ALMA emission. Meanwhile, the continuum emission of HerBS-184 is clumpy and extended, although the lack of a counter-image likely rules out lensing. 
HerBS-170 is not seen in \textit{HST} imaging, although some faint \textit{Spitzer} emission appears visible at the position of HerBS-170. The spatial association of nearby objects does not suggest any lensing association. The continuum emission is clumpy and extended, particularly in Band 5 ($\approx 10 \times{} 12$~kpc; see Table~\ref{tab:intrinsicProperties}). As there is no nearby lensing galaxy seen for HerBS-170, and no counter-image is seen, HerBS-170 is thus likely not lensed.


\section{Spectral line properties}
\label{sec:SpectralLineProperties}
\begin{table}
    \centering
    \caption{Detected emission lines}
    \label{tab:summaryDetections}
    \begin{tabular}{lcr}\hline\hline
    Line & Det. & Sources \\ \hline
    % \multicolumn{3}{c}{\bf HerBS-81A z = 3.160} \\
     \multicolumn{3}{c}{\bf 54 Emission lines} \\ \hline
CO (3--2)                       & 3 & 87; 93; 184 \\
CO (4--3)                       & 5 & 22; 36; 81A; 93; 155 \\
CO (5--4)                       & 5 & 21; 25; 41A; 42A; 42B \\
CO (6--5)                       & 7 &  41A; 81B; 86; 87; 93; 106; 170  \\
CO (7--6)                       & 5 & 25; 81A; 81B; 86; 106 \\
CO (8--7)                       & 2 & 36; 155 \\
CO (9--8)                       & 4 & 21; 41A; 42A; 42B \\
CO (10--9)                      & 3 & 25; 36; 41A \\
CO (11--10)                     & 2 & 21; 42A \\
CO (13--12)                     & 1 & 41A \\
CO (18--17)                     & 1 & 41A \\
$[\textsc{C\,i}]$ $370 \mu m$   & 2 & 93; 106  \\
$[\textsc{C\,i}]$ $609 \mu m$   & 3 & 93; 104; 106 \\
$[\textsc{N\,ii}]$ $205 \mu m$  & 2 & 170; 184 \\
$[\textsc{O\,i}]$ $145 \mu m$   & 1 & 41A \\
H$_2$O ($2_{0,2} - 1_{1,1}$)    & 2 & 41A, 184 \\
H$_2$O ($2_{1,1} - 2_{0,2}$)    & 1 & 25 \\
H$_2$O ($3_{1,2} - 3_{0,3}$)    & 1 & 25 \\
H$_2$O ($3_{2,1} - 3_{1,2}$)    & 2 & 81A; 170 \\
H$_2$O$^+$ ($1_{1,1} - 0_{0,0}$)& 2 & 22; 25 \\ \hline
    \multicolumn{3}{c}{\bf 12 Absorption lines} \\ \hline
CH$^+$ (1$-$0)                  & 3 & 21; 22; 36 \\
CH$^+$ (2$-$1)                  & 1 & 36 \\
OH$^+$ ($1_{0} - 0_{1}$)        & 1 & 22 \\
OH$^+$ ($1_{1} - 0_{1}$)        & 4 & 21; 41A; 42A; 104 \\
OH$^+$ ($1_{2} - 0_{1}$)        & 3 & 21; 22 \\
\hline
    \end{tabular}
\end{table}
The ANGELS observations have revealed a multitude of lines, consisting of 54 emission and 12 absorption lines, summarized in Table~\ref{tab:summaryDetections}.
%: i) 54 emission lines with 38 CO emission lines covering the levels J=3-2 to J=18-17, seven ${\rm H_2O}$ and H$_2$O$^+$ emission lines, two \nii{} 205$\mu$m lines, one \oi{} 145$\mu$m line, and five \ci{}(2-1) emission lines; ii) 12 molecular absorption lines with eight from OH$^+$ and four from CH$^+$. In this section, we discuss the CO Spectral Line Energy Distribution (SLEDs), the molecular absorption lines tracing in- and outflow activity, briefly analyse the detection of the \nii{} and \oi{} emission lines and present the results of the stacked spectrum. 
In this section, we discuss the CO Spectral Line Energy Distributions (SLEDs; Section~\ref{sec:COSLEDs}), the in- and outflow distribution of our sample seen in absorption lines (Section~\ref{sec:absorptionLines}), a brief investigation of the atomic lines (Section~\ref{sec:AtomicLines}), and a stacked spectrum (Section~\ref{sec:stackedSpectrum}). These investigations show the wealth of data from this modest project, {\color{referee} and detailed examination of each subject is planned for subsequent studies.}

\subsection{CO Spectral Line Energy Distributions}
\label{sec:COSLEDs}
The rotational transitions of Carbon Monoxide (CO) have proved a particularly good tracer of the internal structure of DSFGs {\color{referee}(e.g., \cite{Rizzo2024} and references therein)}. CO is the second-most abundant molecule in the Universe after molecular Hydrogen (H$_2$). However, unlike the a-polar H$_2$, it emits brightly from gas clouds fueling star-formation and is therefore a useful probe of the available star-forming gas \citep[e.g.,][]{carilli2013}. 
% DSFGs often appear to contain enough gas to continue star-formation (i.e., the gas-depletion timescale) for another 30 to 300~Myr \citep[e.g.,][]{Hagimoto2023}. 
Starting at its ground state at $115.270$~GHz, CO transitions scale linearly in frequency with higher rotational transitions. These so-called higher-$J$ transitions are sensitive to denser, warmer gas, and a comparison between the brightness of different CO transitions can illuminate the properties of the ISM. Recent comparisons of unresolved low-to-mid $J$ CO line ratios, often called SLEDs, find a large diversity among DSFGs \citep{Harrington2021,Sulzenauer2021,Hagimoto2023}, ranging from near-thermalized excitation to low-excitation conditions similar to the Milky Way \citep{Fixsen1999}.

\begin{figure*}
    \centering
\includegraphics[width=0.245\textwidth]{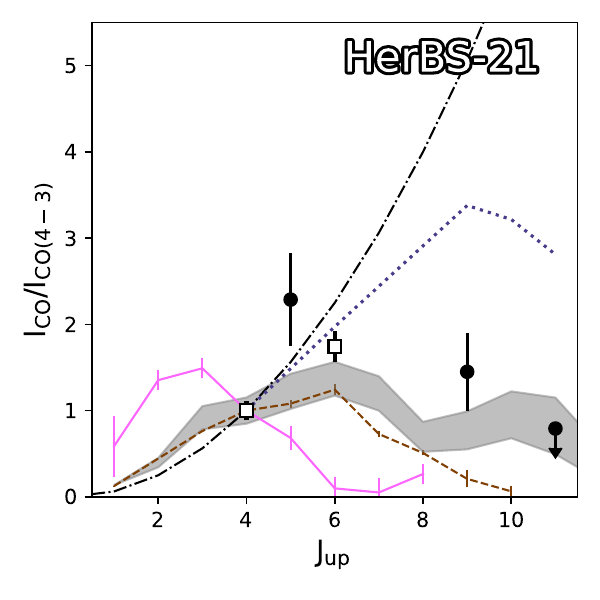}
\includegraphics[width=0.245\textwidth]{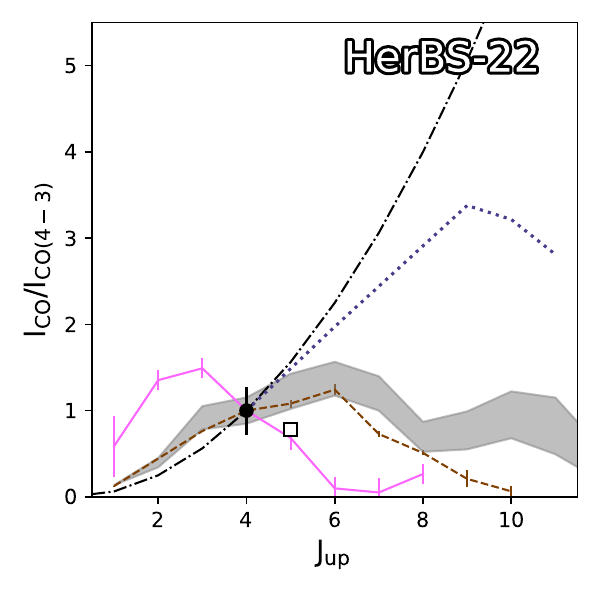}
\includegraphics[width=0.245\textwidth]{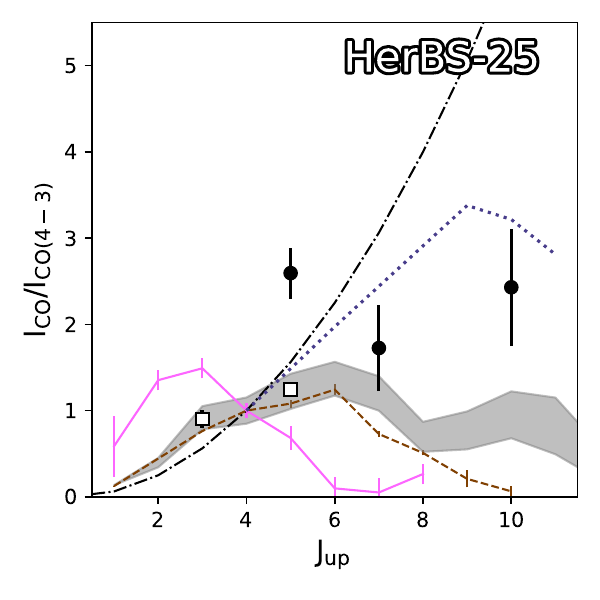}
\includegraphics[width=0.245\textwidth]{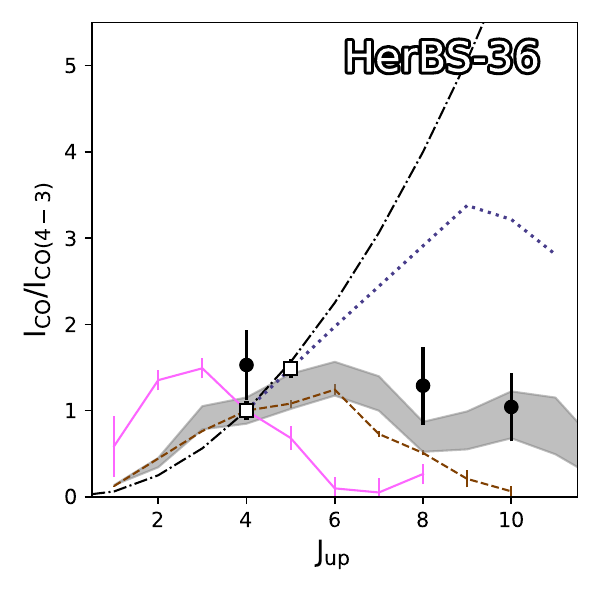}
\includegraphics[width=0.245\textwidth]{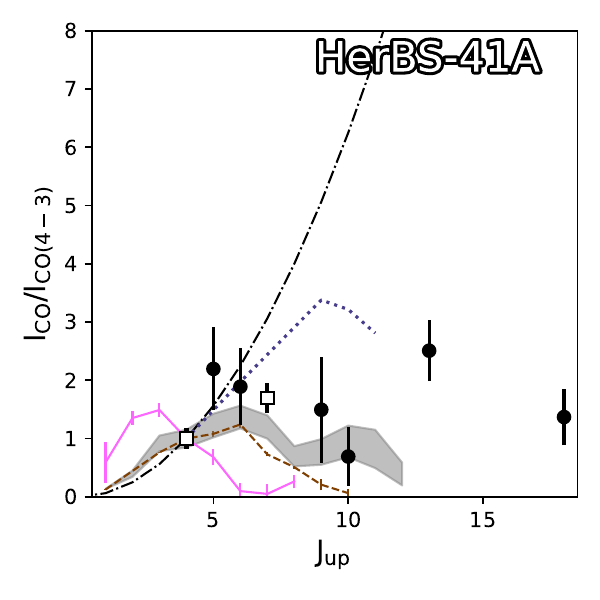}
\includegraphics[width=0.245\textwidth]{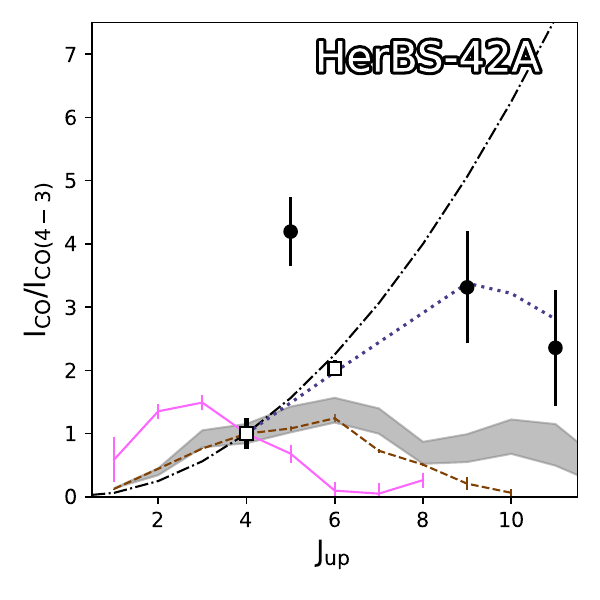}
\includegraphics[width=0.245\textwidth]{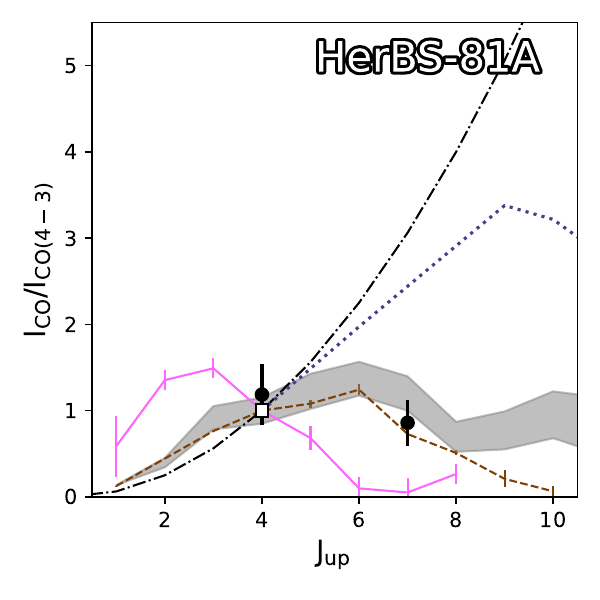}
\includegraphics[width=0.245\textwidth]{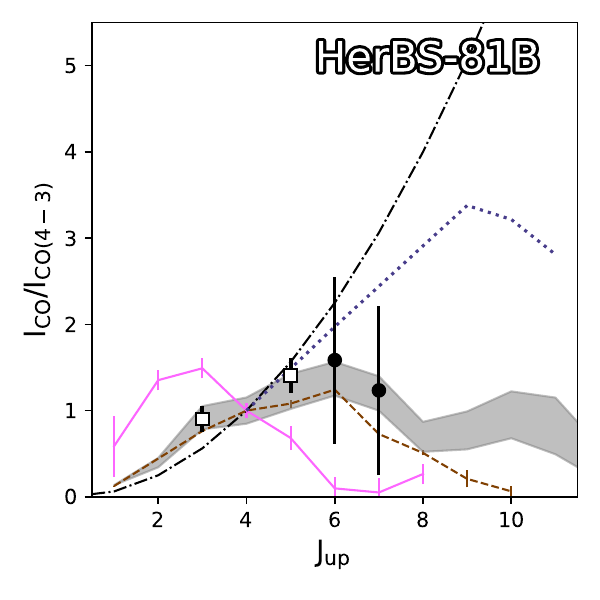}
\includegraphics[width=0.245\textwidth]{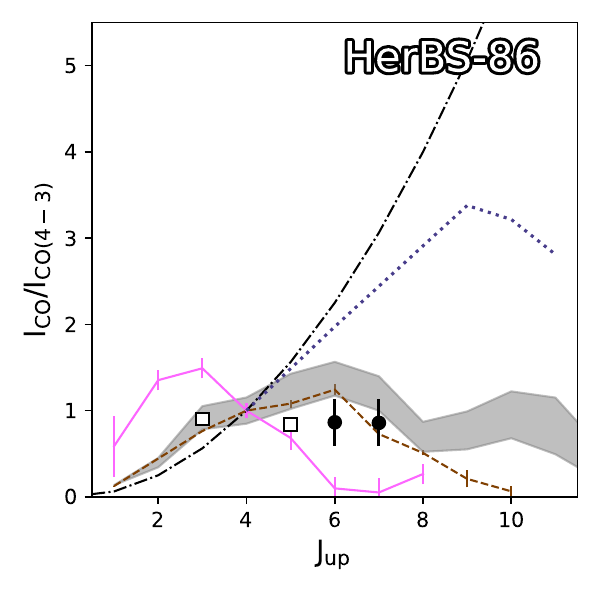}
\includegraphics[width=0.245\textwidth]{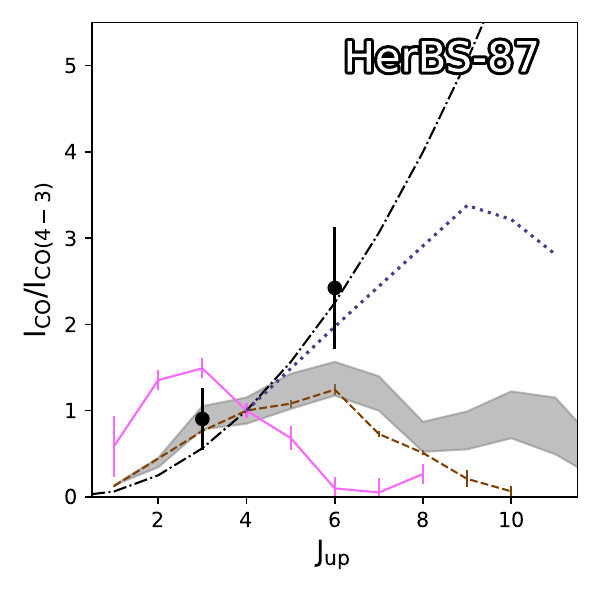}
\includegraphics[width=0.245\textwidth]{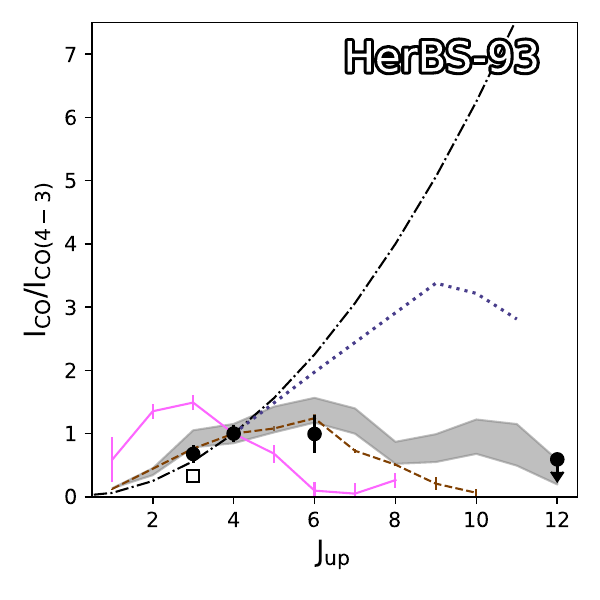}
\includegraphics[width=0.245\textwidth]{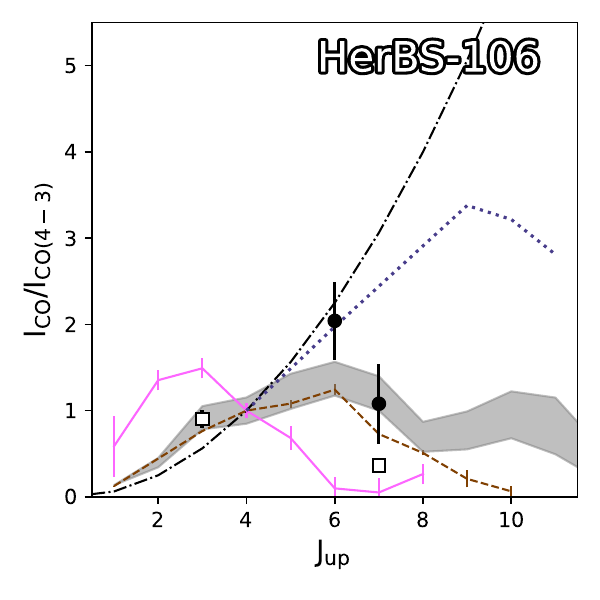}
\includegraphics[width=0.245\textwidth]{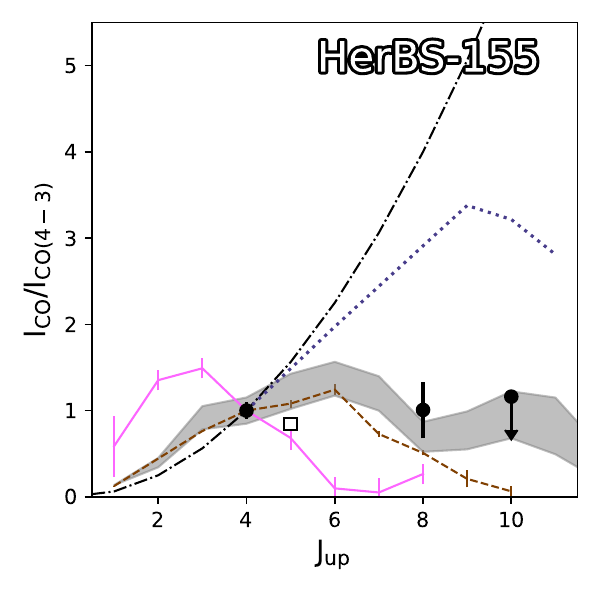}
\includegraphics[width=0.245\textwidth]{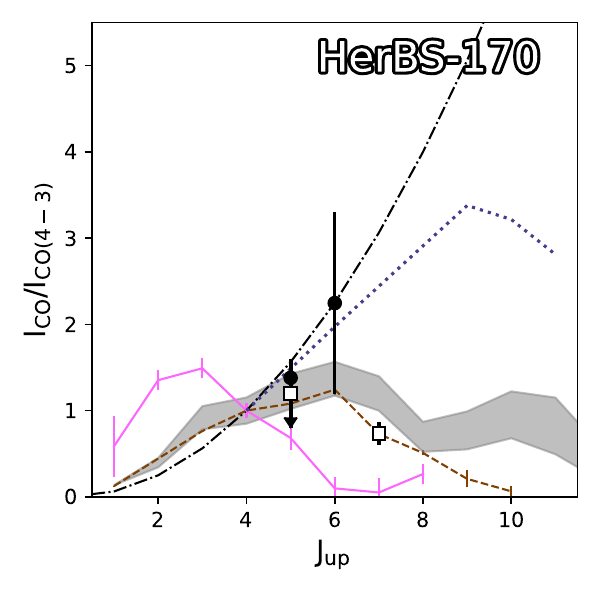}
\includegraphics[width=0.245\textwidth]{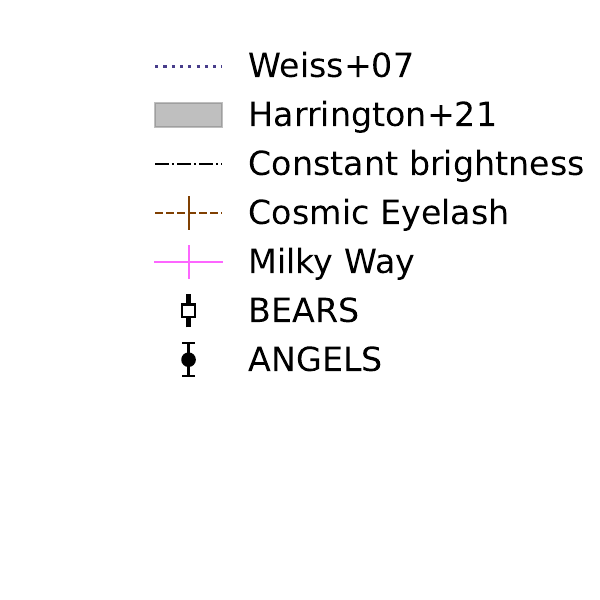}
    \caption{The CO SLEDs of all sources including the new CO line observations from the ANGELS survey and previous values from the BEARS survey \citep{Urquhart2022,Hagimoto2023} show the velocity-integrated line intensities of the CO lines normalized against the CO(4--3) line. If the CO(4--3) line is not available, we use the lowest-$J$ CO line and the stacked CO template from \citet{Harrington2021} to derive the expected CO(4--3) emission line. We compare CO SLEDs against the {\it thermalized} constant-brightness profile ($\propto J^2$; i.e., optically-thick in all lines), a highly-excited quasar \citep{Weiss2007}, the stacked profile by \citet{Harrington2021}, the Cosmic Eyelash \citep{swinbank2010,ivison2010}, and the Milky Way \citep{Fixsen1999}. 
    }
    \label{fig:cosleds}
\end{figure*}

The ANGELS observations provide detections and upper limits on both mid-$J$ ($J = 3 - 7$) and high-$J$ transitions ($J = 8 - 18$). This large spread in the CO $J$ levels allows us to qualitatively investigate the presence of dense gas in some of these sources.
We show the CO SLEDs in Figure~\ref{fig:cosleds} for the sources with CO lines from ANGELS. We normalize the CO SLEDs against the CO(4--3) observations, and if they are not available, we use the average CO SLED template of \citet{Harrington2021} to translate from the lowest-available $J$ transition to CO(4--3), and subsequently normalize against that transition. {\color{referee} The choice to normalize to CO(4--3) -- which is the most frequently detected CO line for our sample -- will best allow us to see variations between the CO SLEDs. Similarly, since the galaxies used to derive the CO SLED template from \cite{Harrington2021} are likely similar to the DSFGs in this {\it Herschel}-selected sample, the CO SLED should be an accurate representation of the typical behaviour of the ANGELS galaxies. } 

{\color{referee} 
Although the BEARS and ANGELS CO line fluxes agree on the whole, several sources show discrepancies in CO line fluxes, even when observed in the same CO transition. Appendix Figure~\ref{fig:compareBEARS_ANGELS} shows the velocity-integrated flux densties for the six galaxies with line measurements of the same CO transition in both BEARS and ANGELS, as well as the fluxes extracted from the resolved maps using the same method as BEARS. Careful re-examination of the BEARS data \citep{Urquhart2022,Hagimoto2023} and ANGELS data suggests these discrepancies could be caused by the different apertures and different approaches taken in the flux extraction. Firstly, in this paper we match the apertures to extract flux despite the additional noise that arises from increased apertures, while previous studies use a curve-of-growth approach that stops velocity-integrated flux density measurements at a $5 \sigma$ threshold, which appears to bias the observed fluxes low. 
{\color{referee3} The comparison between the $5 \sigma$ and the ANGELS methods} on the complete resolved data finds only a $4 \pm 8$~per cent systemic difference between the two methods (with modestly-larger fluxes in the ANGELS methodology), although there appears a large (45~per cent) variation for each line between the methods. This effect could thus influence the analysis of individual SLEDs. 
Secondly, we also note that there are consistently-found discrepancies between unresolved and resolved studies \citep{JvM1995,Czekala2021}, such as the flux estimates from ALMA large program ALPINE when compared to the resolved CRISTAL study \citep{Posses2024}, or the flux estimates reported in resolved observations of bright AGNs \citep{Novak2019}. The origin of these issues arise from the deconvolution technique in \textsc{TCLEAN} that produces a combined image of background noise (in units of Jy~/~dirty~beam) and signal (assuming a Gaussian point-spread function in units of Jy~/~clean~beam). 
As a consequence, the fluxes from resolved maps might produce a different total flux estimate than those from unresolved observations.
Regardless of this caveat, our subsequent interpretation does not depend solely on the BEARS data and we have confidence in our ANGELS-to-ANGELS CO SLED analysis presented below.}

As seen for other samples \citep{yang2017,Canameras2018,Stanley2023}, the CO SLEDs of sources vary strongly, with several sources showing evidence of large reservoirs of warm, dense gas (i.e., HerBS-25, -41A, -42A, -87, and -170), with HerBS-87 tentatively showing steep SLED features similar to the broad absorption line quasar APM 08279+5255 \citep{Weiss2007}. 
Meanwhile, the other sources (i.e., HerBS-21, -93) follow the trend reported in \cite{Harrington2021}. Some of these sources were not observed in high-$J$ CO transitions (HerBS-22 and -86), which impedes us to extend the interpretation towards all sources. Unfortunately, the non-comprehensive nature of this survey means that the current statistics are limited to only a handful of sources. 

The ANGELS observations are not deep enough to detect all lines, but they do indicate the broad statistics of the ANGELS {Pilot sample, \color{referee} such as the lensing fraction and the ratio of sources with high-$J$ CO lines.} The prevalence of dense, warm gas thus appears diverse across distant dusty galaxies, with $\sim 36 \pm 13$~per cent (using a binomial distribution; \citealt{Gehrels1986}) of sources showing such evidence, which could indicate the presence of an AGN \citep[e.g.,][]{vanderwerf2010,Rosenberg2015}. A visual inspection of Figure~5 of \cite{Harrington2021} seems to suggest they have a similar ratio of sources with extreme high-$J$ CO SLEDs. Broadly, their $\chi^2$-minimized turbulence models of the CO SLEDs report 5 sources with no emission at $J_{\rm up} = 10$, while 4 sources have bright CO(10--9) emission. 14 ANGELS sources fall between these two categories, indicating there is some dense gas, but that the sample is not dominated by dense gas.
Since the SLEDs with emission from higher-$J$ CO emission are distributed between lensed and unlensed sources, there does not appear a large effect of differential lensing \citep{serjeant2012,Serjeant2024}, and we note that a thorough characterization of the gas in ANGELS sources requires more extensive modelling beyond the scope of this paper (e.g., \citealt{Harrington2021}).

\subsection{Absorption lines}
\label{sec:absorptionLines}
These dusty galaxies probably lie at the centre of overdensities from quantum-fluctuations left over from the Big Bang \citep{Chiang2013,Chiang2017,Casey2016}. As the Universe expanded, filaments funneled gas onto these central nodes \citep[e.g.,][]{Springel2005,Dekel2009}, allowing for the formation of massive galaxies. This feeding subsequently triggers the onset of feedback mechanisms (\citealt{Man2018}) via Active Galactic Nuclei (AGN) heating \citep{Hlavacek2022hxga.book....5H} and star-formation driven winds \citep[e.g.,][]{Furlanetto2022}. Such dynamics are sometimes seen through wide or asymmetric emission lines, but outflow signatures are difficult to detect against the often-brighter systemic emission in galaxies due to velocity smearing effects and the multi-phase nature of outflows. 

Absorption lines offer a more unbiased and high-fidelity \citep{Spilker2018,Spilker2020a} approach at detecting gas dynamics inside galaxies, by searching for the shadow of an absorption line against the bright continuum of dusty sources. Since the absorption feature originates between the bright continuum-emitting regions and us (the observers), any velocity offset immediately reveals whether the gas is in- or outflowing. Consequently, absorption lines are an important tracer of the gas cycle of distant galaxies. 
To date, a handful of studies have characterized distant dusty galaxies using absorption lines such as the methylidyne cation \citep[CH$^+$;][]{Falgarone2017,Indriolo2018}, ionized hydroxyl \citep[OH$^+$;][]{berta2021,Butler2021,Butler2023,Riechers2021}, and hydroxyl \citep[OH~119$\mu$m;][]{Spilker2018,Spilker2020,Spilker2020a}.\footnote{Systemic emission of these lines, and neighbouring ones, can interfere with a complete interpretation of these lines, although this effect is often minimal.} The CH$^+$ ion is very unstable, formed through strongly endothermic processes, with a dissociation time of around 1~year. As such, it is only observed where it is created, in strongly-shocked regions. Meanwhile, OH$^{+}$ traces more diffuse molecular gas feeding star-formation. 
In total, around forty sources have been observed in sub-mm absorption lines, with around twenty-five sources showing absorption lines at velocities outside of the systemic velocity. The majority of these sources show signs of outflowing gas because of rapid star-formation or AGN feedback.
Meanwhile, inflows of gas are likely occurring as well, although they might be more difficult to image. Observations will need to see {\it down the barrel} of inflowing gas filaments, and as a result large samples tracing the delicate balance of gas in- and outflows are still lacking. Without these necessary observational constraints, hydrodynamical models of the gas flows struggle to benchmark mass loading factors (i.e., outflowing gas over star-formation rate) in the distant Universe, which are crucial to test the duration of star-forming events such as the extreme DSFGs studied in this paper.

The bright continuum emission from these galaxies, particularly in Bands 6 and beyond, allowed the detection of twelve absorption lines across six sources (HerBS-21, -22, -36, -41A, -42A, and -104), where two molecules were seen in absorption (CH$^+$ and OH$^+$). Several transitions of OH$^+$ are seen, namely the strong $(1_1$-$0_1)$ and $(1_2$-$0_1)$ ground-transitions, as well as the weaker $(1_0$-$0_1)$ transition for HerBS-22 \citep[see e.g.,][]{berta2021}. Meanwhile, this is only the second time that CH$^+$(2--1) is seen in the $z > 0.1$ Universe (\citealt{Bakx2024}), although we note that we might miss the $\Delta V < 0$~km/s component in the ANGELS observations. The ANGELS observations also covered multiple other lines that are sometimes seen in absorption, including water and ionized water. The sensitivity limits, and the limits to the bandwidth result in non-detections for the fainter species in absorption, particularly given the fainter continuum emission regions (i.e., below Band 5). Similarly, lower-significance line features such as P-Cygni might be hard to identify, with tentative cases seen in, for example, CH$^+$ in HerBS-21.

\begin{figure}
    \centering
    \includegraphics[width=\linewidth]{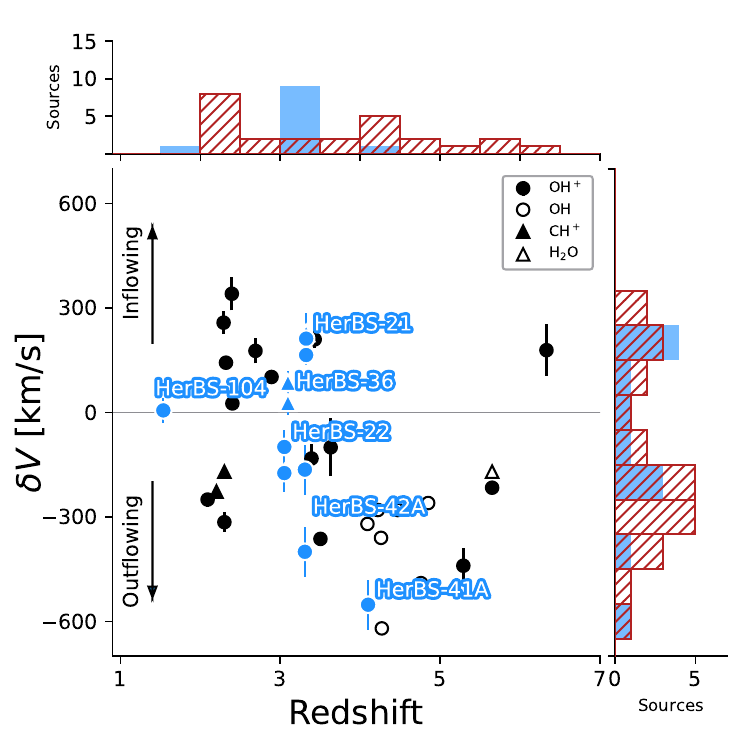}
    \caption{The redshift and relative velocity distribution between absorption lines and their intrinsic velocities from CO observations for the ANGELS sources, shown in blue. Filled triangles indicate CH$^+$ detections and circles indicate OH$^+$ observations. We compare against other high-redshift observations of absorption lines with a velocity shift \citep{Falgarone2017,Indriolo2018,Spilker2018,Spilker2020a,berta2021,Butler2021,Riechers2021}. About half of the sources are seen with inflowing gas, with the other half showing outflowing gas with velocities between $\pm 300$~km/s. The sources with multiple transitions (HerBS-21 and -22) find similar systematic velocity offsets for both their tracers, suggesting that the shocked gas traced by CH$^+$ and the more diffuse molecular gas traced by OH$^+$ originate from similar regions.
    }
    \label{fig:absorptionLines}
\end{figure}

We show an overview of the absorption lines in Figure~\ref{fig:absorptionLines} as a function of their redshift and relative velocity compared to the ANGELS-detected CO transitions. The sources are compared against other absorption line studies of $z > 1$ sources \citep{Falgarone2017,Indriolo2018,Spilker2018,Spilker2020a,berta2021,Butler2021,Riechers2021}. Three of the sources (-21, -36 and -104) have evidence of inflowing gas in their absorption lines, although only the relative velocities of HerBS-21 are outside of the error margins. There appears to be a relation between the velocity offset with redshift, transitioning between in- to outflowing gas at $z = 3$. However, since that this collation of data was obtained with different line tracers and observing differently-selected sources having a range intrinsic velocity widths, larger samples of observations with similar observational parameters (beam size, observation depth, tracers) are necessary before such investigations become more robust.  %Larger samples of observations with similar observation parameters (beam size, observation depth, tracers) are necessary before such investigations are more robust.

HerBS-21, -22, -36 and -42A have multiple molecules seen in absorption. Even though these can be sensitive to vastly different phases of the ISM and even the circum-galactic medium, they indicate similar velocity offsets relative to the host galaxy. This suggests that the shocked regions seen by CH$^+$ of HerBS-21 and HerBS-22 lie close to large volumes of low-density molecular gas traced by OH$^{+}$. The shocked gas in HerBS-36 seems to be red-shifted relative to the bulk molecular gas reservoirs, which could indicate rapid inflows encountering the strong star-forming environments. Given the large velocity widths of these sources (Table~\ref{tab:sources}), these are likely dynamically-complex systems, and the small relatively velocity offsets ($ |\delta V| < 500$~km/s) between the absorption and emission lines could obfuscate a more intricate interaction of the in- or outflowing gas and the host galaxy. 
In order to secure a good interpretation of these sources, deeper and higher-resolution observations on absorption lines in these sources are thus important. A more thorough investigation of in- and outflowing gas masses are beyond the scope of this paper, and they are strongly dependent on geometry arguments \citep[e.g.,][]{Butler2023}.

\subsection{Atomic lines}
\label{sec:AtomicLines}
Large gas clouds surround the ionizing emission of bright young O- and B-stars and shield the outer layers of the highest-energy photons. As a result, high-ionization states of specific elements (e.g., hydrogen, helium, oxygen and nitrogen) mostly emit from the central regions of these photo-dissociation regions (PDRs; \citealt{Tielens1985}), i.e., H$_{\rm II}$ regions. The PDR model of the ISM of galaxies in the distant Universe suggests an onion-like distribution of gas around ionisation sources such as bright, massive (O \& B) stars, where gas temperature decreases as the column density of neutral hydrogen increases. Lines from forbidden transitions such as atomic oxygen \oi{} and singly-ionized nitrogen \nii{} can trace these higher ionization rates and can be important cooling lines of galaxies \citep{Stacey2011,Vogelsberger2020}. Current studies of FIR atomic lines typically focus on the low-redshift Universe with now-defunct observatories such as \textit{Spitzer}, \textit{Herschel} \citep{Cormier2015,DiazSantos2017,Zhang2018} and SOFIA \citep{Ura2023}, or require high redshifts ($z > 4.5$; \citealt{Lee2019,Lee2021,LeFevre2020,Bouwens2022Rebels,Bakx2023glass}) to shift these lines in more favourable atmospheric transmission windows. The local studies often only detect a small fraction of the targeted sources, resulting in difficulties in interpreting the sample as a whole \citep{Bonato2014}, while the most distant searches probe a population different than the DSFGs located at the peak of cosmic evolution \citep{Madau2014}. ANGELS offers to increase the number of DSFGs in the cosmic noon with atomic line observations to better characterize the PDR conditions in the most violent star-bursts across cosmic time.

The highest frequency bands (Band 7 \& 8) covered atomic lines for four sources, namely \niil{} for HerBS-170 and HerBS-184 (in Band 7) and HerBS-81B (in Band 8), and \oil{} for HerBS-41A (in Band 7). The \nii{} emission line is detected in both HerBS-170 and HerBS-184. In the case of HerBS-81, only part of the \nii{} emission line was covered ($V > 50$~km/s) and hence not detected in the current data. The \oi{} emission line was detected in HerBS-41A, although we note that the total flux estimate might be underestimated since we probe only the $\Delta V < 150$~km/s region.
% In Figure 10, you should add ID141 (Cheng et al. 2020 ApJ 898, 33).  For your information, for ID141, muL_IR = 8.5+-0.3 10^13 Sun and mu L_NII = 5.43+_ 0.7 10^8 Sun

% In this study, the higher-frequency Bands were able to target three atomic lines, namely \niil{} for HerBS-170 and -184, and \oil{} for HerBS-41A. Finally, the Band~8 observations of HerBS-81 cover only the V~$> 50$~km/s region of the \nii{} line, but no emission is detected in the spectrum. Meanwhile, both nitrogen lines are detected in HerBS-170 and -184. The \oi{} emission for HerBS-41A is also not detected, and suggests it is too faint to be detected.

\begin{figure}
    \centering
    \includegraphics[width=0.35\textwidth]{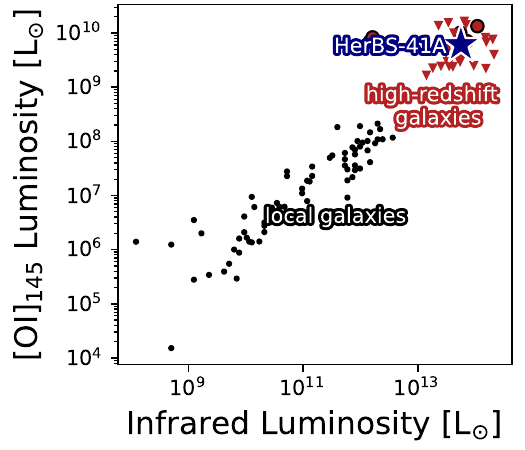}
    \includegraphics[width=0.35\textwidth]{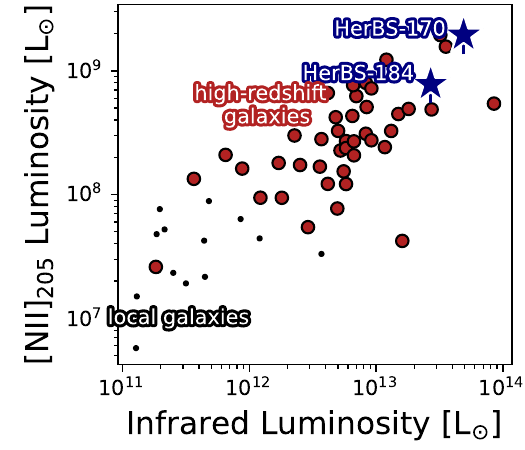}
    \caption{The luminosity scaling relation between \oi{} (top) and \niil{} (bottom) for local galaxies (normal star-forming systems, starbursts, and active galactic nuclei; \citealt{Brauher2008}), and high-redshift galaxies (for \oi{} \citealt{HerreraCamus2018}; for \nii{} \citealt{Cheng2020,Cunningham2020} and references therein). HerBS-41A is detected in \oi{} with the ANGELS observations, and the detections of \nii{} in HerBS-170 and HerBS-184 places them as the brightest Nitrogen emitters in the Universe observed to date. We note that the \oi{} flux estimate for HerBS-41A might be an underestimate since we are only sensitive to emission from the $\Delta V < 150$~km/s. There is no evidence of far-infrared line deficits, often seen in \cii{} high surface star-formation rate galaxies. }
    \label{fig:atomicLines}
\end{figure}

Figure~\ref{fig:atomicLines} shows the observed \oi{} and \nii{} line luminosity scaling relations of the three detected ANGELS sources, compared to local \citep{Brauher2008} and distant galaxies \citep{HerreraCamus2018,Cheng2020,Cunningham2020} from other surveys. We note that we do not scale our luminosity for the magnification of HerBS-41A, since we do not have lensing models of our spectral line. Our sources lie on the bright end of the scaling relations, and HerBS-170 qualifies as the brightest Nitrogen emitter observed to date. There is no evidence for a far-infrared line deficit, which appears to affect high star-formation surface-density galaxies \citep[e.g.,][]{Rybak2019}. As we build catalogues of atomic lines, these lines become invaluable to provide strong constraints of the gas densities \citep[e.g.,][]{Doherty2020}, ionisation parameters \citep{Hagimoto2023} and metallicities \citep[e.g.,][]{Tadaki2022} to characterize this diverse sample of DSFGs. 

\subsection{Stacked spectrum}
\label{sec:stackedSpectrum}
In an effort to reveal fainter line transitions, line stacking experiments combine the emission from multiple galaxies together for a higher-fidelity signal \citep{Spilker2014,fudamoto2017,Birkin2021,Chen2022,Hagimoto2023,Reuter2023}. These fainter lines can reveal unique phases of the ISM. For example, hydrogen cyanide traces dense, star-forming gas \citep{Riechers2006,Riechers2010,Oteo2017,Bethermin2018,Canameras2021,Rybak2022,Yang2023}, and the presence of carbon and oxygen isotopologues are a measure of metal enrichment history  \citep{Henkel2010,Bethermin2018,Zhang2018,Maiolino2019}. These lines are often faint and thus require long observation times. The observations of multiple sources can be combined through stacking, although these often are done only at the wavelengths where spectral scans searched for the redshifts, i.e., the 3~ and 2~mm observed wavelengths. Consequently, faint lines in the 1~mm regime remain poorly studied, even though many molecular species exist in this wavelength range. Consequently, even if individual observations are not able to directly detect individual lines, stacking is thus an important tool, particularly when samples become large enough to enable A/B comparison tests.

\begin{figure*}
\centering
    \includegraphics[width=0.65\paperheight]{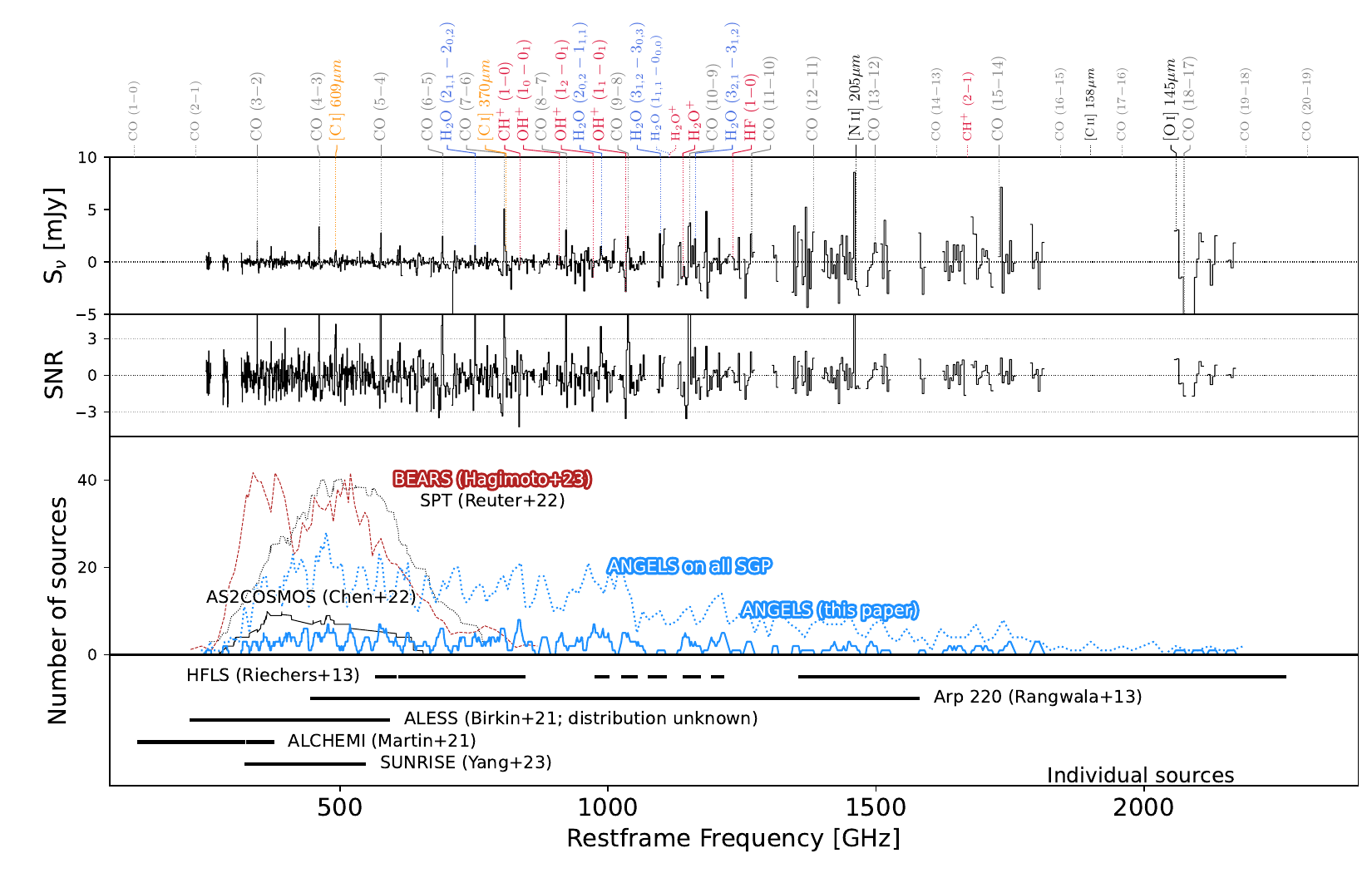}
    \caption{The composite spectrum based on the ANGELS observations. {\it The top panel} shows the noise-weighted stacked spectrum of an archetypal $z = 2.5$ L$_{\rm IR} = 10^{13}$~L$_{\odot}$ galaxy. The brightest spectral line is the \nii{} line at around 10~mJy (see Sect.~\ref{sec:stackedSpectrum}). 
    {\it The middle panel} shows the signal-to-noise ratio of the spectrum.
    {\it  The bottom panel} shows the number of sources that contribute to each of the spectral line in a solid \textit{blue line}. Similar stacking experiments \citep{reuter20,Chen2022,Hagimoto2023} are shown as histograms, and the predicted coverage of the full SGP sample is shown in \textit{a blue dashed line}. Follow-up studies of individual sources \citep{Rangwala2014,Riechers2013,Birkin2021,Martin2021,Yang2023} are shown as solid lines. While the coverage of the ANGELS sources in this paper becomes sparse at higher frequency, the full survey aims to detect at least a few sources out to 2300 GHz.}
    \label{fig:stackedSpectrumLowerHalf}
\end{figure*}

Figure~\ref{fig:stackedSpectrumLowerHalf} shows the stacked spectrum from the ANGELS observations using the same method as \cite{Hagimoto2023}. 
In short, we extract the complete spectra from the regions where the band~7 emission exceeds $2 \sigma$ surrounding the sources using an additional smoothing of one beam. We decide to stack the spectrum of each galaxy with the goal of representing a single archetypal galaxy at $z = 2.5$ (approximately the mean of this sample; \citealt{Urquhart2022}) with an infrared luminosity of $10^{13}~\mathrm{L}_{\odot}$. This involves normalizing all spectra to the same luminosity and to a common redshift. Consequently, we require a constant $L'$ across all redshifts, and normalize the spectrum accordingly:
\begin{equation}
S_{\nu, \rm common} = S_{\rm \nu} \left(\frac{D_\mathrm{L}(z_{\rm source})}{D_\mathrm{L}(z_{\rm common})}\right)^2 \frac{1+z_{\rm common}}{1+z_{\rm source}}, \label{eq:commonRedshift}
\end{equation}
where $D_\mathrm{L}$ refers to the luminosity distance at redshift $z$, and $z_{\rm common}$ is set to 2.5. This factor accounts for the cosmological dimming for each spectral line, as well as for the redshift-dependence of the flux density unit.

{\color{referee2}
We use the {\it Herschel}-based infrared luminosities as discussed in Section~\ref{sec:ContSnapshotsResults}, where we reweight the fields by their respective Band~7 fluxes (Bendo et al. in prep.; see Table~\ref{tab:infraredLuminosities}), since the sources are most reliably detected at this frequency. When compared to the method in \cite{Hagimoto2023}, where the weighting was performed based on the infrared luminosities derived from the ALMA Band~4 continuum, there appears to be a marginal ($< 15$~per cent) effect from source to source when compared to a fixed constant dust temperature.
}
%We then normalize the infrared luminosity of each galaxy to $10^{13}~\mathrm{L}_{\odot}$, which are derived directly from the Band~7 continuum emissions of the ANGELS sources (Bendo et al. in prep.) since the sources are the most clearly detected in this band.} We do not use the {\it Herschel} photometry because many sources have multiples in the same field. We assume an average dust temperature of 35~K and a dust emissivity index $\beta_{\rm dust} = 2$ (which is consistent with the dust spectral energy distributions measured by \citealt{Bendo2023}). 
Each of the weighted spectra are then added by their variance-weighted fluxes{\color{referee}, which prioritizes the highest signal-to-noise ratios (SNR) of the lines.} {\color{referee}As discussed in detail by \cite{Hagimoto2023} and \cite{Reuter2023}, this means that the brightest sources are still predominantly determining the spectrum. It explains the lack of effect in line SNR when changing the (intrinsic) luminosity or dust-mass weightings as shown in \cite{Spilker2014}. {\color{referee2} However, the relative line ratios within the composite spectrum do strongly vary as different weighting schemes are used \citep[see e.g. the discussion in][]{Reuter2023}, and are important to account for in future composite spectra. }%Since our current galaxy luminosity estimate is based on a fixed dust temperature, our luminosity-scaling at $10^{13} \rm L_{\odot}$ is equivalent to a dust-mass weighting at $2 \times{} 10^{9} \rm M_{\odot}$.  }

The top panel of Figure~\ref{fig:stackedSpectrumLowerHalf} shows the rest-frame composite spectrum using the above-mentioned weighting scheme. The middle panel shows the signal-to-noise ratio of the observations, while the lowest panel shows the total number of spectral lines seen per rest-frame frequency, and compares it to other and future stacking experiments. 

All CO lines with $J_{\rm up} < 11$ are seen in the stacked spectrum. Only two out of five of the \ci{}(1--0) lines are detected individually, but the \ci{}(1--0) line is clearly detected in the stacked emission. The line profile of \ci{}(1--0) appears broader, and could remain undetected because of its potentially more extended nature \citep{Valentino2020CI}. Both H$_2$O ($2_{1,1}$--$2_{0,2}$) and (2$_{0,2}$ - 1$_{1,1}$) are detected in the stacked spectrum, in line with the individual detections and previous stacking experiments \citep{Yang2016,Zavala2018,Hagimoto2023,Reuter2023}. The brightest stacked spectral line, \nii{}, is detected, and stands out in the stacked spectrum. Three molecular absorption lines are seen at the $-3 \sigma$ level, CH$^+$(1--0), OH$^+$ $(1_1$-$0_1)$, and $(1_2$-$0_1)$, which are also seen in individual observations, while the fainter OH$^+$ feature $(1_0$-$0_1)$ remains undetected. Importantly, the velocity difference relative to the systemic redshift of absorption lines are large. Detecting such features in a line stack is thus surprising. In future observations, larger samples will be able to more fully characterize fainter lines, and combine the resolved (ANGELS) and unresolved (BEARS) data to better characterize the 200 - 800~GHz regime, while short-wavelength observations fill in a contiguous composite spectrum well into the THz regime, probing atomic lines such as \ciil{} and \oiiil{}{\color{referee2}, as well as provide a composite CO SLED out to high-$J$ transitions as explored in previous work \citep{Hagimoto2023,Reuter2023}. }

\section{Resolved properties of ANGELS sources}
\label{sec:ResolvedStudies}
{\color{referee} The ANGELS observations have provided a high-resolution view on a sample of sixteen {\it Herschel} fields, revealing 54 emission and 12 absorption lines, while providing meaningful upper limits on another 27 lines.} The beam sizes of the observations range between 0.5 to 0.1~arcseconds, enough to resolve the internal structure of these distant sources. In the following, we will use this resolved data to investigate the properties of the star-formation in the ANGELS sources.

\subsection{The resolved star-formation law in ANGELS sources}
\label{sec:resolvedKS}
The Schmidt-Kennicutt relation between the star-formation surface density and molecular gas mass surface density ($\Sigma_\mathrm{SFR}$--$\Sigma_\mathrm{H_2}$; \citealt{Schmidt1959,Kennicutt1998}) is a diagnostic of the star-formation mode. High-redshift dusty starbursts typically have higher star-formation rates (SFRs) relative to their molecular gas compared to normal star-forming galaxies (e.g., \citealt{Casey2014}), which results in a shorter depletion time scale ($\sim 100$~Myr). The reasons for this boost in star formation are often hypothesised to be connected with recent merger events \citep{Sanders1988,Barnes1991,Hopkins2008}, although several other theories have been posited, including secular evolution and violent disc collapse \citep{Cai2013,Gullberg2019,Hodge2019,Rizzo2020,Fraternali2021,Zavala2022}.

Figure~\ref{fig:ksresolved} shows the resolved Schmidt-Kennicutt relation (\citealt{Schmidt1959,Kennicutt1998}) for the ANGELS sources and reference samples of low-redshift galaxies (\citealt{Kennicutt1989,delosReyes2019,delosReyes2019b,Kennicutt2021}), high-redshift star-forming galaxies (\citealt{Tacconi2013}) and high-redshift DSFGs (\citealt{Hatsukade2015,C.C.Chen2017} and references therein). The parent sample of BEARS is shown in {\it dark red circles} \citep{Hagimoto2023}. 
These modestly-resolved observations show more extreme star-forming environments in the mixed lensed- and unlensed sample than found for unlensed DSFGs. We use a CO-H$_2$ conversion factor ($\alpha_{\rm CO}$) value of $4.0$ from the recent study of \citet{Dunne2021,Dunne2022}, adjusting the reference sample of DSFGs accordingly for a fair comparison, although we note that this molecular gas conversion can vary sharply from source to source \citep{Harrington2021}. Similarly, we use an IR luminosity-to-SFR factor of $1.73 \times 10^{-10} \rm \,M_{\odot} yr^{-1}/L_{\odot}$ valid for a Salpeter 1-100~M$_{\odot}$ initial mass function. 
We calculate the position of each galaxy using a pixel-by-pixel comparison between the dust continuum emission from Band 7 to the lowest-$J$ ANGELS-observed CO transition. Since pixels within each beam are not statistically independent, we subsequently down-sample to account for this effect.

The higher-resolution Band 7 dust continuum is smoothed to the same beam size as the Band~3, 4 or 5 CO emission using the images as they appear in the sky. Then we compare the maps on a pixel-by-pixel basis to provide a distribution of the sources, and while the selection itself could be affected by differential lensing, the position of these sources on the star-formation law should not be affected by the lensing as surface brightness is preserved during a lensing event. We use pixels with either the moment-0 CO or dust emission beyond $3 \sigma$ within a radius including all of the source flux, as shown in the Appendix figures. As a lower limit, we demand that the pixel in the other image has at least $1 \sigma$ emission, and ensure that the majority ($> 80$~per cent) of pixels are not excluded from this additional constraint. 
We show these pixel-per-pixel comparison of the star-formation law in the Appendix Figures~\ref{fig:individualKS} and \ref{fig:individualKSzoomin} to investigate the individual properties of the sources, and for all sources where available, compare directly to the CO and \ci{}-derived star-formation and gas surface densities from the moderately-resolved study in \cite{Hagimoto2023}. 
These show a relatively wide distribution of star-formation or gas surface densities, varying by up to an order of magnitude as seen in previous studies as well \citep{Ikeda2022}. {\color{referee3} Note that the sharp lines in resolved Schmidt-Kennicutt relations in Figure~\ref{fig:individualKSzoomin}, particularly for HerBS-25 and -106, come from the $> 1 \sigma$ criterion. Deeper observations of these sources could reveal lower star-formation rate and gas mass surface densities on the source outskirts. This could shift their average position on the star-formation diagram downwards by up to $\sim 0.2~\rm dex$ based on the current distribution of the $> 1 \sigma$ emission shown in Figure~\ref{fig:individualKSzoomin}. This further demonstrates the internal variation of the star-formation law within galaxies, and the need for efficient fueling of gas within galaxies to sustain these high star-formation rates for durations in excess of $100$~Myrs. }

The difference between the moderately resolved study (\citealt{Hagimoto2023}) and the ANGELS observations mostly lie along the constant depletion times (i.e., the diagonal lines). This indicates that the discrepancy between these studies could originate from less accurate, and dependent size estimations. Additionally, CO line tracers between CO(3--2) and CO(6--5) are used for the molecular gas mass estimates, which add additional uncertainties in the conversion to the ground-state, i.e., to the luminosity of CO(1--0). This can explain the additional offset in the x-direction ($\Sigma_{\rm H_2}$) for HerBS-86 and -106, where CO(6--5) was the lowest-$J$ transition. Reassuringly, there do not appear to be significant y-axis offsets, indicating that the star-formation rate estimates are in agreement (both this study and \cite{Hagimoto2023} use ALMA dust continuum measurements).

\begin{figure}
    \centering
    \includegraphics[width=\linewidth]{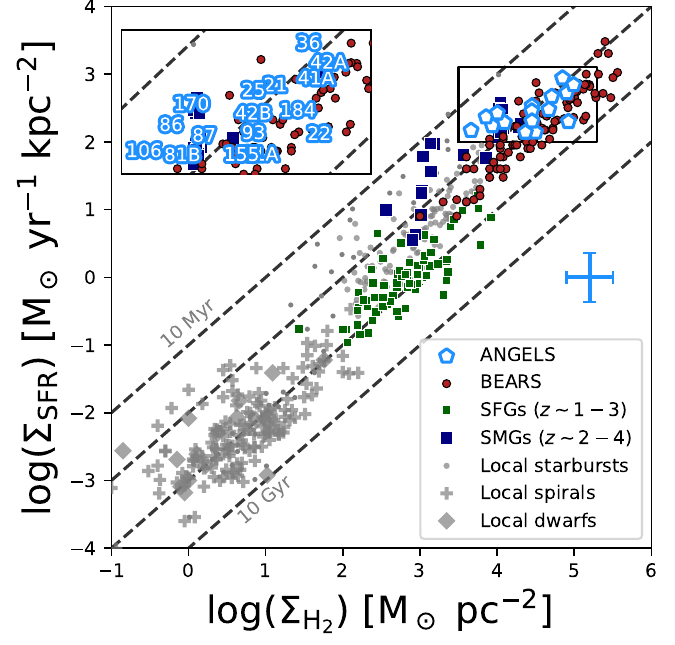}
    \caption{The resolved Schmidt-Kennicutt star-formation law of the ANGELS sources with observations of CO emission lines. We compare the sources against reference samples of low-redshift galaxies (\citealt{Kennicutt1989,delosReyes2019,delosReyes2019b,Kennicutt2021}; {\it grey points, plus points, and diamonds}), high-redshift star-forming galaxies (\citealt{Tacconi2013}; \textit{green squares}) and high-redshift DSFGs (\citealt{Hatsukade2015,C.C.Chen2017} and references therein; {\it blue squares}). The parent sample of BEARS is shown in {\it dark red circles} \citep{Hagimoto2023}. We use an $\alpha_{\rm [CI]}$ value of $17.0$ and an $\alpha_{\rm CO}$ value of $4.0$ from the recent study of \citet{Dunne2021,Dunne2022}. We adjust the reference sample of DSFGs ({\it blue squares}) from \citet{C.C.Chen2017} from the initially-assumed $\alpha_{\rm CO} = 0.8$ to $4.0$ for a fair comparison.
    These sources lie in the upper-right region of the star-formation law, with similar depletion times as the resolved observations of unlensed sources \citet{C.C.Chen2017}.
    The sources have similar or slightly smaller volumes of star-forming gas than unresolved studies with similar surface star-formation rates. This could indicate that the dense star-forming regions deplete faster than the moderately-resolved studies, and require efficient fueling from more gas-rich regions to maintain their high surface star-formation rates. 
    }
    \label{fig:ksresolved}
\end{figure}

The distribution of the sources is similar to the moderately-resolved pilot sample of BEARS ({\it red points}), with larger gas and SFR surface densities than those found by previous observations for unlensed sources \citep{C.C.Chen2017}. There appears to be enough gas available for around 100 megayears or one gigayear of star-formation in these sources. The extended gas reservoirs seen in even these short snapshots thus indicate that the spatial distribution of gas and dust can play an important role in the evolution of dusty star-forming galaxies \citep[see e.g.,][]{Ikeda2022}.
The order of magnitude spread in the pixel-per-pixel surface gas densities and surface star-formation rates indicates that DSFGs require efficient feeding of gas from the gas-rich regions into star-forming regions to ensure that the high star-formation rates can be sustained for the depletion timescales indicated by un- or marginally-resolved observations \citep[e.g.,][]{Hagimoto2023}.

\subsection{Star-formation surface densities of ANGELS sources}
\label{sec:eddingtonLimits}
When many stars are formed in a small environment, star-formation feedback becomes a more important contributor of the evolution of galaxies. Figure~\ref{fig:EddingtonLimit} compares the intrinsic sizes of the ANGELS sources against their intrinsic infrared luminosities. By comparing their star-formation surface densities to other galaxies \citep{bothwell2013,Spilker2016,Kamieneski2024} and to expected limits of star-formation rate \citep{Andrews2011}, we find a large variation across our sample of sources. 
We use the Sersic profiles for the sources when available from the lensing reconstruction (seen by crosses; Table~\ref{tab:intrinsicProperties}), where the Sersic indices are free to vary between 0.5 and 5. The non-lensed sources are fitted using the {\tt casa} task {\tt IMFIT}, which assumes a Sersic index of 0.5 that appears to match the emission profiles well based on the residual images. These unlensed sources are broadly larger in size (shown in pentagons instead of crosses). We calculate the per-source luminosity using the \textit{Herschel}-photometry, which better accounts for the total star-formation rate in the total system, but could cause discrepancies between the $\Sigma_{\rm SFR}$ estimates from the resolved star-formation law and this discussion ({\color{referee2}see Sec.~\ref{sec:ContSnapshotsResults}}). In the case of any multiplicity, we distribute the flux based on the weighted flux densities of the Band~7 continuum emission.

The bulk of sources reside between the 10 and 100~M$_{\odot}$/yr/kpc$^2$ regime, similar to other sources seen with {\it Herschel} and SPT. Broadly, three sources are close to the {\it forbidden} regime of star-formation, of which two are in the HyLIRG regime. Although likely the bulk of the sub-mm emission of these sources is due to star-formation \citep{bakx18}, an obscured AGN could be contributing to the emission and produce these compact-but-bright sources. In total, eight sources are in the $L_\mathrm{IR}>10^{13}~\mathrm{L_\odot}$ regime. 
The more diffuse galaxies, HerBS-159A and B (see Section~\ref{sec:spatiallyExtendedSources}) lie in the $<10$~M$_{\odot}$/yr/kpc$^2$ regime, where typically more normal star-forming galaxies are found \citep[e.g.,][]{Tacconi2013}. Note that the radius of these sources is extracted on a tapered image ({\it uv}-taper of 0.5~arcsec), resulting in larger errors, but regardless their radii are larger than average. 
The diverse and complex nature of ANGELS sources is apparent from these resolved ALMA observations. In the subsequent Section, we compare these rich results against the known literature of DSFGs. 

\begin{figure}
    \centering
    \includegraphics[width=\linewidth]{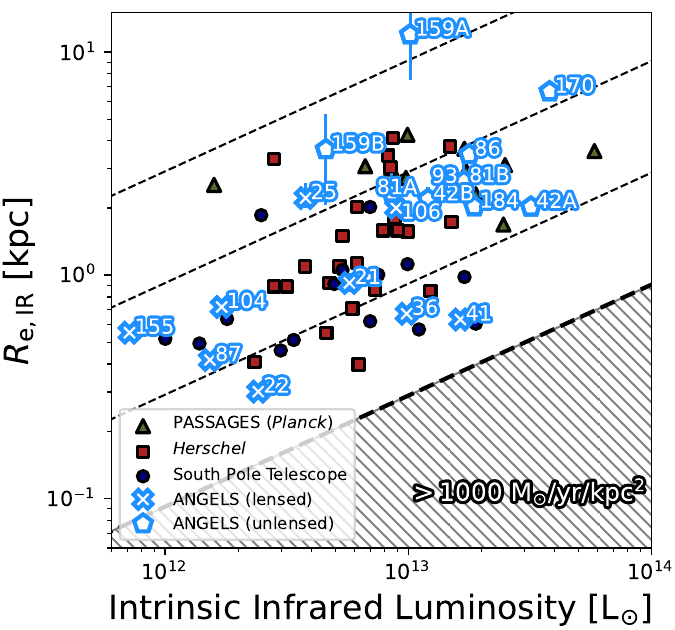}
    \caption{The radius of the sources in proper angular distance as a function of their luminosity for the ANGELS sources and other distant galaxies \citep{bussmann13, Spilker2016,Kamieneski2024}. We use the intrinsic sizes for all lensed sources (see Table~\ref{tab:intrinsicProperties}). The hatched region indicates the forbidden regime for a stable star-forming system according to \citet{Andrews2011}, with increasing diagonal dashed lines indicating star-formation surface densities of 1000, 100, 10 and 1~M$_{\odot}$/yr/kpc$^2$. The majority of the sources lie around 100~M$_{\odot}$/yr/kpc$^2$, with the surface star-formation rates of HerBS-22, -41A and -36 approaching the limits of star formation. The diffuse sources (see Section~\ref{sec:spatiallyExtendedSources}) HerBS-159A and -B appear to be more extended with low star-formation rate densities on the order of $\approx 10$~M$_{\odot}$/yr/kpc$^2$.
    }
    \label{fig:EddingtonLimit}
\end{figure}

\section{Nature of ANGELS sources}
\label{sec:NatureOfANGELSSources}
The ANGELS sources are effectively randomly sampled from the BEARS survey, with a selection based solely by their proximity to each other on scales beyond the scales of the cosmic web (see Figure~\ref{fig:ANGELSFIELDS}){, as is discussed further in \cite{negrello2017}}. The guiding principle behind the observations was solely to maximize the number of lines that we could target within a single tuning modulo the atmosphere. As a result, these observations provide an unbiased sampling of the properties of high-redshift, bright {\it Herschel} sources. In this section, we investigate the properties of ANGELS sources in an effort to better characterize the bright end of {\it Herschel} sources.

\subsection{Lensing among the HerBS sources}
\label{sec:lensingAmongHerBSSources}
The total sample of sixteen {\it Herschel} sources turned out to consist of 26 separate ALMA sources. Out of these, nine ALMA sources are apparent lenses from the observations. Although an additional number of sources might still be lensed (HerBS-42A, -159A), we exclude these in the statistics of these sources. 
{\color{referee} The lensing fraction of 56 $\pm$ 13~per cent (assuming a binomial distribution; \citealt{Gehrels1986}) is on the lower end of theoretical predictions (e.g., 70~per cent; \citealt{negrello2007}), with direct comparisons to galaxy evolution models \citep{Cai2013,bakx18,Bakx2020Erratum} expected even higher fractions ($\sim 85$~per cent), in line with subsequent comparisons to near-infrared imaging ($\sim 82$~per cent; \citealt{bakx2020}). }
Lensed galaxies allow the probing of populations that would remain under the detection capabilities of facilities, with magnifications up to several tens \citep{dye2015,Dye2018,Spilker2016,Amvrosiadis2018,Zavala2018,Rizzo2020,Kamieneski2024}. The properties of these sources appears diverse, with over two order of magnitudes spread in the source radius and associated star-formation surface density (Figure~\ref{fig:EddingtonLimit}). The comparison samples between the lensed and unlensed ANGELS sources are selected from different areas and depths. As a result, larger scale surveys will select towards more rare geometries of lensing where strongly-boosting caustics intersect with bright central regions resulting in brighter fluxes. Four of the sources in ANGELS are also present in the sample from \cite{negrello2017}, namely HerBS-21, 22, 25, and 42A/B. Two sources (HerBS-36 and 41A) also meet the \cite{negrello2017} selection criterion of $S_{500} > 100$~mJy, and are also lensed. All four common sources have lensing rank C (unclear lensing nature). HerBS-42-A/B is a double system and the individual $S_{500}$ fluxes fall below the 100~mJy criterion, confirming the prediction by \cite{negrello2007} that most of the $S_{500} > 100$~mJy sources -- also those with C rank -- are likely strongly lensed.

The {\it HerBS} sample is selected at the transition from mostly lensed surveys \citep{Harrington2016,Spilker2016,negrello2017} to unlensed systems found often in ground-based surveys such as LABOCA \citep[e.g.,][]{weiss2009} and SCUBA(-2) \citep[e.g.,][]{Simpson2019,Garratt2023}. Among the lensed sources, differential lensing\footnote{Due to sharp variations in the lensing magnification of different regions of a galaxy, the observed, lensed line and continuum fluxes of galaxies could give a warped perspective on the intrinsic properties of galaxies. This effect is exacerbated by the initial selection effects, which might favour the selection of galaxies with strong amplifications of star-forming regions.} \citep{serjeant2012,Serjeant2024} could more strongly affect the lensed samples, resulting in a more biased and inhomogeneous survey at the higher magnifications. Meanwhile, the lensing magnifications of our survey ($\mu = 2 - 28$ with a median $\mu = 8.9$) are surprisingly similar to those seen of the SPT ($\mu = 2 - 33$ with a median $\mu = 6.3$; \citealt{Spilker2016,reuter20}) and {\it Planck} ($\mu = 2 - 28$ with a median $\mu = 7$; \citealt{Kamieneski2024}), and the effect of bias could be similar across our samples. We find that the majority of these lensed systems are small (0.3--3~kpc) central dusty systems, similar to the sources found in ALESS \citep{Hodge2013,Karim2013MNRAS.432....2K}. The line profiles of these galaxies do not appear to indicate any source multiplicity or merging state contrary to the idea that all DSFGs are mergers as often seen in the local Universe \citep{Sanders1988,Barnes1991,Hopkins2008}. Instead, central dusty star-bursts could be driving the bulk of the lensed sources seen in {\it Herschel} and other surveys 
\citep{Barro2016,Hodge2016, Cibinel2017,Ikarashi2017ApJ...849L..36I,Tadaki2017ApJ...834..135T,Gullberg2019,Pantoni2021}, where galaxies are building up a compact central dense stellar component and change their morphology from disk-dominated to bulge-dominated systems, perhaps in late-stage mergers and/or violent disk instabilities \citep{Zolotov2015}. This could be a bias due to the lensing selection -- where compact sources are more susceptible to be strongly lensed, either because of their prevalence or their high star-formation surface densities, than extended merging systems -- or an effect of differential lensing. Only deep observations tracing the more extended emission in the lower-magnification regions would be able to differentiate between these two scenarios.

\subsection{Hyperluminous galaxies}
\label{sec:hyperluminousGalaxies}
The high-resolution observations revealed the lensing properties of the ANGELS sources. {\color{referee2}Eleven} sources remain above the Hyper-Luminous Infrared Galaxy (HyLIRG; $> 10^{13} L_{\rm IR}$ in Fig.~\ref{fig:EddingtonLimit}) regime., although HerBS-159A is potentially lensed. This is higher than typically observed in other samples \citep[c.f.,][]{reuter20,Liao2024}, likely because of the fainter selection of the {\it HerBS} sources relative to other surveys. Unlike the lensed sources (sizes between $300 - 2000$~pc with a median size of 600~pc), unlensed systems appear to be more extended (0.3-3~kpc), such as HerBS-93, -170 and -184. The extended morphology of HerBS-170 further makes its HST-dark nature surprising, as it requires a large, homogeneous dust screen to obscure this emission. Source confusion affects many of the {\it apparently-brightest Herschel} galaxies \citep{Bendo2023}, however several of the {\it intrinsically-brightest} sources also come from fields with multiples, such as HerBS-41A, -42A, -81A. This means that source confusion only little affects the ability of {\it Herschel} to find HyLIRGs in the slightly-fainter flux regimes ($S_{500} = 50 - 100$~mJy; \citealt{ivison16,Oteo2017}).

Four sources showcase the extreme end of star-formation in this sample: HerBS-36, -41A, -42A, and -81A. Their CO SLEDs indicate similar star-forming conditions as found in other intense star-forming systems, such as Planck-selected systems. %The \oi{} emission is a dominant cooling line for the neutral atomic gas, together with \cii{} (a rough luminosity ratio of \cii{}/\oi{}~$\sim 5$ is expected; \citealt{Litke2023}). The bright \oi{} emission of HerBS-41A lies on the expected scaling relation for local  and high-$z$ galaxies (Figure~\ref{fig:atomicLines}). 
The surprising discovery of inflowing gas seen in a shocked tracer for HerBS-36 could further suggest that inflows play an important role in boosting star-formation rates \citep{berta2021}, with part of the source close to the limits of star formation \citep{Andrews2011}, which can thus provide a ready source for shocks. 

The morphology of these systems furthermore appears clumpy, with modest amounts of the flux density ($\approx 10 - 20$~per cent for HerBS-93, -170 and -184) and thus associated star-formation originating from these clumps. Such structures could be essential for the formation of bulges and of wider star-formation across these intense starbursts \citep{Hodge2019,Rujopakarn2019, Gullberg2019}. Contrary to this, observations of lensed ULIRGs \citep{Dye2018,Rizzo2020} often find more stable rotators, although SDP.81 is an obvious counter-example (\citealt{dye2015,Rybak2015MNRAS.451L..40R,Tamura2015}). An important consideration here is also that lensing could affect our picture of the clumpiness of these sources, both through differential lensing and $uv$-plane issues addressed in \cite{Gullberg2019}. HerBS-41A and -93 show velocity widths of approximately 1000~km/s, which indicates that they are already very massive systems (${\rm \sim 5 \times{} 10^{11} M_{\odot}}$; \citealt{Hagimoto2023}) without obvious indications of merging. The discrepancy between smaller compact lensed ULIRGs and the larger weakly- or unlensed clumpy HyLIRGs is an important avenue for future investigations, as we find five of our sources to show these extended clumpy morphologies, with an incidence rate of roughly $35 \pm 25$~per cent. 

Particularly the lack of Hubble counterpart for HerBS-170 poses an interesting test for the clumpy morphology of dust. At a rest-frame wavelength of 2100~\AA{}, these Hubble observations are sensitive to direct starlight. With a total star-formation rate of $3000 - 7000 \rm M_{\odot}/yr$ depending on the IMF, the dust screen needs to extend across the system to obscure both the small-scale clumps (a few 100~$M_{\odot}/yr$) and the central star-forming region. HerBS-170 has roughly 4 orders of magnitude higher dust mass than the optically-selected $z = 6 - 8$ REBELS sources \citep{Ferrara2022}, whose dust screen size extends about thirty times further ($\sim 1000$ times in terms of area). Using eq. 11 from \cite{Ferrara2022}, the optical depth of HerBS-170 is thus roughly ten times higher than REBELS with $A_V = 0.1 - 0.4$ (\citealt{Inami2022}). Meanwhile, the REBELS observations are only moderately deeper ($1 \sigma_{1500} = 0.06 \mu$Jy and $1 \sigma_{2100} = 0.1 \mu$Jy, respectively), although REBELS lie at higher redshifts. The rest-frame optical dust obscuration of HerBS-170 is thus, as expected, much higher than optically-selected sources, but perhaps more importantly, the distribution of this dust must be coincident with star-formation across the entire extended system, suggestive of a remarkably-homogeneous dust screen or of star-formation occuring in separate but similarly-obscured clumps across the system at 10~kpc separation \citep[see e.g., ][]{Cochrane2021}.

While the dust properties in DSFGs are quite similar (dust emissivity index, gas-to-dust ratio; \citealt{Bendo2023,Hagimoto2023}), the CO excitations and other line ratios suggest a diverse group \citep{Daddi2015A&A...577A..46D,Valentino2020CO,Birkin2021, Hagimoto2023,Stanley2023}, indicating that there are multiple ways to form a DSFG \citep[][]{QuirosRojas2024}.
The morphological and kinematic properties of these HyLIRGs seems to indicate a potential pathway where star-forming clumps are associated with gaseous disks that are unstable (Toomre $Q < 1$) due to high turbulence \citep{Krumholz2012ApJ...745...69K,Harrington2021}. This would be contrary to smaller central dusty star-bursts that are rotationally-dominated \citep{Dekel2009} seen mostly through lenses \citep{Dye2018,Rizzo2020} and ground-based surveys \citep{Hodge2013,Karim2013MNRAS.432....2K}. The diversity of models to explain DSFGs \citep{Baugh2005MNRAS.356.1191B,Dave2010MNRAS.404.1355D,Narayanan2009MNRAS.400.1919N,narayanan2015} should thus accurately reflect the multiple pathways of starbursts and their transition to quenched galaxies \citep{Toft2014,Barro2016}.

\subsection{Spatially-extended sources}
\label{sec:spatiallyExtendedSources}
HerBS-159A \& B are extended sources, which made the emission fall below the detection criteria of our observations. By tapering the data (see Section~\ref{sec:ContSnapshotsResults}) we find $\sim 3 - 5 \sigma$ continuum emission visible across Bands 5, 6, 7 and 8 for these galaxies. 
This system is of particular interest, as HerBS-159B is one of the CUBS (Companion sources with Unusually Bright lineS) sources in \cite{Hagimoto2023}. The CO SLED of this system is in excess of the maximum expected behaviour from the equipartition theorem between CO excitations; i.e., it is {\it super-thermalized}, and cannot be explained by linear-scaling effects such as differential lensing \citep{Serjeant2024}. Out of the total sample of 46 sources within the CO studies in \cite{Hagimoto2023}, only four CUBS sources are found. If there is any intrinsic physical difference between these sources, their physical interpretation thus offers insight into a unique galaxy process within the DSFG phase, with an occurrence rate of $\sim 8$~per cent. Under the assumption of a single DSFG phase with a duration of $\approx 200$~Myr (i.e., their average gas depletion time; \citealt{Hagimoto2023}), this could thus point to a galaxy phase duration of $\sim 20$~Myr.

It is still unclear how galaxies would achieve such a galaxy phase, however it is possible to physically provide an {\it super-thermalized} CO SLED through optical depth effects (see the modelling in \citealt{Hagimoto2023}). These could be triggered by a merger phase or by the presence of an obscured AGN. Unfortunately, HerBS-159B does not have any observations of spectral emission lines within ANGELS that further provide constraints on the system properties. 

Instead, the fact that the HerBS-159 system requires tapering to reveal the component galaxies is an important hint to their galaxy evolutionary phase. 
Similar galaxies have been reported in different studies. \cite{Tadaki2020} reported on such a galaxy with extended emission, missed in the untapered data, while looking at 85 massive galaxies with ALMA. 
\cite{Sun2021} investigated a sample of lensed sources with {\it Herschel} through ALMA observations. Two sources out of a sample of 29 have similar features to the HerBS-159 system. They report that these sources lack the central star-bursts dominating the majority of galaxies, similar to the morphologies that are seen in the lensing reconstructions for our sources (Fig.~\ref{fig:lensingreconstruction}). This is not in agreement with a bright central component associated with a far-infrared luminous AGN.

\textit{HST} imaging revealed that HerBS-159A is likely a lensed system with extended emission across multiple arcseconds. Regardless, this means that the Einstein rings also carry a certain thickness, implying extended emission in the source plane. No obvious lensing morphology is seen for HerBS-159B (the CUBS source; \citealt{Hagimoto2023}), with no evidence of a nuclear emitting region. The fact that all CUBS sources were seen as multiples suggests this could be an important component to the formation of this phase. The early merging, pre-coalescence phase thus seems the most likely explanation for this phase. This early phase, also occurring on the scale of $\sim$ten megayears. Rare fields such as HerBS-159 thus require detailed further study to find out how they fit within the emerging paradigm of DSFGs, and might reflect an important part in merging history still missing from models.

\section{ANGELS as a line survey technique}
\label{sec:ANGELSasAsurvey}
The ANGELS pilot aimed to investigate the use of Bands 3 through 8 in line observations through quick snapshots. In this Section, we summarize the results of this method, and provide caveats and suggested improvements towards expansion of this survey to larger samples of bright sources.

\subsection{ANGELS as a redshift machine!?}
\label{sec:angelsasaredshiftmachine}
Three sources did not yet have spectroscopic redshifts at the start of the ANGELS observations: HerBS-87, -104, and -170 (see Table~\ref{tab:sources}). 
The BEARS observations revealed single line detections \citep{Urquhart2022}, but not at sufficient depth to exclude different redshift solutions \citep{Bakx2022}. HerBS-87 had a single line detected at 160.96~GHz. HerBS-104 had a line detected at 90.91~GHz. The quality of the emission lines seen for HerBS-170 was below the significance to conclusively state the redshift, with lines at 111.15 and 156.13~GHz. 

ANGELS observations revealed additional lines to ensure the robust redshifts of each of these three sources. Of course, these ANGELS tunings did not optimize towards lines for these three sources, but we use the method from \cite{Bakx2022} to estimate the chance of a single line, or better yet, a robust redshift estimate from even blind observations. In total, we find a roughly 50~per cent chance to derive a robust redshift from six tunings (Bands 3 through 8) on a typical DSFG-like redshift distribution, and a $> 75$~per cent chance of detecting a single line for each source targeted. ANGELS thus offers a good chance of assisting in the redshift completion of the SGP fields. {\color{referee} In this pilot study, thirteen galaxies (+ three neighbouring galaxies) had spectroscopic redshifts, and the choice for observing windows was driven solely by the expected lines in our observing window for these sixteen galaxies. However, in a scenario where much fewer spectroscopic redshifts are known, preference to lower frequency observations could be given when using tools similar to Appendix Figure~\ref{fig:optiTuneGraph}, as they relatively cover a larger redshift region for the same observing window. }

\begin{figure}
    \centering
    \includegraphics[width=\linewidth]{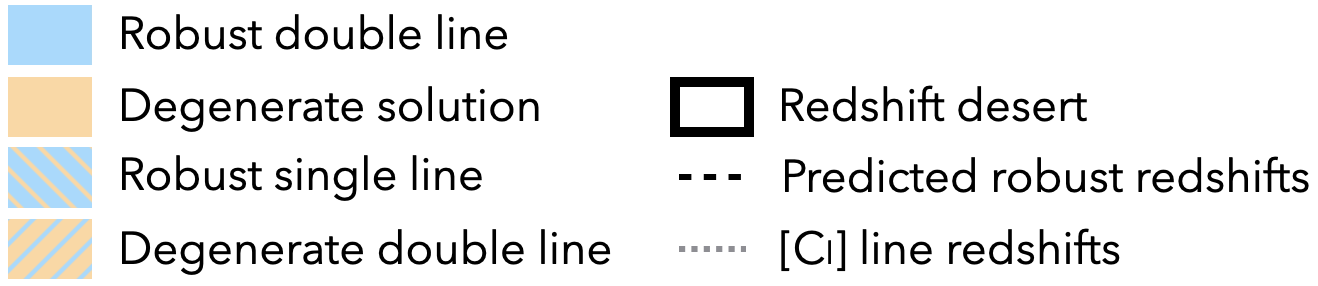}
\includegraphics[width=\linewidth]{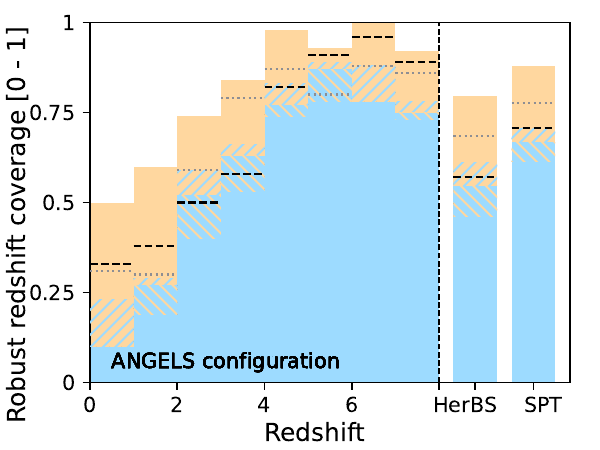}
    \caption{\color{referee} We use the method from \citet{Bakx2022} to calculate the redshift efficiency of the ANGELS method across a blind sample of galaxies - without any prior line confirmations. These show a large probability for robust redshift identification for sources without known redshifts beyond $z > 2$ ({\it blue regions}). The HerBS (\citet{bakx18,Bakx2020Erratum}) and SPT sources (\citealt{reuter20}; the two right-most histograms) show the predicted capability to find robust and single-line ({\it orange region}) redshifts for a sample with the same redshift distribution as the HerBS and SPT samples. The blue regions indicate the chance of finding more than one spectral line, while orange regions suggest a single spectral line will be found in a tuning. {\it Hatched blue fill} indicates cases where one can identify the redshift robustly with even a single spectral line, while \textit{hatched orange fill} indicates the situation where redshift degeneracies remain down to a $5 \sigma$ uncertainty in $z_{\rm phot}$. The {\it grey dotted lines} indicate the robust redshifts if we include the possibility of detecting atomic lines such as \ci{}, and the {\it black dashed lines} indicate the robust redshifts if we include the possibility of single-line robust redshifts and for cases where there still remains a degeneracy between multiple CO line solutions (see \citet{Bakx2022} for details). The three sources with ANGELS-confirmed redshifts already had a single line detection, and ANGELS thus offers a good chance of assisting in the redshift completion of the SGP fields. }
    \label{fig:RSG_ANGELS}
\end{figure}

\subsection{Efficiency compared to overhead-limited surveys}
ANGELS has offered a high-resolution view on galaxies across a wide spectroscopic scope using ALMA. The project was uniquely enabled by the bright nature of {\it Herschel} sources with yet high enough surface density to enable efficient scheduling (see Figure~\ref{fig:ANGELSFIELDS}).
In total, 94 spectral lines were targeted across 28 different line transitions. On average, each source is observed in five transitions, with HerBS-41A being targeted in as many as 12 lines. 66 lines were detected, resulting in a detection ratio of 70\%. In Bands 3 and 4, eleven transitions were targeted, and respectively 8 and 9 of these transitions were detected. Band 5 targeted 19 lines, and we detected 14. Most lines were targeted in Band 6, where out of 27 lines, 19 were detected. Band 7 had 17 lines targeted and 11 were detected, and Band 8 targeted 9 lines, with five lines detected.

At only 6.5~hours of observing time, the ANGELS survey targeted one line every four minutes, and detected a line every six minutes. Since each band required around one hour (and Band 8 required 1.5 hours), the Band~6 tunings were most efficient, at one line detected {every three minutes}. This is faster than even efficient blind redshift surveys of sub-mm bright sources such as BEARS, which detected around 150~spectral lines across roughly 30 hours of observation time (12 min. per line). We note that this method also used the optimized observations of clusters of sources (see Fig.~\ref{fig:ANGELSFIELDS}) to increase the efficiency. Similarly, the $z$-GAL NOEMA survey \citep{Cox2023,Berta2023,Ismail2023} managed to detect 318 emission lines for 135 \textit{Herschel}-selected galaxies, resulting in a detection rate of about one line every 30 minutes. As a rough estimate, every targeted observation requires between twenty and thirty minutes of overheads such as calibration steps, reducing even surveys of bright sources to an efficiency of one line per every thirty minutes. 
As such, clustering is an important tool for future spectral surveys, and the ANGELS method demonstrated here offers at the very least a doubling, but even an order-of-magnitude increase in the efficiency of line surveys with ALMA on bright targets.

\subsection{Caveats with the ANGELS method}
We note several short-comings of this method, which could be used as caveats for future attempts to do efficient line surveys with ALMA. 
Firstly, the ANGELS method is efficient, but does not provide the homogeneity of targeted observations with ALMA. In fact, these observations are not deep enough to guarantee a detection, and if these observations were set to guarantee a detection, the efficiency boost of ANGELS would reduce significantly. Detection bias can strongly influence the scientific consensus surrounding important topics, such as the gas depletion time or ionization state of galaxies \citep[e.g.,][]{Bonato2014}. Lines that remain below the detection limit could be teased out using line stacking. Here we note an important caveat of line stacking, which reflects only the behaviour of a group of galaxies (see the Section below). Since DSFGs are found to be a diverse population, line stacking alone cannot reveal more than population-wide behaviour, although once large-enough samples of data are explored, AB tests (i.e., comparing one subsample directly to another) in stacking might provide a solution to this problem.

Secondly, there exist a few more variables that can be optimized towards future ANGELS observations. The optimized tuning set-up for each cluster was determined on pre-defined clusters of galaxies. Sources on the edges of these clusters could switch between clusters, providing potentially more efficient or targeted observations to increase the total number of detected spectral lines. 
Meanwhile, the parametrization of the number of lines is based on a fixed set-up of spectral windows, which provide the widest bandwidth for all Bands except for Band 6. Band 6 has a variable intermediate frequency which has not been further optimized during these ANGELS observations, and would require higher-order optimization steps to fully explore.
The method for finding the tunings to detect as many lines as possible (Appendix Fig.~\ref{fig:optiTuneGraph}) does not account for the chance to detect lines that were already observed in previous (BEARS) observations. The method also did not highlight that several sources, such as HerBS-159, did not have any detected spectral lines in their observations. 
Figure~\ref{fig:optiTuneGraph} does account for an estimate of the atmospheric windows. However, it is hard to notice where the spectral lines exactly fall within these windows, and the effect of poor transparency is hard to properly take into account. Line species or even line types are clumped together, and as the sensitivity of receivers change with frequency, these graphs do not fully represent the difficulty of detecting specific species.

Lastly, although this has not turned out to be a major issue with our observations, these graphs do not account for the change in primary beam sensitivity with increasing frequency. As several of our sources were known multiples, we are unable to resolve some sources across all bands.

\subsection{Future perspectives on complete \textit{Herschel} samples}
Figure~\ref{fig:ANGELSFIELDS} offers an alluring peek on the potential of ANGELS on the complete sample of southern {\it Herschel} sources. The current observations target 16 sources in just 6.5~hours. To cover the full 88~southern {HerBS} sources, we would require around thirty hours. Scaling up the number of detected lines, we could expect to detect $\sim 350$~lines, exceeding even the most successful high-$z$ large programmes with ALMA and NOEMA \citep[e.g.,][]{LeFevre2020}. On top of expanding the catalogues of high-$J$ CO SLEDs, in- and outflowing lines, and atomic lines, a full-scale SGP survey would enable a comprehensive stacked spectrum into the Terahertz regime, a crucial next step until space-based missions such as the Origins Space Telescope \citep{Meixner2019} take flight. The ability to characterize 88 sources, notwithstanding multiples (roughly an additional 30\% of sources) would enable sample comparisons on the order of 10~\%, and could better test the claims made on this smaller sample of ANGELS pilot souces.

These pilot observations use the ALMA Bands with relatively-good atmospheric transmission. Observations in Bands 9 and 10 would be technically difficult because of the atmospheric transmissivity, {\color{referee} but would also offer a unique view on the short-wavelength dust continuum and on bright atomic lines, which are important diagnostics of the ISM \citep[e.g., \oi{}, \oiii{}, \cii{}, \nii{} and \niii{}; ][]{RamosPadilla2023}.} The initial {\it Herschel} observations are to a greater or lesser extent affected by source confusion \citep{Bendo2023}, and a per-band view on these (and all southern) ANGELS sources would reveal the dust temperature in a pixel-by-pixel fashion, find any internal properties such as strong nuclear emission \citep{Tsukui2023}, and provide a solid foundation to study the complete dust emission of these dusty galaxies. 

One of the main aims of ALMA development is the wideband sensitivity upgrade \citep{Carpenter2023}. The goal of ALMA in 2030 is to double the receiver bandwidth, with a final goal of a quadrupling of the bandwidths. Here we estimate the effect this would have on the efficiency of the ANGELS method beyond 2030.
Appendix Figure~\ref{fig:optiTuneGraph} is the {\it tool} used to derive the optimum tunings. As a mathematical abstraction, this graph depicts a convolution of the available lines (i.e., a set of delta functions at their redshifted, observed wavelengths) with the observable bandwidths across Bands~3 through 8. These graphs would look like flat lines for both extremes in bandwidths; a limitless spectral coverage would not require (or allow) optimization, while a very narrow spectral windows would quickly reduce the probability of finding lines across multiple sources in a single set-up. The current set-up sees at least a few lines per Band, so as the bandwidths increase from their current state, the peak values would increase. However, the troughs would increase faster still, reducing the relative effect of optimization (i.e., the difference between the best and worst tuning set-ups). 

This does not mean that the ANGELS method becomes redundant. Firstly, these pilot observations were not deep enough to detect all targeted lines, particularly across the fainter sources. By increasing the integration time of our short snapshots to even a modest 5~minutes per source, the number of sources in a cluster identified by the ALMA Observing Tool would reduce rapidly (i.e., within 10 degrees on the sky). This would leave fewer lines per selected set-up, again increasing the importance of optimization. Secondly, this pilot aims to target {\it as many lines as possible}. More science-case guided observations (instead of proof-of-principle) might want to target only specific line transitions such as CO(1--0), atomic lines or \ci{}. With an increasing bandwidth, it becomes easier to design experiments that are able to widely target specific lines while also increasing the serendipity of the observations. Instead of designing the observations to place the targeted spectral line in the middle of the wider bandwidth, the ALMA tool should consider the possibility to pick up lines from multiple sources without the need for changing the local oscillator and the subsequent recalibration. 
Another such use-case might overcome the current limit on the resolution. The ANGELS strategy does not work with the ALMA configurations with a baseline greater than 3~km (C43-7 and beyond), which require sources to be within 1 degree on the sky instead of 10. Although several sources are on the sky closer than this (particularly in lower-flux selections), this would make the current version of the ANGELS method difficult to execute on a large sample.
Finally, these observations provide a showcase towards the careful inspection of the data cubes of observations without the explicit goal to detect lines. As ALMA approaches ultra-wide data cubes, careful exploration of ancillary lines in the spectra becomes more important, in particular when tools such as Machine Learning become better able to assist in the full exploitation of ALMA data \citep{Guglielmetti2023}.

\section{Conclusions}
\label{sec:conclusions}
The ANGELS observations have efficiently shared calibrators to facilitate a spectral survey of sixteen sources in just 6.5 hours of observation time. These observations revealed  
\begin{itemize}
\renewcommand\labelitemi{\small \textbf{$\blacksquare{}$}}
\item ALMA targeted 94 lines and detected 66 of these lines. The majority (54) were seen in emission, with an additional twelve lines seen in absorption.
\item The dust continuum observations across six bands revealed the intrinsic nature of the sample, and for two diffuse sources, tapering was required to image them at $> 5 \sigma$ significance. Of the 26 targets that have been observed, 21 sources have spectroscopic redshifts that enabled detailed studies of their emission and absorption lines. 
\item The CO SLEDs show a variation of sources, with extreme SLEDs shown for a small fraction (36\%) of sources, which suggests a fraction of DSFGs are dominated by dense, warm gas.
\item {\color{referee} The observations of 12 absorption line features show one rare case of inflowing gas through a red-shifted molecular absorption lines, and two cases of strongly outflowing gas through blue-shifted emission. The sources that are seen in multiple molecular absorption lines have similar systematic velocity offsets, suggesting that the absorption features are sensitive to similar regions of the circum-galactic and internal gas.}
\item 
    Atomic lines have been detected in three sources, namely: the \oil{} emission line in HerBS-41A, and the \niil{} emission line seen in HerBS-170 and HerBS-184, with HerBS-170 being the strongest \nii{} emitter found to date. 
    \item A stacked spectrum offers a preliminary view on the capabilities of a large sample, which will be able to characterize DSFGs out to $\approx 2000$~GHz.
    \item A preliminary bimodal picture of DSFGs is appearing, where lensed sources are indicative of typically central dusty star-bursts, found in other surveys to be rotationally stable, while the weakly- or unlensed HyLIRGs could represent turbulent massive star-forming events driven by unstable galaxy kinematics and clumpy gas flows.
\end{itemize}

Exploitation of the complete southern Herschel bright sources samples could enable a large line survey of $\sim$350 molecular and atomic lines, providing robust statistics with a fidelity of $\sim$10 per cent. This increase in fidelity will help characterize the diversity of DSFGs at \textit{cosmic noon}, in order to investigate the origins of the most violent star-forming events in the Universe. Similarly, studies with short-wavelengths will enable a complete characterization of the dust, target short-wavelength emission lines, and allow for a more accurate separation of the unresolved \textit{Herschel} fluxes. With the advent of the WSU \citep{Carpenter2023}, this method will become even more efficient -- and arguably important to include in most observations -- as the bandwidths of ALMA receivers and capabilities of the correlators increase in the near future. 

\section*{Acknowledgements}
{\color{referee2} The authors kindly thank the anonymous referee for their insightful comments and suggestions to improve this manuscript.}
Financial support from the Knut and Alice Wallenberg foundation is gratefully acknowledged through grant no. KAW 2020.0081.
This work was supprted by NAOJ ALMA Scientific Research Grant Numbers 2018-09B and JSPS KAKENHI No. 17H06130, 22H04939, and 22J21948, 22KJ1598.
SS was partly supported by the ESCAPE project; ESCAPE - The European Science Cluster of Astronomy and Particle Physics ESFRI Research Infrastructures has received funding from the European Union’s Horizon 2020 research and innovation programme under Grant Agreement no. 824064. SS also thanks the Science and Technology Facilities Council for financial support under grant ST/P000584/1. SU would like to thank the Open University School of Physical Sciences for supporting this work. 
This paper makes use of the following ALMA data: ADS/JAO.ALMA\#2016.2.00133.S, 2018.1.00804.S, 2019.1.01477.S, and 2021.1.01628.S. 

%%%%%%%%%%%%%%%%%%%%%%%%%%%%%%%%%%%%%%%%%%%%%%%%%%
\section*{Data Availability}
The data underlying this article will be shared on reasonable request to the corresponding author.



\bibliographystyle{mnras}
\bibliography{1_example} % if your bibtex file is called example.bib

\section*{Affiliations}
{\noindent
\small
\textit{$^{1}$Department of Space, Earth, \& Environment, Chalmers University of Technology, Chalmersplatsen 4 412 96 Gothenburg, Sweden \\
$^{2}$Department of Physics, Graduate School of Science, Nagoya University, Aichi 464-8602, Japan \\ 
$^{3}$National Astronomical Observatory of Japan, 2-21-1, Osawa, Mitaka, Tokyo, Japan\\
$^{4}$Centre for Extragalactic Astronomy, Durham University, Department of Physics, South Road, Durham DH1 3LE, UK\\
$^{5}$UK ALMA Regional Centre Node, Jodrell Bank Centre for Astrophysics, Department of Physics and Astronomy, \\University of Manchester, Oxford Road, Manchester M13 9PL, UK\\
$^{6}$Hiroshima Astrophysical Science Center, Hiroshima University, 1-3-1 Kagamiyama, Higashi-Hiroshima, Hiroshima 739-8526, Japan \\
$^{7}$School of Physical Sciences, The Open University, Milton Keynes, MK7 6AA, UK\\
$^{8}$Departamento de Fisica, Universidad de Oviedo, C. Federico Garcia Lorca 18, E-33007 Oviedo, Spain\\
$^{9}$Instituto Universitario de Ciencias y Tecnologas Espaciales de Asturias (ICTEA), C. Independencia 13, E-33004 Oviedo, Spain\\
$^{10}$INAF, Osservatorio Astronomico di Padova, Vicolo Osservatorio 5, I-35122 Padova, Italy\\
$^{11}$Institut d’Astrophysique de Paris, Sorbonne Universit\'{e},UPMC Universit\'{e} Paris 6 and CNRS, UMR 7095, 98 bis boulevard Arago, F-75014 Paris, France\\
$^{12}$European Southern Observatory, Alonso de Córdova 3107, Vitacura, Casilla 19001, Santiago de Chile, Chile\\
$^{13}$European Southern Observatory (ESO), Karl-Schwarzschild-Strasse~2, D-85748 Garching, Germany\\
$^{14}$School of Cosmic Physics, Dublin Institute for Advanced Studies, 31 Fitzwilliam Place, Dublin D02 XF86, Ireland\\
$^{15}$Institute for Astronomy, University of Edinburgh, Royal Observatory, Blackford Hill, Edinburgh EH9 3HJ, UK\\
$^{16}$ARC Centre of Excellence for All Sky Astrophysics in 3 Dimensions (ASTRO 3D)\\
$^{17}$School of Earth and Space Exploration, Arizona State University, Tempe, AZ 85287-6004, USA\\
$^{18}$Department of Astronomy, University of Cape Town, Private Bag X3, Rondebosch 7701, Cape Town, South Africa\\
$^{19}$INAF, Instituto di Radioastronomia-Italian ARC, Via Piero Gobetti 101, I-40129 Bologna, Italy\\
$^{20}$I. Physikalisches Institut, Universit\"at zu K\"oln, Z\"ulpicher Strasse 77, D-50937 K\"oln, Germany\\
$^{21}$Research School of Astronomy and Astrophysics, Australian National University, Cotter Road, Weston Creek, ACT 2611, Australia\\
$^{22}$Leiden Observatory, Leiden University, PO Box 9513, NL-2300 RA Leiden, Netherlands\\
$^{23}$INAF - Osservatorio Astronomico di Roma, via di Frascati 33, I-00078 Monte Porzio Catone, Italy\\
$^{24}$Institut de Radioastronomie Millimétrique (IRAM), 300 Rue de la Piscine, 38400 Saint-Martin-d’H\`{e}res, France\\
$^{25}$Department of Physics \& Astronomy, University of California, Irvine, 4129 Reines Hall, Irvine, CA 92697, USA\\
$^{26}$School of Physics and Astronomy, Cardiff University, The Parade, Cardiff, CF24 3AA, UK\\
$^{27}$Department of Astronomy, School of Science, SOKENDAI (The Graduate University for Advanced Studies), 2-21-1 Osawa, Mitaka, Tokyo 181-8588, Japan \\
$^{28}$Department of Astronomy, The University of Tokyo, 7-3-1 Hongo, Bunkyo, Tokyo 113-0033, Japan\\
$^{29}$Department of Physics and Astronomy, University of British Columbia, 6224 Agricultural Road, Vancouver, BC V6T 1Z1, Canada\\
$^{30}$Astrophysics Branch, NASA—Ames Research Center, MS 245-6, Moffett Field, CA 94035, USA\\
}}

%%%%%%%%%%%%%%%%%%%%%%%%%%%%%%%%%%%%%%%%%%%%%%%%%%

%%%%%%%%%%%%%%%%% APPENDICES %%%%%%%%%%%%%%%%%%%%%

\appendix

\section{Graph to optimize the observing bands}
Figure~\ref{fig:optiTuneGraph} shows the tool used to define the optimum observing conditions for efficiently targeting as many spectral lines as possible. 

\begin{figure*}
    \centering
    \includegraphics[width=1.2\linewidth,angle
    =90]{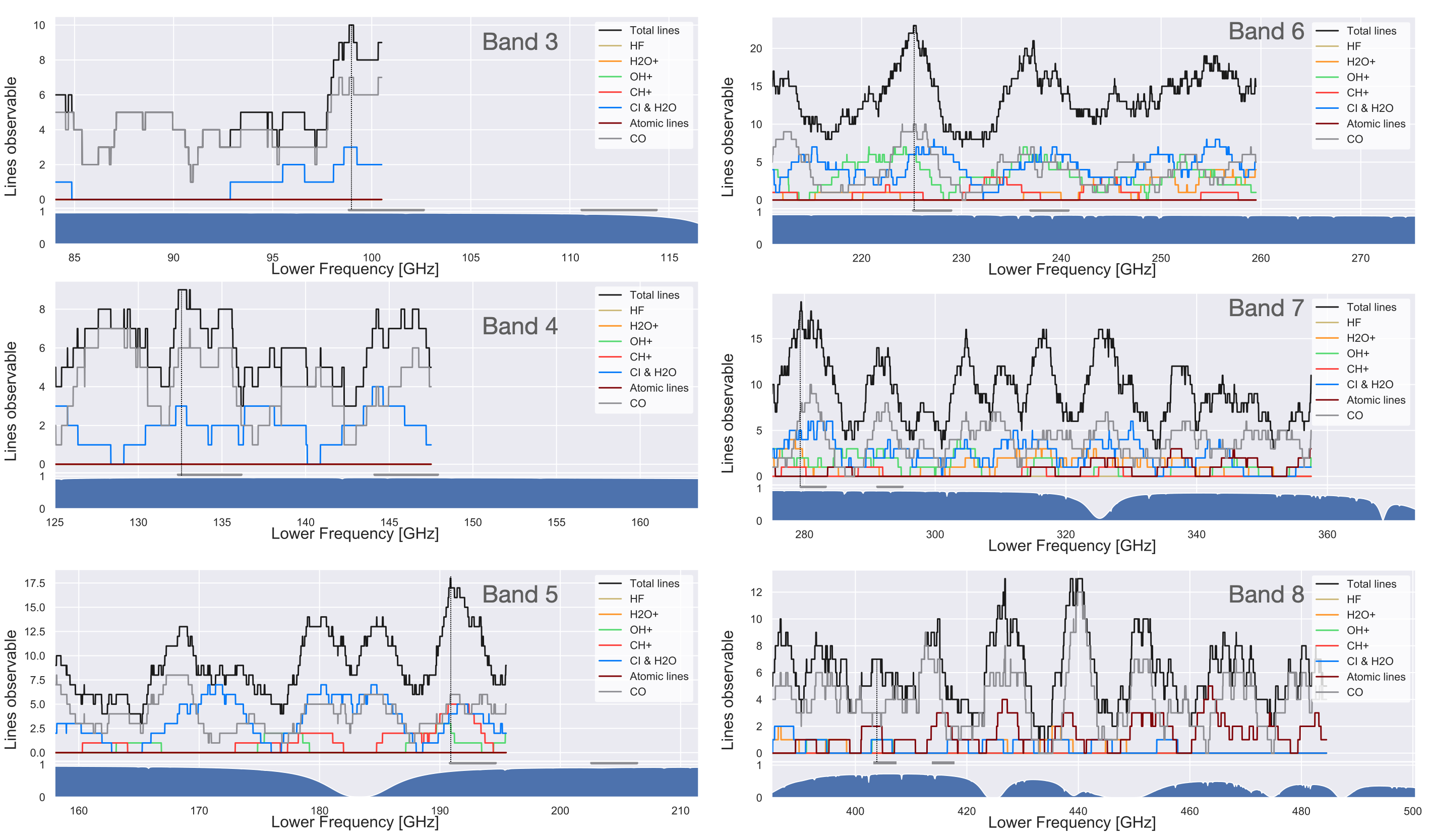}
    \caption{These six graphs were used to find the optimum tuning of each of the bands to target the most lines given a fixed spectral set-up. The \textit{x}-axis shows the frequency set-up by indicating the lowest frequency of the spectral windows, and the lower- and upper-sidebands are separated by 8~GHz. The grey bar indicates the selected frequency set-up, and the dashed line indicates the total number of expected lines that are targeted with the observations.
    The optical transparency of the atmosphere is indicated by the bottom panel of each graph, between 1 (fully transparent) and 0 (fully opaque), in order to evaluate the effect of the optical transmission while targeting as many lines as possible. 
    Notable absorptions include the water feature in Band 5, a central absorption component in Band 7, and the majority of Band 8. As such, we did not always choose the highest position in the top panel for our observations. }
    \label{fig:optiTuneGraph}
\end{figure*}

\section{Sources without spectroscopic redshifts}
In this section we report the seven sources without spectroscopic redshifts that were included in the field-of-view of ALMA. We present their continua in Figure~\ref{fig:offsetSources}. 
\label{sec:offsetSources}
\begin{figure*}
    \centering
    \includegraphics[width=\textwidth]{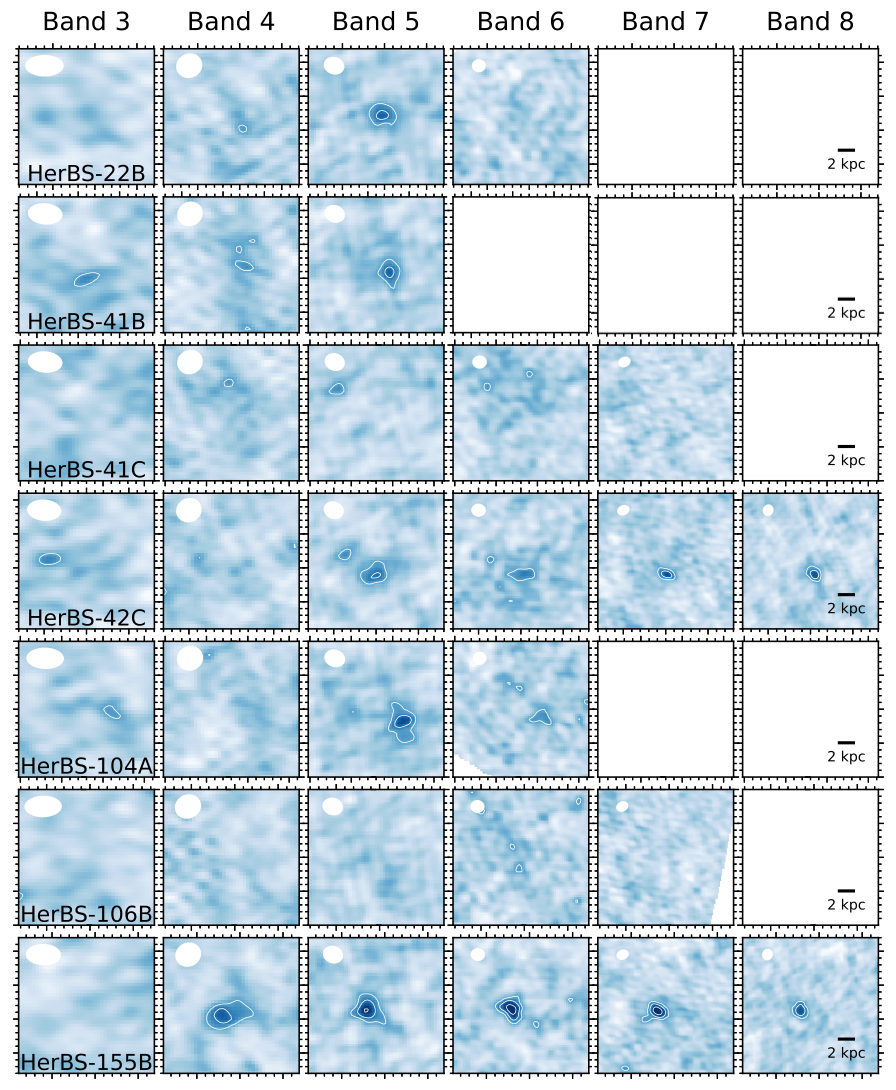}
    \caption{These sources are not the main targets in the ANGELS data, but were found through unresolved BEARS observations \citep{Bendo2023}. They do not have spectroscopic redshifts, and the majority are not completely covered by the primary beam ({\sc pb-cor} $ < 0.3$) for the highest frequencies. Here, we show the images before the primary beam correction to better showcase the source across the bands.}
    \label{fig:offsetSources}
\end{figure*}

\section{Spectral line profiles}
\label{sec:AppendixSpectra}
The tabulated measured properties of the spectral lines are shown in Table~\ref{tab:spectralLineData}. All detected lines are shown in Figures~\ref{fig:ANGELSSpectra}.

\begin{onecolumn}
\begin{longtable}{lccccc}
\caption{Line tables} \label{tab:spectralLineData} \\
\hline \hline
Line & Frequency 		& Band 	&  SdV 		& dV 	& V - V$_{\rm CO}$		  	\\
 	 & [GHz]	 		& 		&  [Jy km/s] & [km/s]  & [km/s] 	 	\\
\hline
\endfirsthead

\multicolumn{6}{c}%
{{\tablename\ \thetable{} -- {\it continued from previous page}}} \\
\hline \hline
Line & Frequency 		& Band 	& SdV 		& dV 	& V - V$_{\rm CO}$ 		  	\\
 	 & [GHz]	 		& 		& [Jy km/s] & [km/s] 	 & [km/s]  	\\ \hline 
\endhead

\multicolumn{6}{r}{{\it Continued on next page}} \\ \hline \hline
\endfoot

\hline \hline
\endlastfoot

\multicolumn{6}{c}{\bf HerBS-21 z = 3.323} \\ 
 $[\textsc{C\,i}]$ $609 \mu m$ & 113.847 & band 3 &  $< 1.54$ &  \\
 CO (5--4) & 133.251 & band 4 &  14.18 $\pm$ 3.34 & 521 $\pm$ 92 & -23 $\pm$ 39 \\
 CH$^+$ (1$-$0) & 193.171 & band 5 &  -1.21 $\pm$ 0.75 & 360 $\pm$ 169 & 205 $\pm$ 71 \\
 OH$^+$ ($1_{2} - 0_{1}$) & 224.799 & band 6 &  -6.1 $\pm$ 1.45 & 439 $\pm$ 78 & 189 $\pm$ 33 \\
 OH$^+$ ($1_{1} - 0_{1}$) & 238.982 & band 6 &  -2.54 $\pm$ 0.95 & 177 $\pm$ 50 & 142 $\pm$ 21 \\
 CO (9--8) & 239.851 & band 6 &  8.99 $\pm$ 2.82 & 642 $\pm$ 152 & 13 $\pm$ 64 \\
 CO (11--10) & 293.311 & band 7 &  4.91 $\pm$ 1.4 & 416 $\pm$ 89 & -109 $\pm$ 37 \\

\multicolumn{6}{c}{\bf HerBS-22 z = 3.051} \\
 CO (4--3) & 	113.820 & band 3 &	 11.72 $\pm$ 3.25 & 568 $\pm$ 118 & 34 $\pm$ 50 \\
 CH$^+$ (1$-$0) & 206.141 & band 5 &  -0.75 $\pm$ 0.36 & 107 $\pm$ 39 & -138 $\pm$ 15 \\
 OH$^+$ ($1_{0} - 0_{1}$) & 224.428 & band 6 &  -1.84 $\pm$ 0.75 & 178 $\pm$ 55 & -65 $\pm$ 23 \\
 OH$^+$ ($1_{2} - 0_{1}$) & 239.893 & band 6 &  -5.14 $\pm$ 1.0 & 470 $\pm$ 0 & -140 $\pm$ 44 \\
 H$_2$O$^+$ ($1_{1,1} - 0_{0,0}$) & 281.373 & band 7 &  11.89 $\pm$ 3.19 & 941 $\pm$ 163 & 869 $\pm$ 93 \\

\multicolumn{6}{c}{\bf HerBS-25 z = 2.912} \\
 CO (5--4) & 147.324 &	band 4 &  18.96 $\pm$ 2.17 & 226 $\pm$ 19 & 9 $\pm$ 8 \\
 H$_2$O ($2_{1,1} - 2_{0,2}$) & 192.238 & band 5 &  6.35 $\pm$ 2.81 & 478 $\pm$ 160 & -73 $\pm$ 68 \\
 CO (7--6) & 206.220 & band 5 &  12.61 $\pm$ 3.66 & 249 $\pm$ 54 & -34 $\pm$ 23 \\
 H$_2$O ($3_{1,2} - 3_{0,3}$) & 280.512 & band 7 &  5.93 $\pm$ 2.52 & 196 $\pm$ 63 & -106 $\pm$ 26 \\
 H$_2$O$^+$ ($1_{1,1} - 0_{0,0}$) & 291.299 & band 7 &  3.88 $\pm$ 2.7 & 136 $\pm$ 71 & -128 $\pm$ 30 \\
 CO (10--9) & 294.661 & band 7 &  17.76 $\pm$ 4.98 & 941 $\pm$ 185 & 357 $\pm$ 93 \\

\multicolumn{6}{c}{\bf HerBS-36 z = 3.095} \\
 CO (4--3) & 112.619 & band 3 &  6.88 $\pm$ 1.85 & 464 $\pm$ 94 & -18 $\pm$ 40 \\
 CH$^+$ (1$-$0) & 203.926 & band 5 &  -2.23 $\pm$ 0.38 & 283 $\pm$ 36 & 10 $\pm$ 15 \\
 CO (8--7) & 225.208 & band 6 &   5.80 $\pm$ 2.03 & 440 $\pm$ 116 & -4 $\pm$ 49 \\
 CO (10--9) & 281.541 & band 7 &	4.69 $\pm$ 1.79 & 392 $\pm$ 112 & 56 $\pm$ 48 \\
 CH$^+$ (2$-$1) & 407.638 & band 8 &  -5.21 $\pm$ 2.03 & 128 $\pm$ 43 & 64 $\pm$ 16 \\

\multicolumn{6}{c}{\bf HerBS-41A z = 4.098} \\
 CO (5--4) & 113.147 &	band 3 &  5.05 $\pm$ 1.64 & 710 $\pm$ 174 & 12 $\pm$ 74 \\
 CO (6--5) & 135.869 &	band 4 &  4.35 $\pm$ 1.51 & 842 $\pm$ 221 & 248 $\pm$ 93 \\
 H$_2$O ($2_{1,1} - 2_{0,2}$) & 147.515 & band 4 &   $<$ 1.89 \\
 H$_2$O ($2_{0,2} - 1_{1,1}$) & 193.787 & band 5 &  1.31 $\pm$ 0.97 & 647 $\pm$ 363 & -232 $\pm$ 154 \\
 OH$^+$ ($1_{1} - 0_{1}$) & 202.652 & band 5 &  -0.88 $\pm$ -0.47 & 349 $\pm$ 147 & -540 $\pm$ 57 \\
 CO (9--8) & 203.663 & band 5 &  3.44 $\pm$ 2.1 & 790 $\pm$ 363 & -37 $\pm$ 154 \\
 H$_2$O$^+$ ($1_{1,1} - 0_{0,0}$) & 223.531 & band 6 & $< 2.9$ \\
 CO (10--9) & 226.111 & band 6 &  1.59 $\pm$ 1.15 & 940 $\pm$ 524 & -21 $\pm$ 211 \\
 CO (13--12) & 293.944 & band 7 &  5.77 $\pm$ 1.2 & 941 $\pm$ 72 & -200 $\pm$ 89 \\
$[\textsc{O\,i}]$ $145 \mu m$ & 404.081 & band 8 &  10.27 $\pm$ 2.72 & 941 $\pm$ 135 & -481 $\pm$ 106 \\
 CO (18--17) & 406.999 & band 8 &  3.15 $\pm$ 1.11 & 941 $\pm$ 4 & -173 $\pm$ 299 \\

\multicolumn{6}{c}{\bf HerBS-42A z = 3.307} \\
 CO (5--4) & 133.787 & band 4 &	 13.42 $\pm$ 1.75 & 667 $\pm$ 65 & -18 $\pm$ 27 \\
 CH$^+$ (1$-$0) & 193.889 & band 5 &  $ < 1.13$  \\
 OH$^+$ ($1_{1} - 0_{1}$) & 225.634 & band 6 &  -0.4 $\pm$ 0.32 & 342 $\pm$ 203 & -418 $\pm$ 86 \\
 OH$^+$ ($1_{2} - 0_{1}$) & 239.87 & band 6 &   -0.73 $\pm$ 0.5 & 340 $\pm$ 177 & -182 $\pm$ 75 \\
 CO (9--8) & 240.775 & band 6 &  10.6 $\pm$ 2.84 & 647 $\pm$ 134 & -18 $\pm$ 53 \\
 CO (11--10) & 294.401 & band 7 &  7.54 $\pm$ 2.94 & 941 $\pm$ 190 & 255 $\pm$ 153 \\

\multicolumn{6}{c}{\bf HerBS-42B z = 3.314} \\
$[\textsc{C\,i}]$ $609 \mu m$ & 114.085 & band 3 &   $ < 2.28$ \\
CO (5--4) & 133.601 & band 4 &  3.35 $\pm$ 1.58 & 564 $\pm$ 201 & 16 $\pm$ 85 \\
CH$^+$ (1$-$0) & 193.574 & band 5 &   $< 1.67$ \\
CO (9--8) & 240.482 & band 6 &  1.84 $\pm$ 0.74 & 150 $\pm$ 45 & 42 $\pm$ 19 \\
OH$^+$ ($1_{2} - 0_{1}$) & 239.48 & band 6 & $< 3.44$  \\  
 CO (11--10) & 293.923 & band 7 & $<2.90$ \\

\multicolumn{6}{c}{\bf HerBS-81A z = 3.160} \\
CO (4--3) & 110.800 &	band 3 &  6.05 $\pm$ 2.05 & 587 $\pm$ 150 & -63 $\pm$ 63 \\
CO (7--6) & 194.002 &	band 5 &  4.38 $\pm$ 1.69 & 442 $\pm$ 129 & 149 $\pm$ 54 \\
$[\textsc{C\,i}]$ $370 \mu m$ & 194.509 & band 5  & $< 1.32$ \\
H$_2$O ($2_{0,2} - 1_{1,1}$) & 237.482 & band 6  & $< 1.54$ \\
H$_2$O ($3_{2,1} - 3_{1,2}$) & 279.546 & band 7  & 8.32 $\pm$ 4.78 & 762 $\pm$ 336 & -337 $\pm$ 137 \\
% CO (15--14) & 415.641 & band 8  & 

\multicolumn{6}{c}{\bf HerBS-81B z = 2.588} \\
CO (6--5) & 192.976 & band 5  & 3.16 $\pm$ 1.93 & 870 $\pm$ 401 & 61 $\pm$ 167 \\
CO (7--6) & 225.139 & band 6  & 2.46 $\pm$ 1.96 & 932 $\pm$ 563 & 60 $\pm$ 235 \\
$[\textsc{C\,i}]$ $370 \mu m$ & 225.769 & band 6  & $< 1.62$ \\
CO (13--12) & 418.115 & band 8  & $<3.5$\\
$[\textsc{N\,ii}]$ $205 \mu m$ & 407.678 & band 8  & $< 3.87$ \\

\multicolumn{6}{c}{\bf HerBS-86 z = 2.564} \\
 CO (6--5) & 194.045 &	band 5 &  5.36 $\pm$ 1.68 & 532 $\pm$ 126 & -3 $\pm$ 53 \\
 CO (7--6) & 226.320 &	band 6 &  5.32 $\pm$ 1.68 & 446 $\pm$ 106 & -8 $\pm$ 45 \\

\multicolumn{6}{c}{\bf HerBS-87 z = 2.059} \\
CO (3--2) & 113.066 &	band 3 & 5.16 $\pm$ 2.03 & 474 $\pm$ 156 & 70 $\pm$ 64 \\
CO (6--5) & 226.127 &	band 6 &  13.84 $\pm$ 4.06 & 280 $\pm$ 62 & 118 $\pm$ 26 \\

\multicolumn{6}{c}{\bf HerBS-93 z = 2.402} \\
 CO (3--2) & 101.638 &	band 3 &  4.01 $\pm$ 1.53 & 821 $\pm$ 237 & 0 $\pm$ 100 \\
 CO (4--3) & 135.547 &	band 4 &  8.72 $\pm$ 3.22 & 782 $\pm$ 218 & -46 $\pm$ 92 \\
 $[\textsc{C\,i}]$ $609 \mu m$ & 144.668 & band 4 &  4.79 $\pm$ 2.18 & 371 $\pm$ 140 & 203 $\pm$ 52 \\
 CO (6--5) & 203.420 & band 5 &  6.1 $\pm$ 2.08 & 941 $\pm$ 101 & 464 $\pm$ 215 \\
 $[\textsc{C\,i}]$ $370 \mu m$ & 237.987 & band 6 &  2.07 $\pm$ 1.34 & 155 $\pm$ 76 & 499 $\pm$ 3 \\
 CO (12--11) & 406.84 & band 8 &  $<2.67$ \\

\multicolumn{6}{c}{\bf HerBS-104 z = 1.536} \\
 $[\textsc{C\,i}]$ $609 \mu m$ & 194.070 & band 5 & 3.56 $\pm$ 1.0 & 324 $\pm$ 68 & 58 $\pm$ 29 \\
 OH$^+$ ($1_{1} - 0_{1}$) & 407.397 & band 8 &  -2.09 $\pm$ 1.03 & 121 $\pm$ 45 & 64 $\pm$ 19 \\

\multicolumn{6}{c}{\bf HerBS-106 z = 2.369} \\
 $[\textsc{C\,i}]$ $609 \mu m$ & 146.0854 & band 4 &  4.36 $\pm$ 1.78 & 229 $\pm$ 70 & -31 $\pm$ 29 \\
 CO (6--5) & 205.254 &	band 5 &  9.49 $\pm$ 2.12 & 542 $\pm$ 91 & -11 $\pm$ 39 \\
 H$_2$O ($2_{1,1} - 2_{0,2}$) & 223.221 & band 6 &  $< 1.20$\\
 CO (7--6) & 239.507 & band 6 &  5.02 $\pm$ 2.17 & 219 $\pm$ 71 & -14 $\pm$ 30 \\
 $[\textsc{C\,i}]$ $370 \mu m$ & 240.177 & band 6 &  23.39 $\pm$ 5.68 & 941 $\pm$ 116 & -58 $\pm$ 96 \\
 H$_2$O ($2_{0,2} - 1_{1,1}$) & 293.24 & band 7 &  $<4.91$\\

\multicolumn{6}{c}{\bf HerBS-155 z = 3.077} \\
 CO (4--3) & 113.121 &	band 3 & 5.32 $\pm$ 0.52 & 144 $\pm$ 15 & -20 $\pm$ 12 \\
 CH$^+$ (1$-$0) & 204.827 & band 5 &  $<  2.33$  \\
 OH$^+$ ($1_{0} - 0_{1}$) & 222.997 & band 6 &  $<2.82$ \\
 CO (8--7) & 226.150 &	band 6 &  5.36 $\pm$ 1.75 & 212 $\pm$ 52 & -2 $\pm$ 22 \\
 H$_2$O$^+$ ($1_{1,1} - 0_{0,0}$) & 279.51 & band 7 &  $ < 1.14$\\
 CO (10--9) & 282.735 & band 7 &  $< 6.18$ \\

\multicolumn{6}{c}{\bf HerBS-159A z = 2.236} \\
 OH$^+$ ($1_{0} - 0_{1}$) & 280.951 & band 7 &  $ < 2.6$  \\ 

\multicolumn{6}{c}{\bf HerBS-159B z = 2.236} \\
 OH$^+$ ($1_{0} - 0_{1}$) & 280.951 & band 7 &  $< 1.38$ \\

\multicolumn{6}{c}{\bf HerBS-170 z = 4.182} \\
 CO (5--4) & 111.188 & band 3 &  $< 2.07$ \\
 CO (6--5) & 133.426 & band 4 &  3.37 $\pm$ 1.59 & 421 $\pm$ 150 & -15 $\pm$ 63 \\
 H$_2$O ($2_{1,1} - 2_{0,2}$) & 145.079 & band 4 &  $< 2.10$ \\
 H$_2$O ($3_{2,1} - 3_{1,2}$) & 224.344 & band 6 &  4.8 $\pm$ 2.61 & 941 $\pm$ 479 & 43 $\pm$ 87 \\
 HF (1$-$0) & 237.765 & band 6 &  $< 2.3$ \\
 $[\textsc{N\,ii}]$ $205 \mu m$ & 281.873 & band 7 &  4.36 $\pm$ 1.93 & 304 $\pm$ 102 & 51 $\pm$ 43 \\

\multicolumn{6}{c}{\bf HerBS-184 z = 2.507} \\
 CO (3--2) & 98.577 &	band 3 &  4.32 $\pm$ 1.29 & 395 $\pm$ 89 & 6 $\pm$ 37 \\
 CH$^+$ (1$-$0) & 238.118 & band 6 &  $<1.74$ \\
 H$_2$O ($2_{0,2} - 1_{1,1}$) & 281.701 & band 7 &  1.39 $\pm$ 0.96 & 472 $\pm$ 245 & -22 $\pm$ 104 \\
 OH$^+$ ($1_{1} - 0_{1}$) & 294.587 & band 7 & $< 1.74$  \\
 $[\textsc{N\,ii}]$ $205 \mu m$ & 416.629 & band 8 &  4.15 $\pm$ 2.94 & 362 $\pm$ 194 & 4 $\pm$ 82 \\

\end{longtable}
\end{onecolumn}

\begin{figure}
\includegraphics[width=0.5\linewidth]{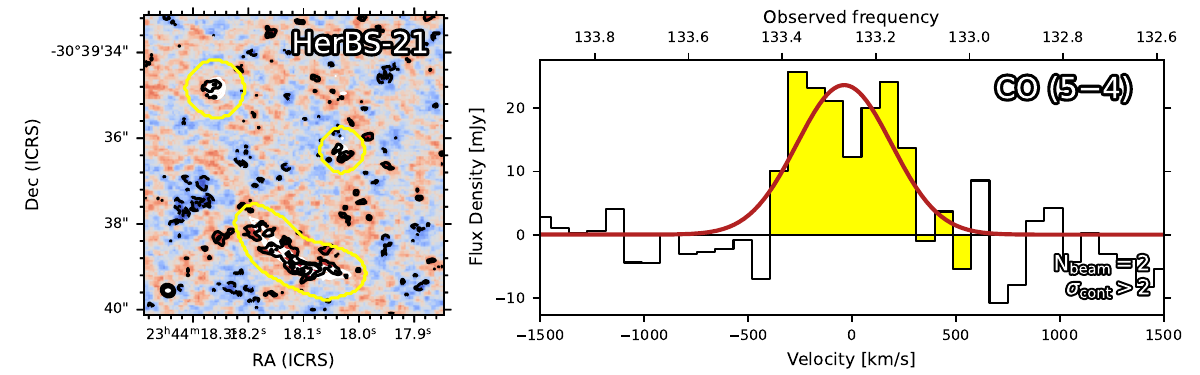}
\includegraphics[width=0.5\linewidth]{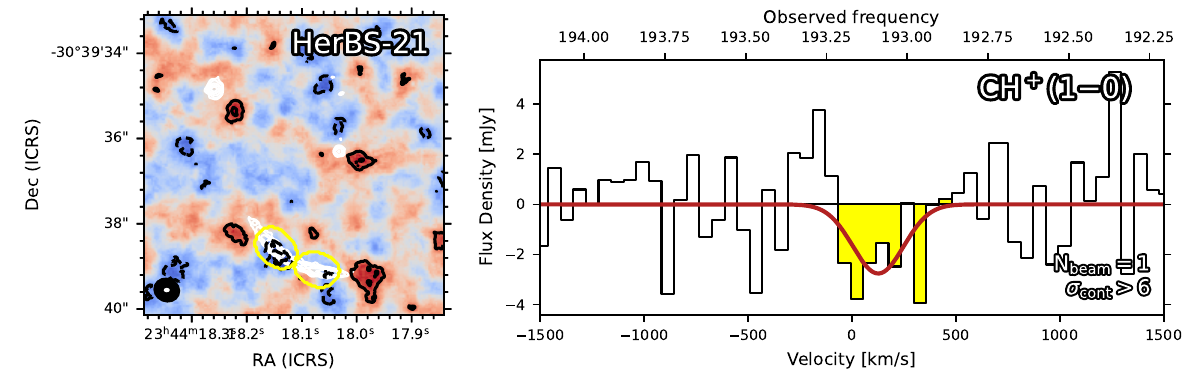}
\includegraphics[width=0.5\linewidth]{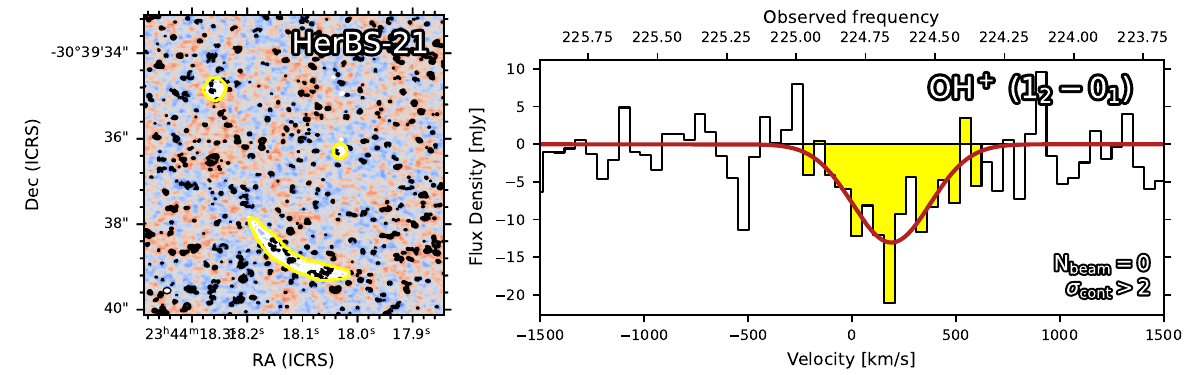}
\includegraphics[width=0.5\linewidth]{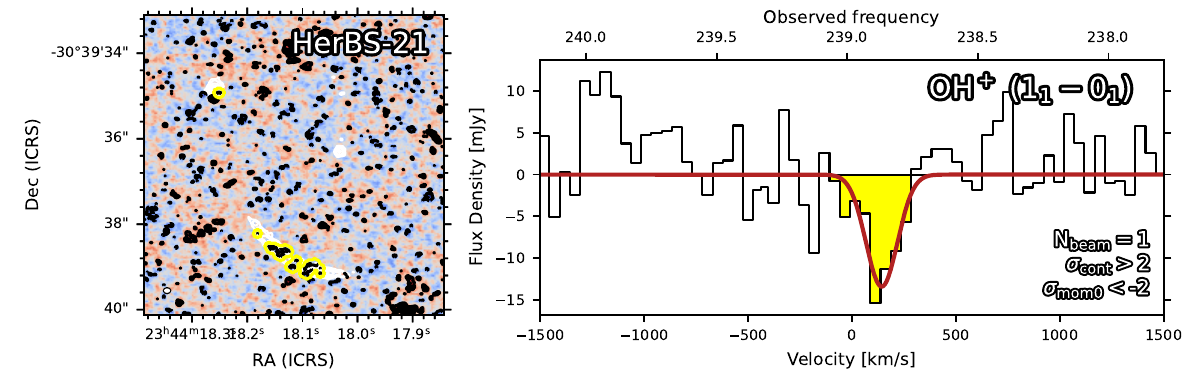}
\includegraphics[width=0.5\linewidth]{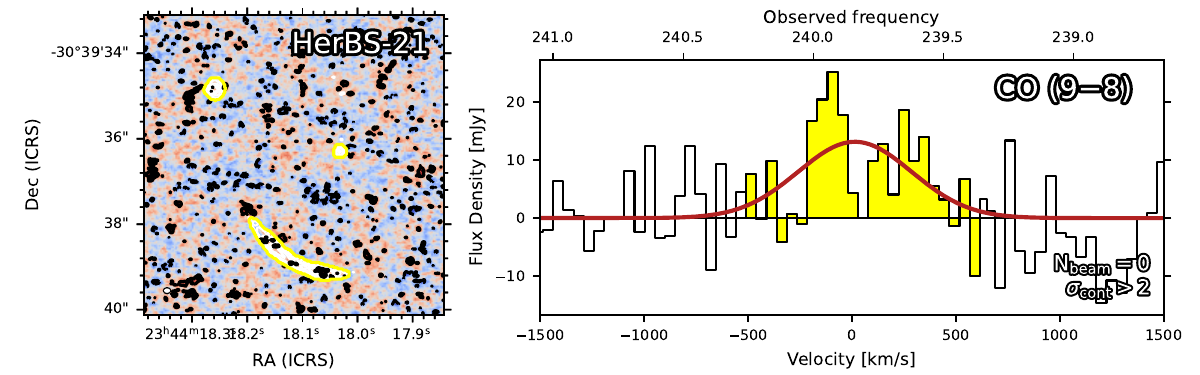}
\includegraphics[width=0.5\linewidth]{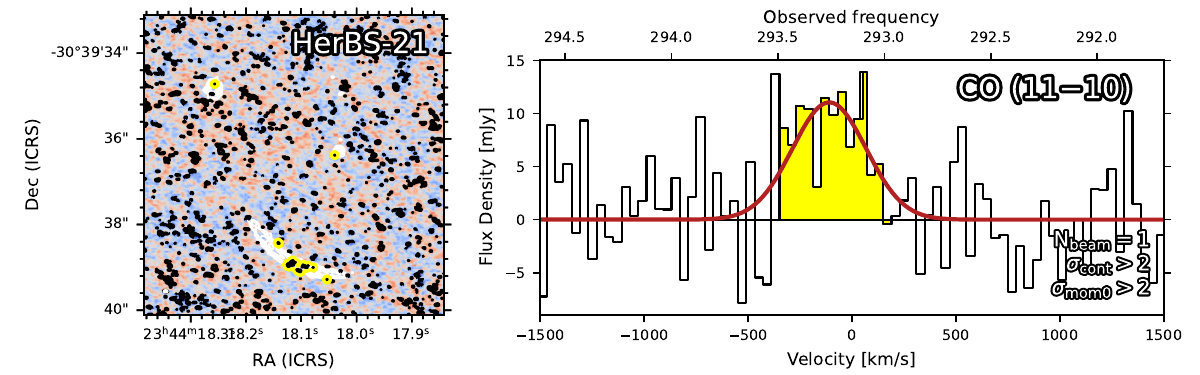}
\caption{The moment-0 (left) and line spectra (right) of HerBS-21 at $z = 3.323$. The white contours show the untapered band~7 image at $2, 3, 5... \sigma$, and the red solid / blue dashed contours show the $+ / - 2, 3, 5... \sigma$ moment-0 line emission. The yellow contour shows the selected aperture based on the selection criteria highlighted in the bottom-right of the spectrum, using an aperture based on the tapered band~7 continuum ($\sigma_{\rm cont}$)  and moment-0 ($\sigma_{\rm mom-0}$) emission, which are subsequently tapered by N$_{\rm beams}$ times the moment-0 beam. Similarly, the moment-0 image is created from the highlighted emission bins. The line properties listed in Table~\ref{tab:spectralLineData} are derived from the line fits.}
\label{fig:ANGELSSpectra}
\end{figure}\addtocounter{figure}{-1}
\begin{figure}
\includegraphics[width=0.5\linewidth]{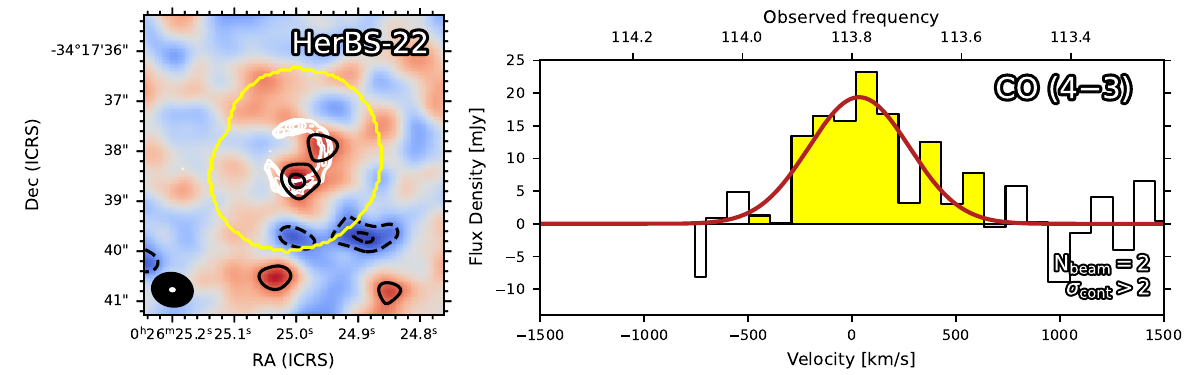}
\includegraphics[width=0.5\linewidth]{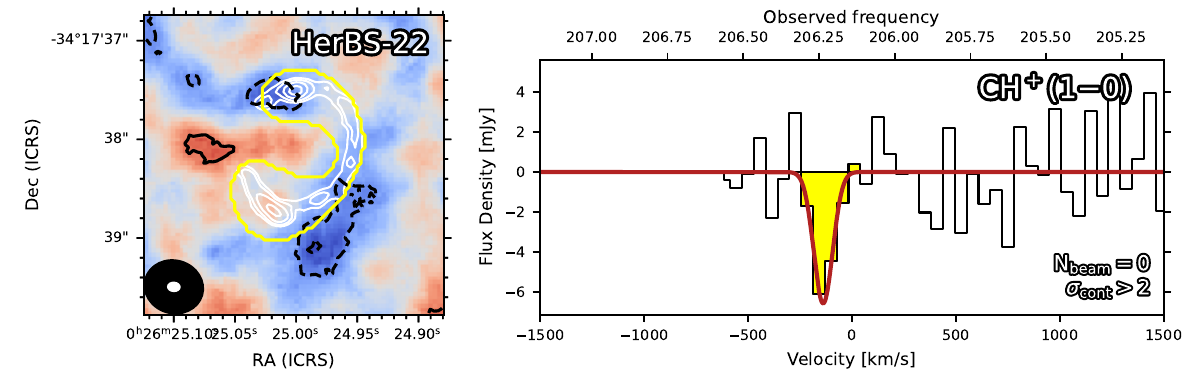}
\includegraphics[width=0.5\linewidth]{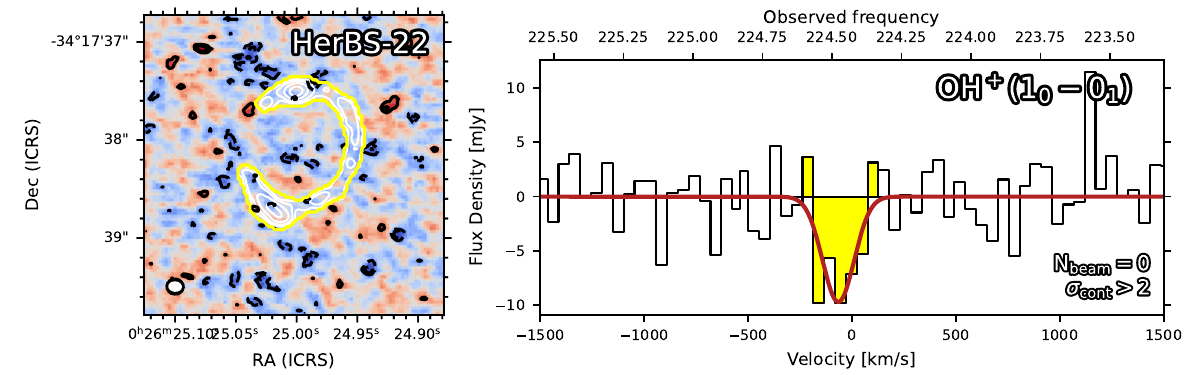}
\includegraphics[width=0.5\linewidth]{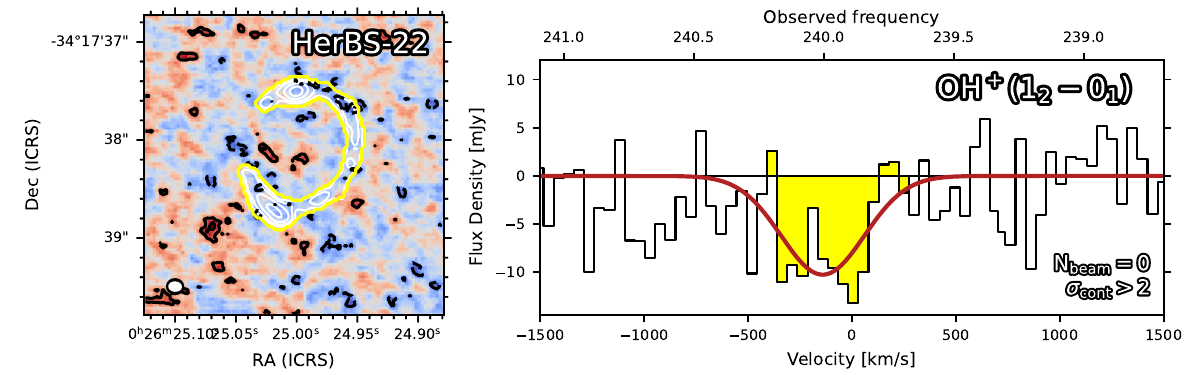}
\includegraphics[width=0.5\linewidth]{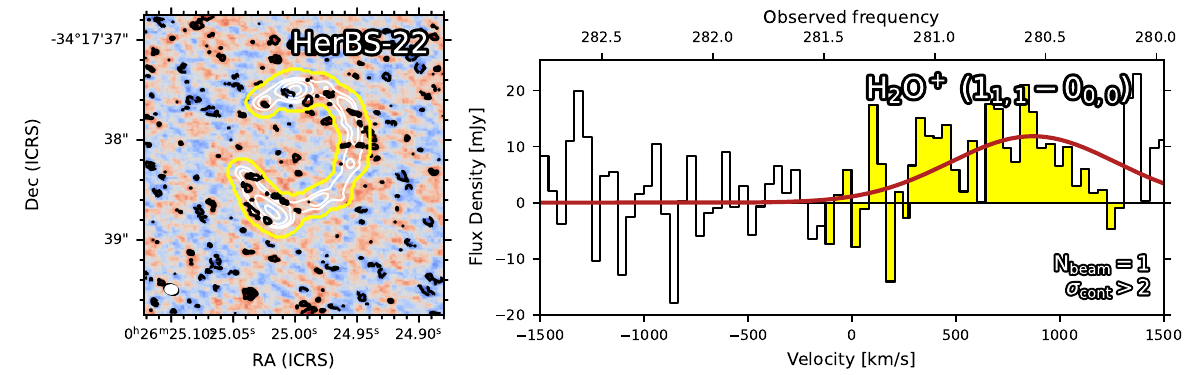}
\caption{The moment-0 (left) and line spectra (right) of HerBS-22 at $z = 3.050$}
\end{figure}\addtocounter{figure}{-1}
\begin{figure}
\includegraphics[width=0.5\linewidth]{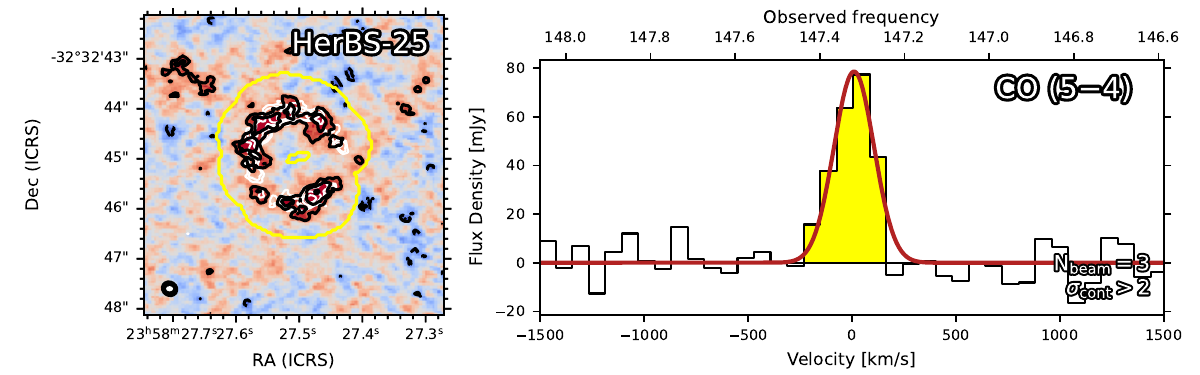}
\includegraphics[width=0.5\linewidth]{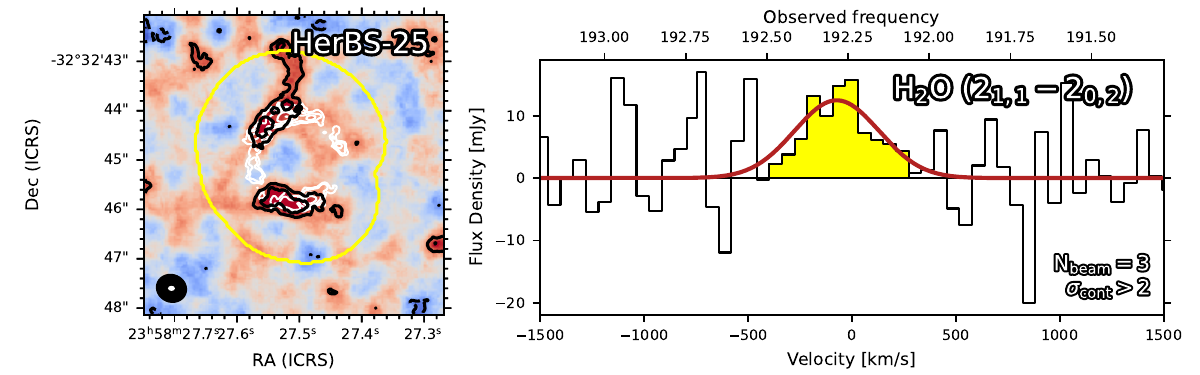}
\includegraphics[width=0.5\linewidth]{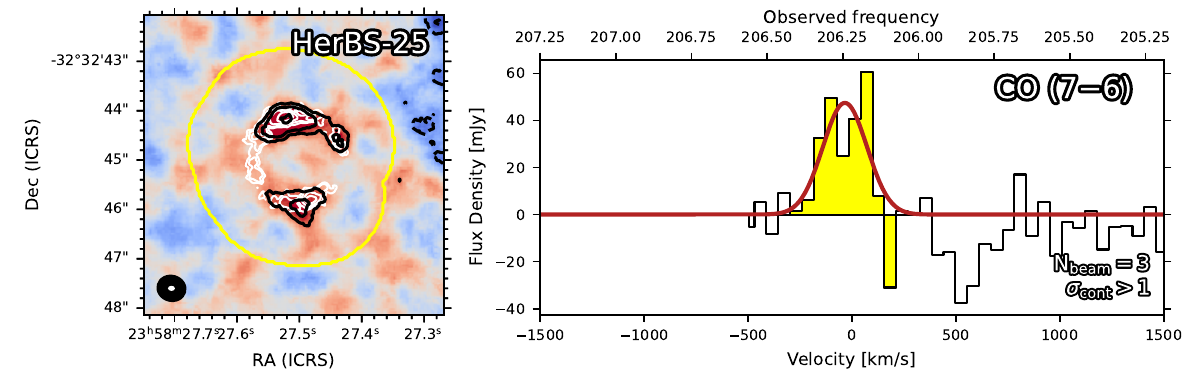}
\includegraphics[width=0.5\linewidth]{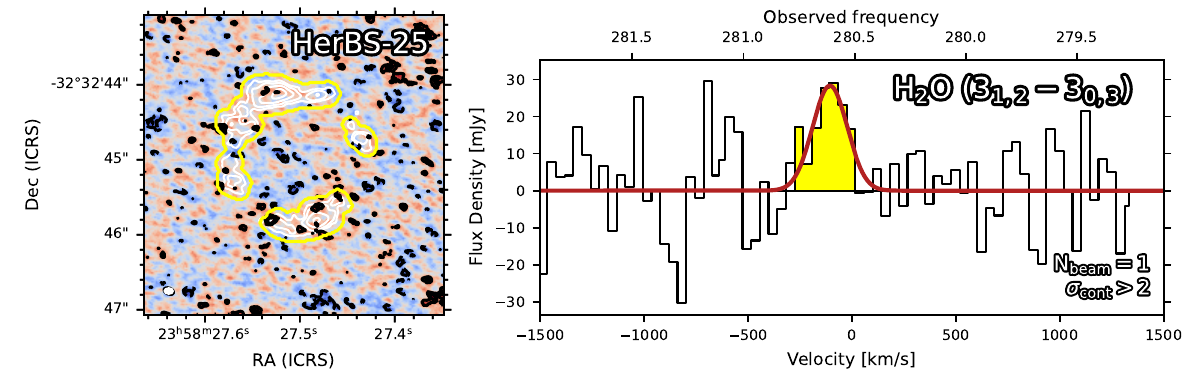}
\includegraphics[width=0.5\linewidth]{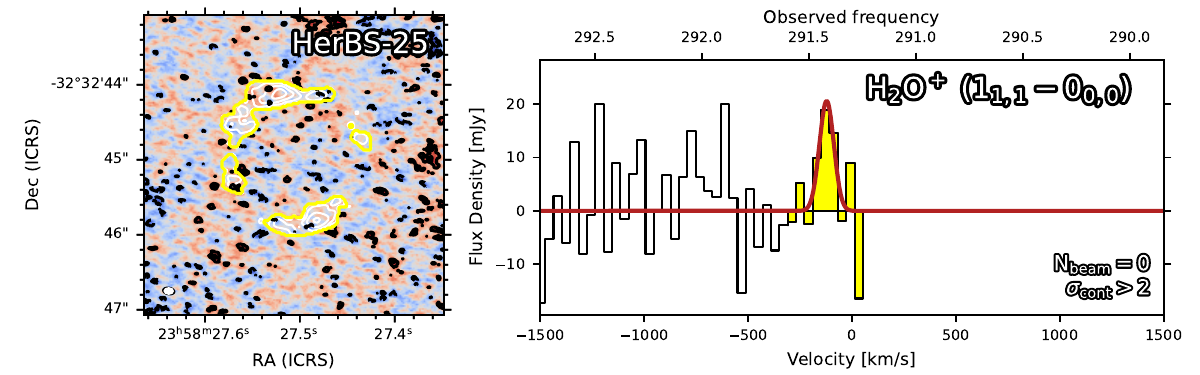}
\includegraphics[width=0.5\linewidth]{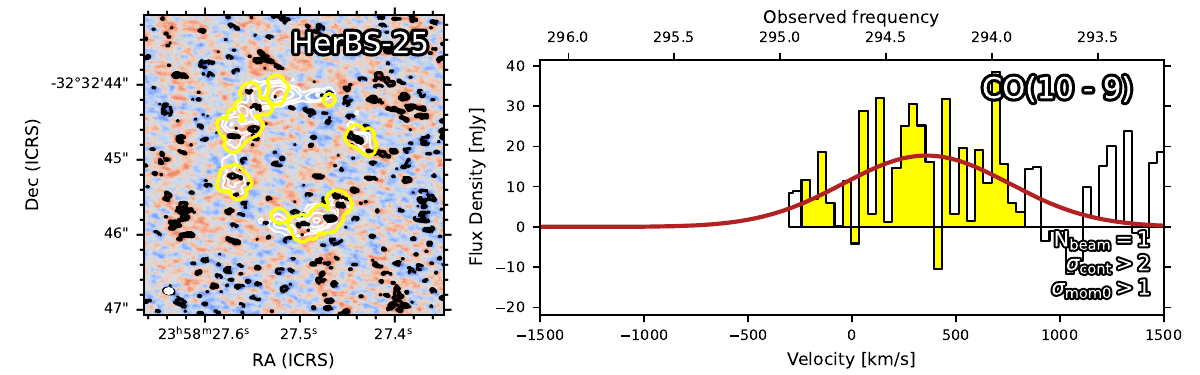}
\caption{The moment-0 (left) and line spectra (right) of HerBS-25 at $z =2.912 $}
\end{figure}\addtocounter{figure}{-1}
\begin{figure}
\includegraphics[width=0.5\linewidth]{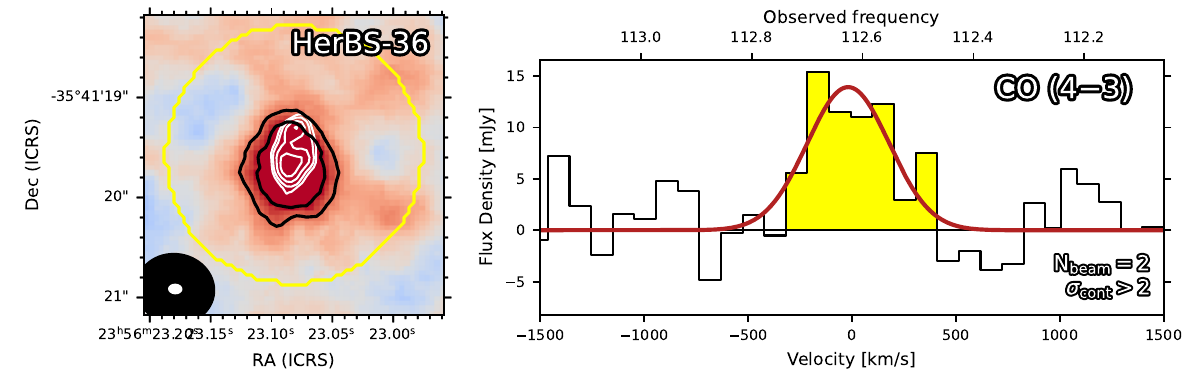}
\includegraphics[width=0.5\linewidth]{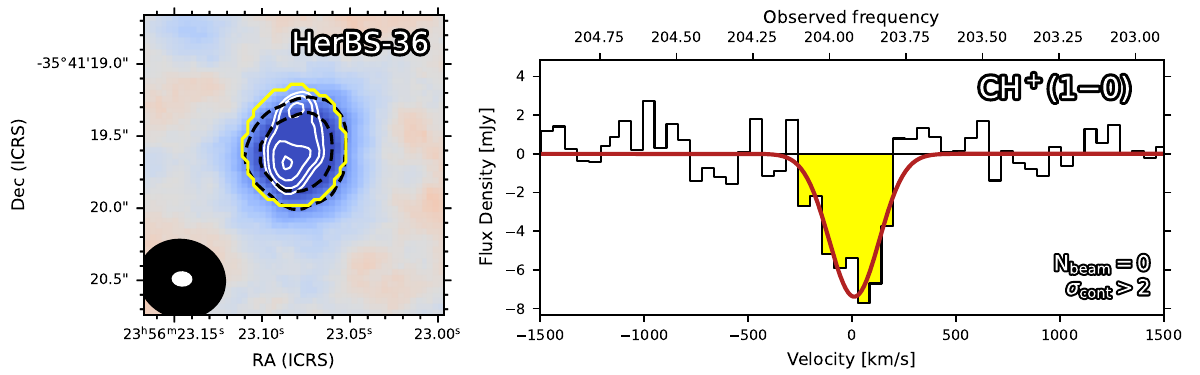}
\includegraphics[width=0.5\linewidth]{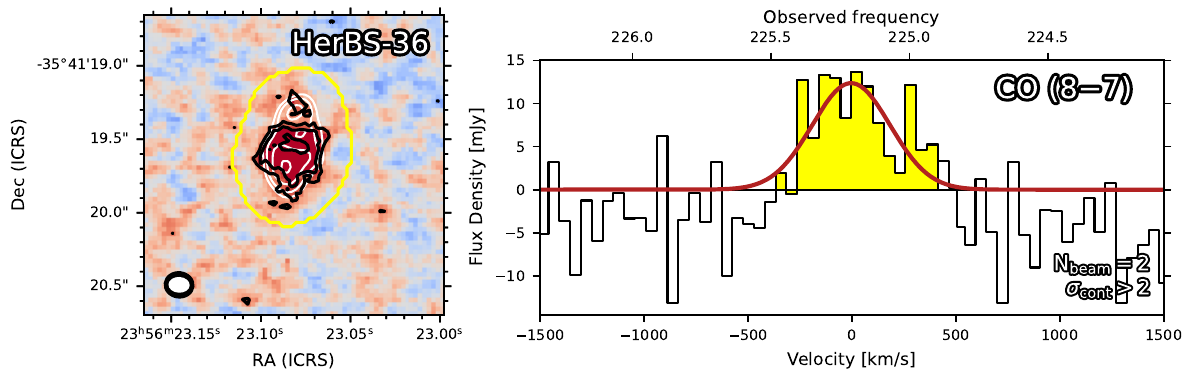}
\includegraphics[width=0.5\linewidth]{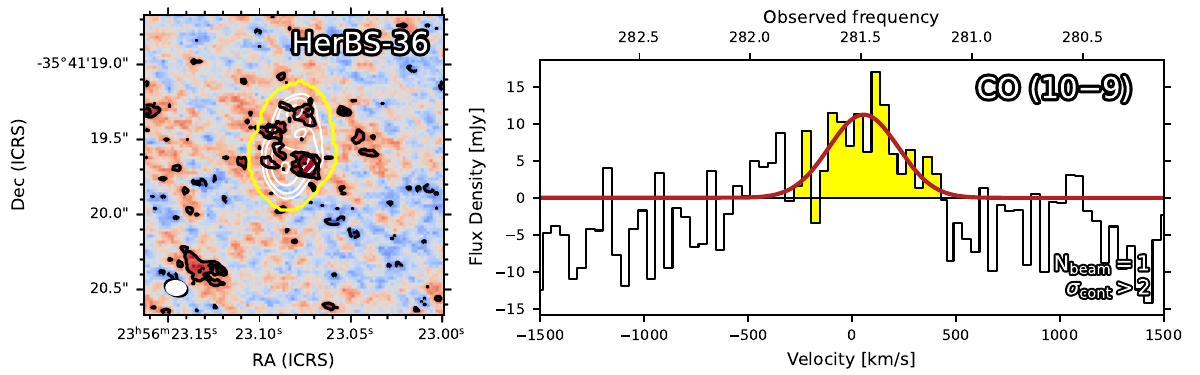}
\includegraphics[width=0.5\linewidth]{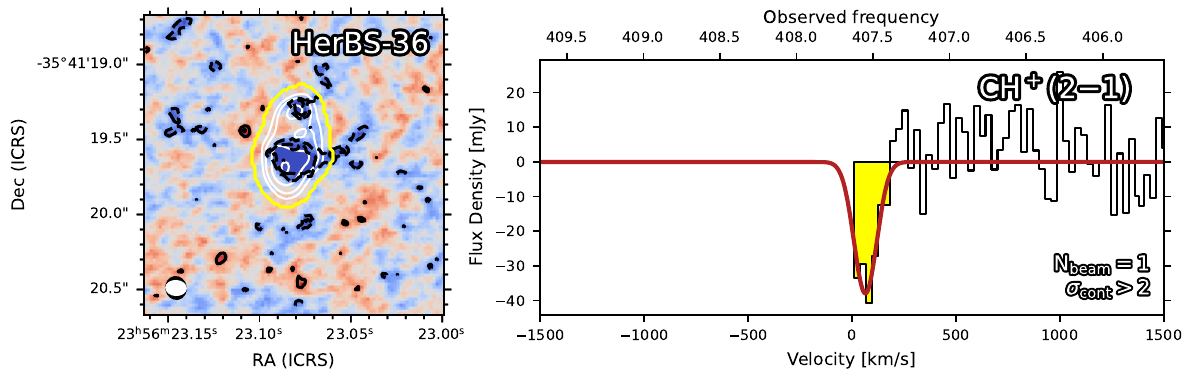}
\caption{The moment-0 (left) and line spectra (right) of HerBS-36 at $z = 3.095$}
\end{figure}\addtocounter{figure}{-1}
\begin{figure}
\includegraphics[width=0.5\linewidth]{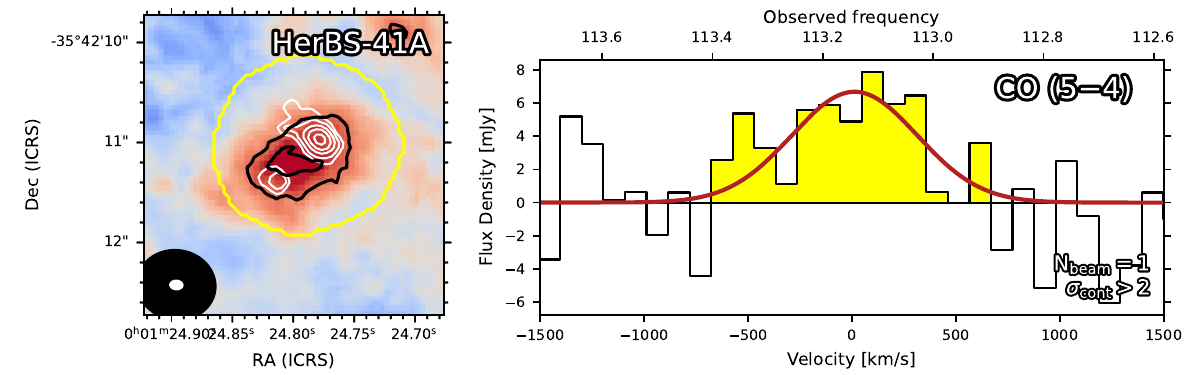}
\includegraphics[width=0.5\linewidth]{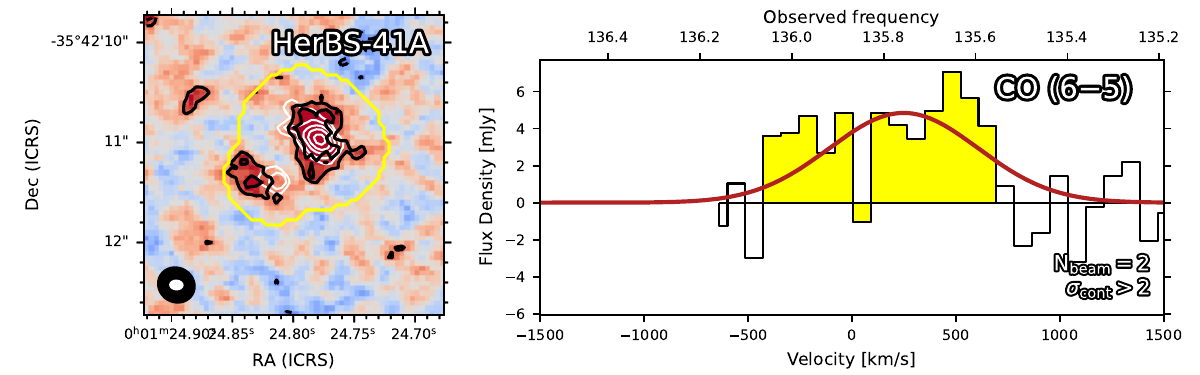}
\includegraphics[width=0.5\linewidth]{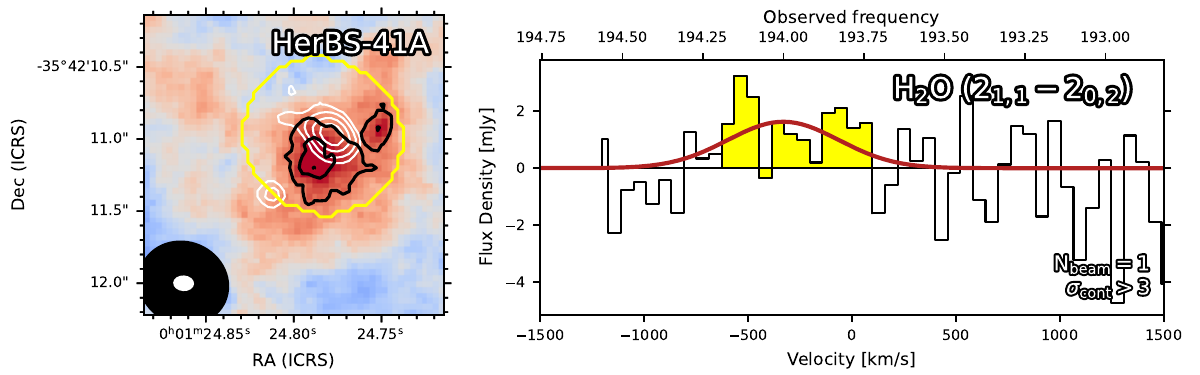}
\includegraphics[width=0.5\linewidth]{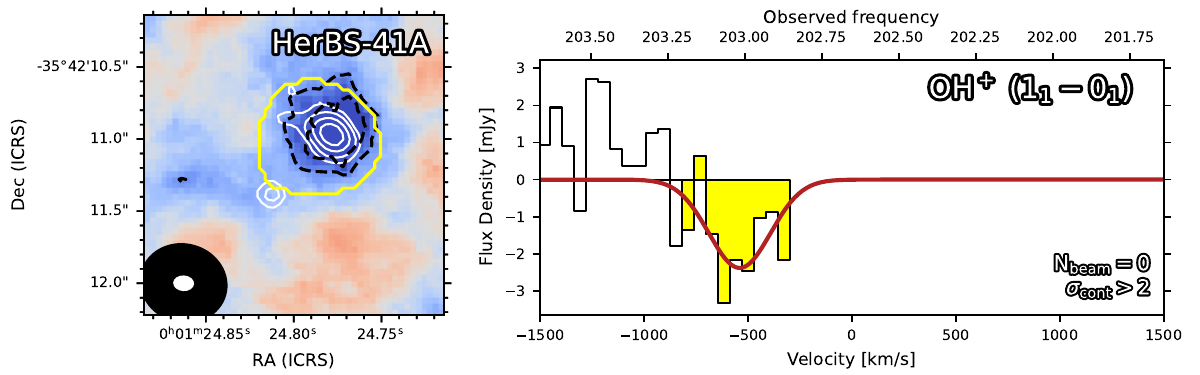}
\includegraphics[width=0.5\linewidth]{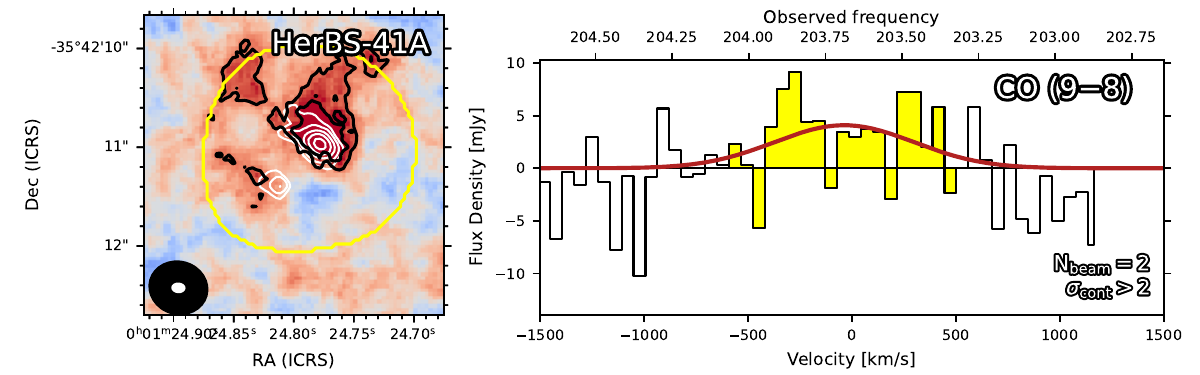}
\includegraphics[width=0.5\linewidth]{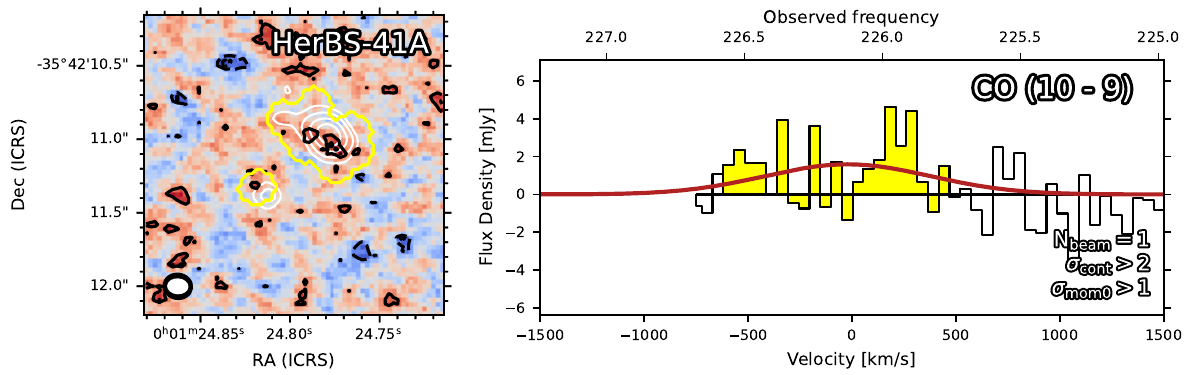}
\includegraphics[width=0.5\linewidth]{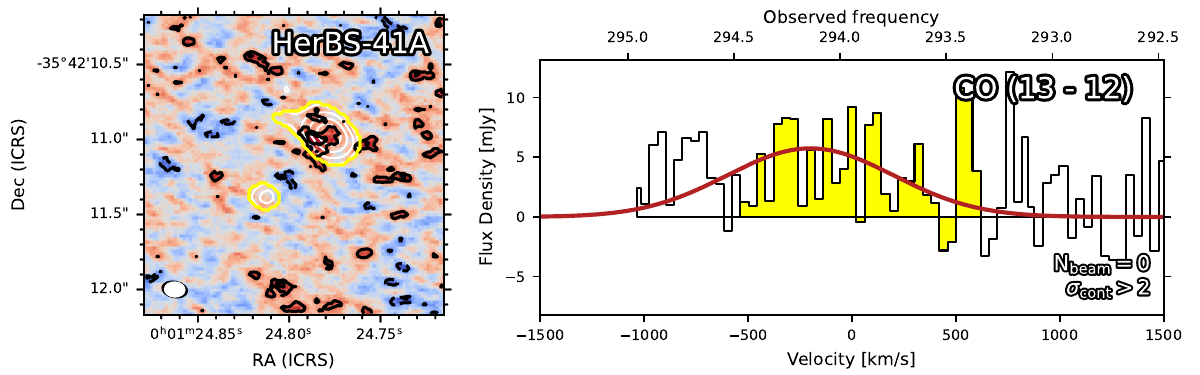}
\includegraphics[width=0.5\linewidth]{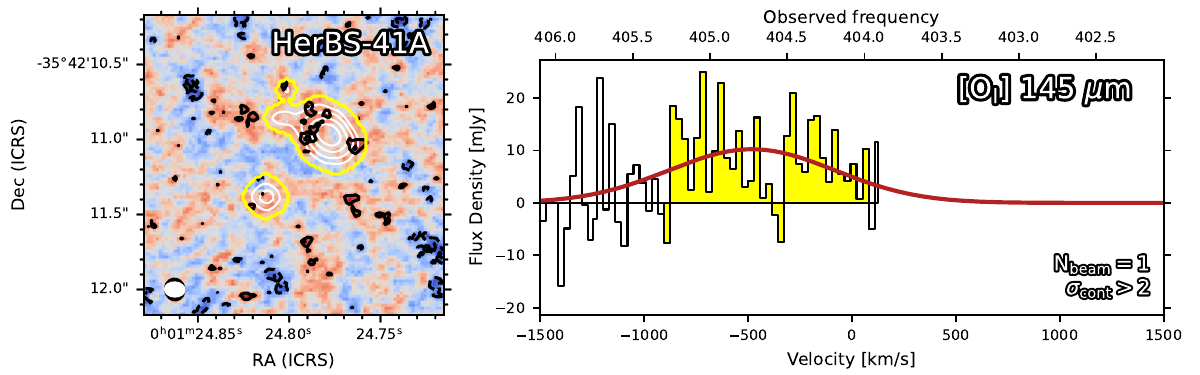}
\includegraphics[width=0.5\linewidth]{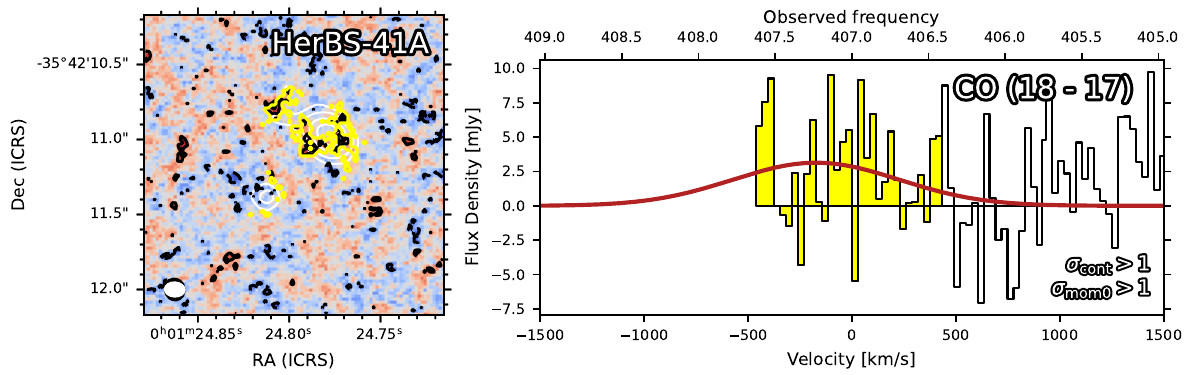}
\caption{The moment-0 (left) and line spectra (right) of HerBS-41A at $z = 4.098$}
\end{figure}\addtocounter{figure}{-1}
\begin{figure}
\includegraphics[width=0.5\linewidth]{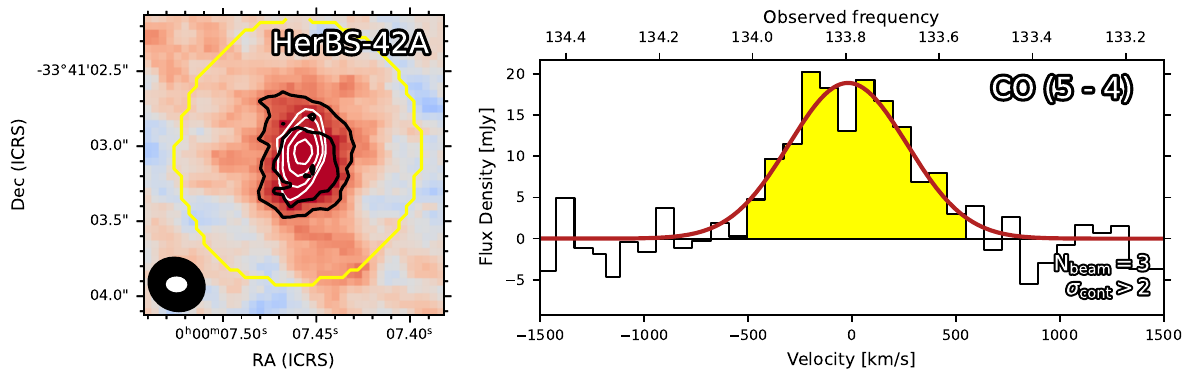}
\includegraphics[width=0.5\linewidth]{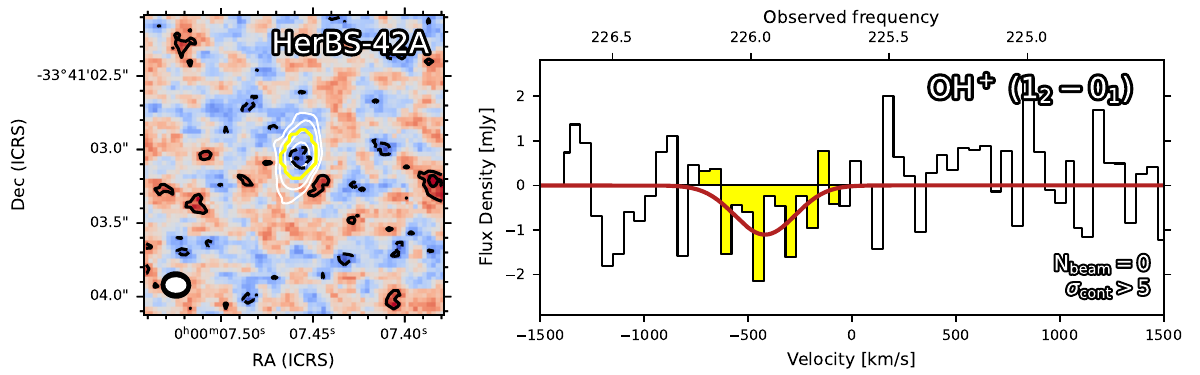}
\includegraphics[width=0.5\linewidth]{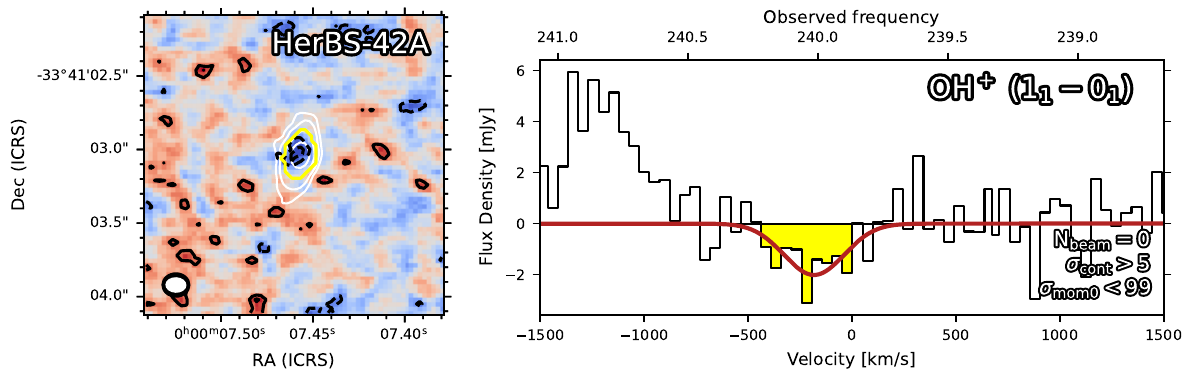}
\includegraphics[width=0.5\linewidth]{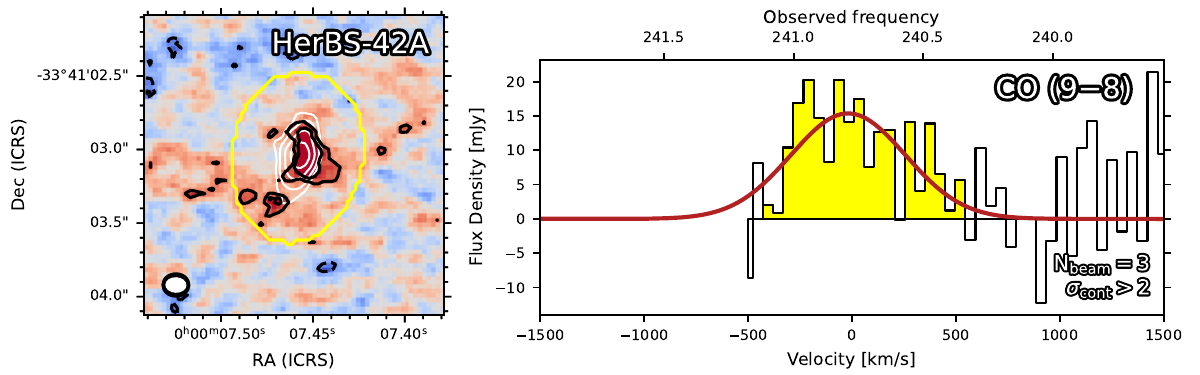}
\includegraphics[width=0.5\linewidth]{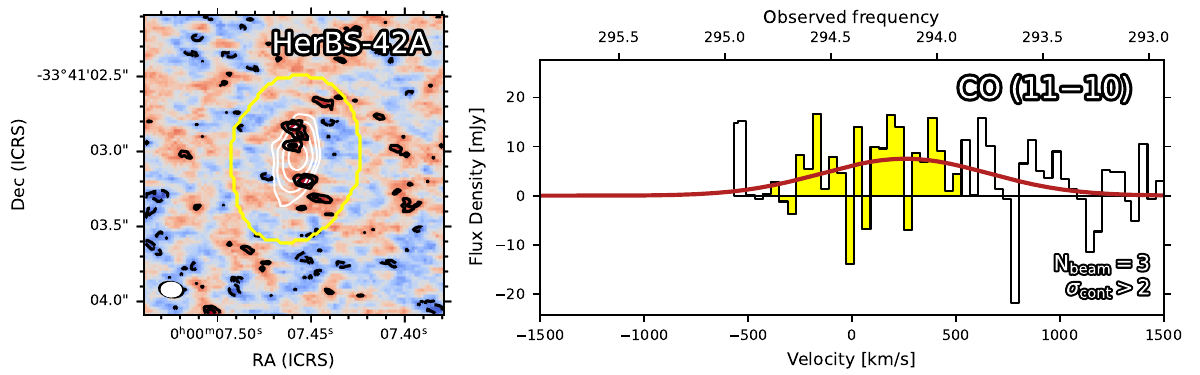}
\caption{The moment-0 (left) and line spectra (right) of HerBS-42A at $z = 3.307$}
\end{figure}\addtocounter{figure}{-1}
\begin{figure}
\includegraphics[width=0.5\linewidth]{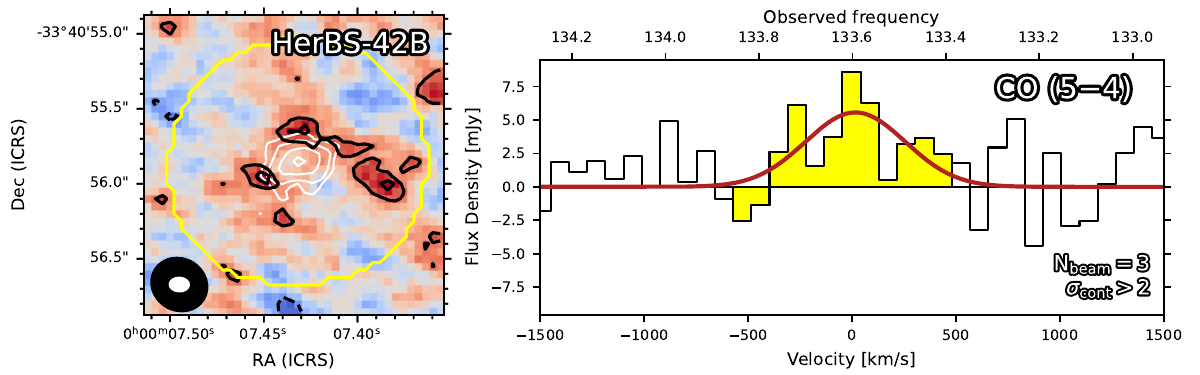}
\includegraphics[width=0.5\linewidth]{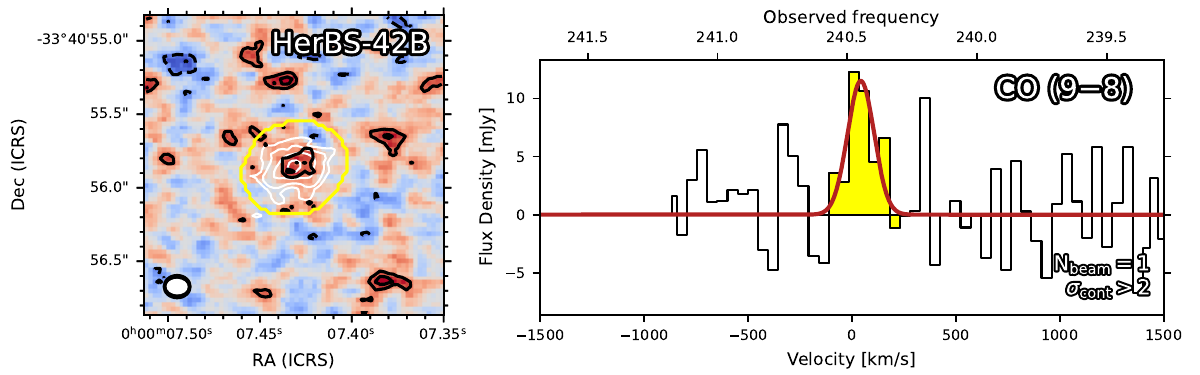}
\caption{The moment-0 (left) and line spectra (right) of HerBS-42B at $z = 3.314$}
\end{figure}\addtocounter{figure}{-1}
\begin{figure}
\includegraphics[width=0.5\linewidth]{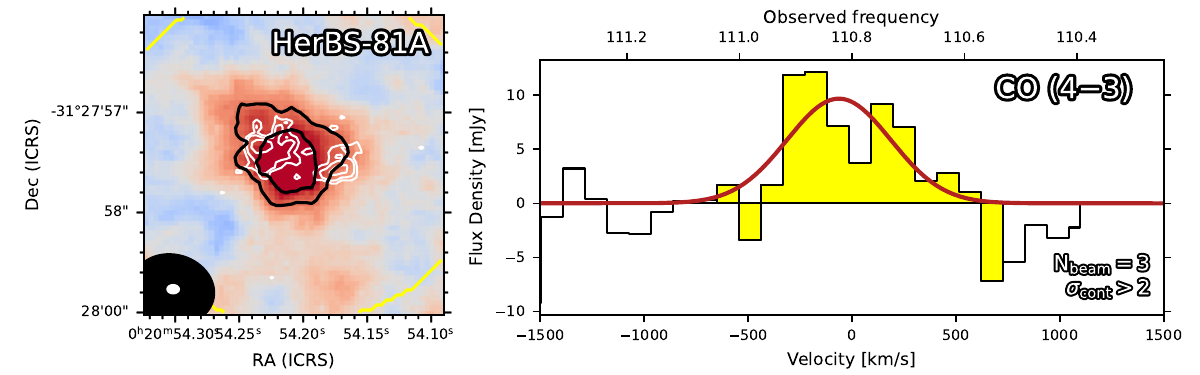}
\includegraphics[width=0.5\linewidth]{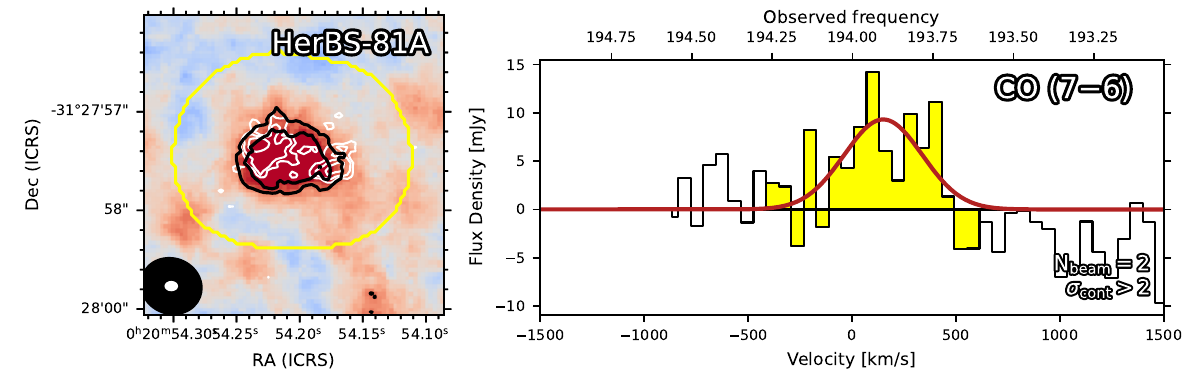}
\includegraphics[width=0.5\linewidth]{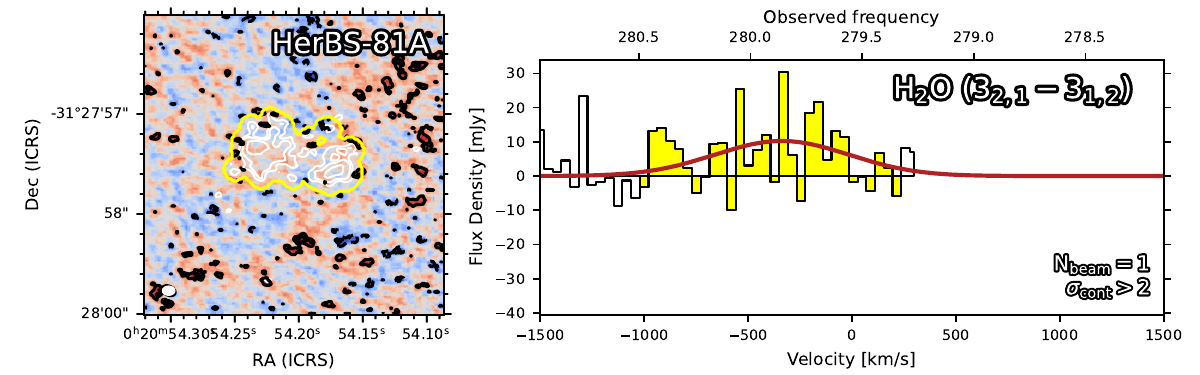}
\caption{The moment-0 (left) and line spectra (right) of HerBS-81A at $z = 3.160$}
\end{figure}\addtocounter{figure}{-1}
\begin{figure}
\includegraphics[width=0.5\linewidth]{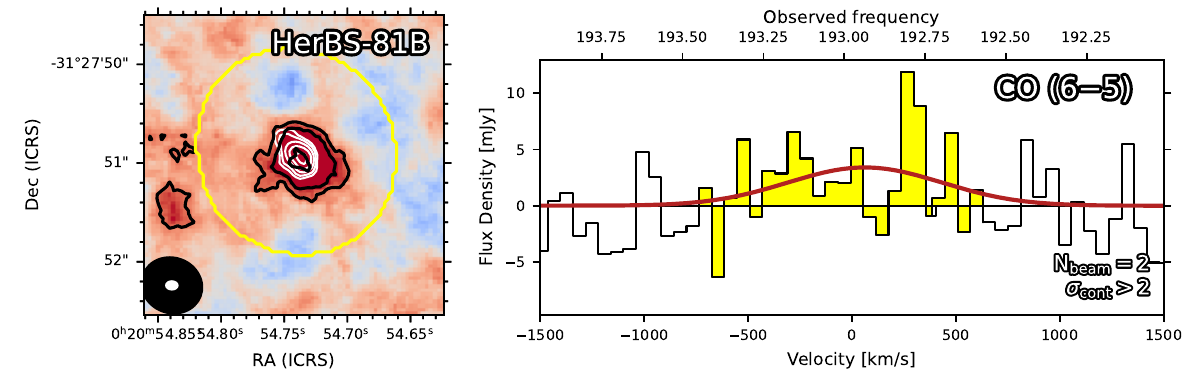}
\includegraphics[width=0.5\linewidth]{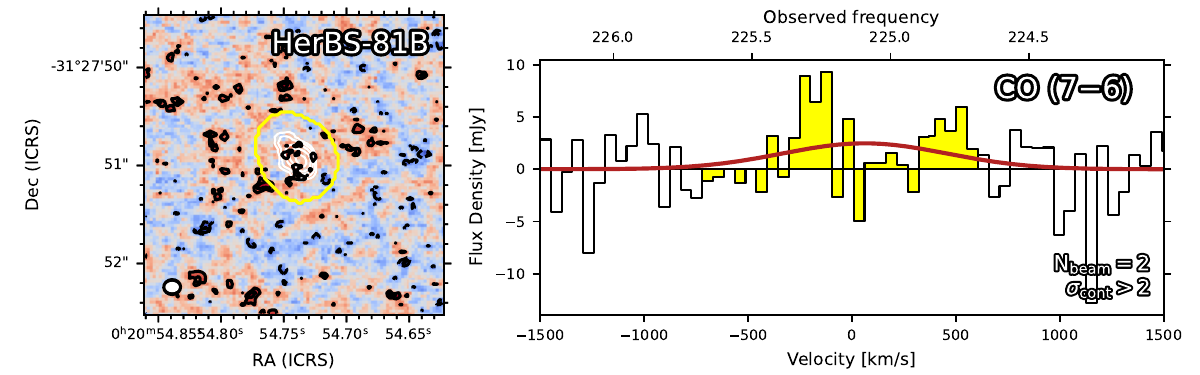}
\includegraphics[width=0.5\linewidth]{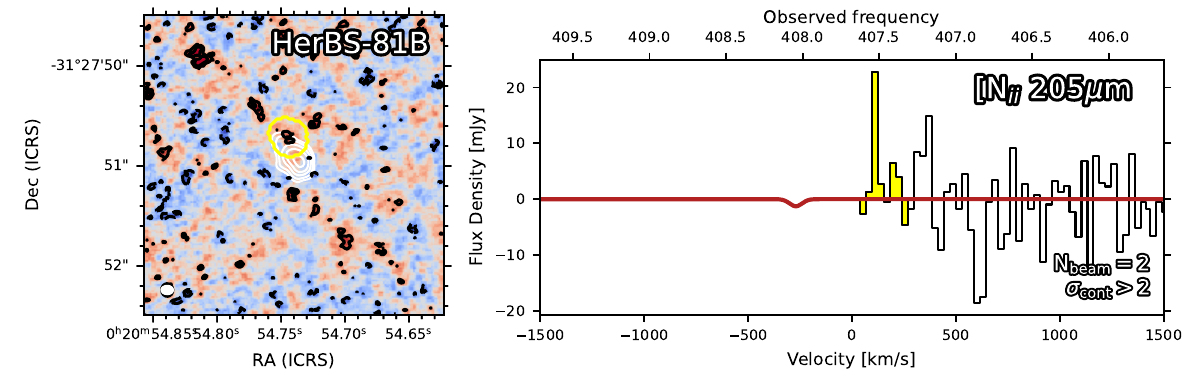}
\caption{The moment-0 (left) and line spectra (right) of HerBS-81B at $z = 2.588$}
\end{figure}\addtocounter{figure}{-1}
\begin{figure}
\includegraphics[width=0.5\linewidth]{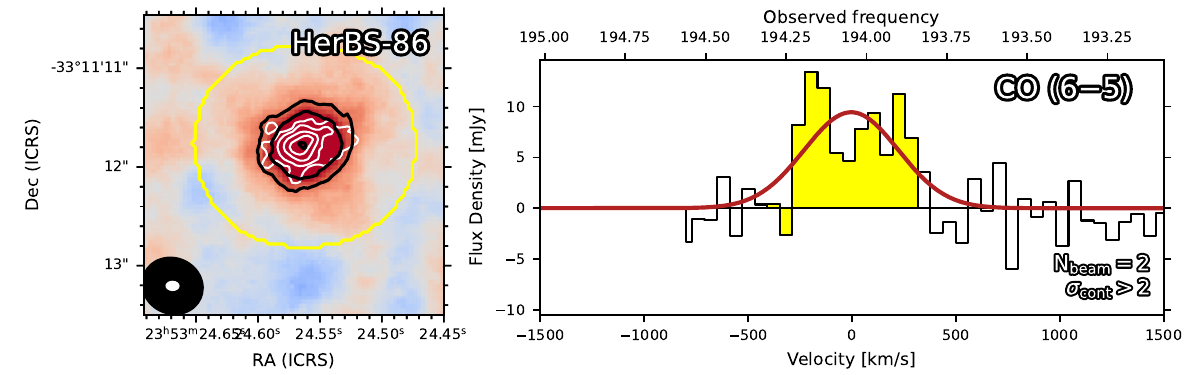}
\includegraphics[width=0.5\linewidth]{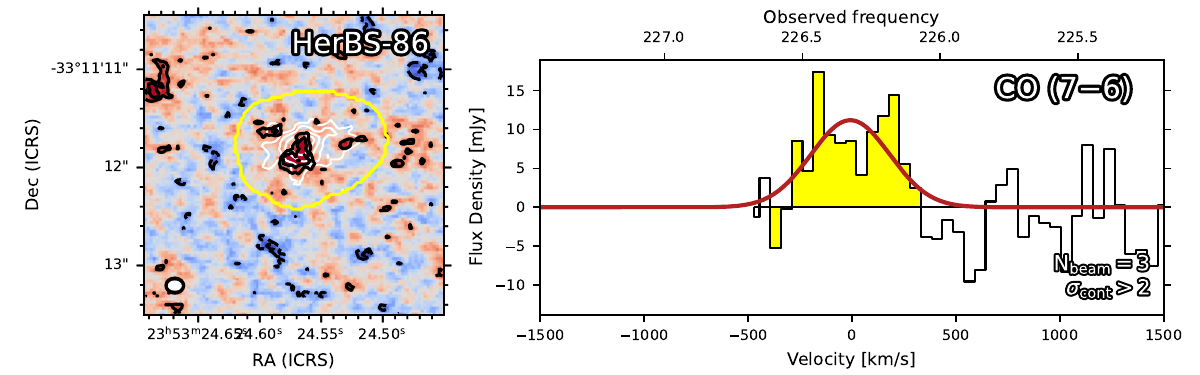}
\caption{The moment-0 (left) and line spectra (right) of HerBS-86 at $z = 2.564$}
\end{figure}\addtocounter{figure}{-1}
\begin{figure}
\includegraphics[width=0.5\linewidth]{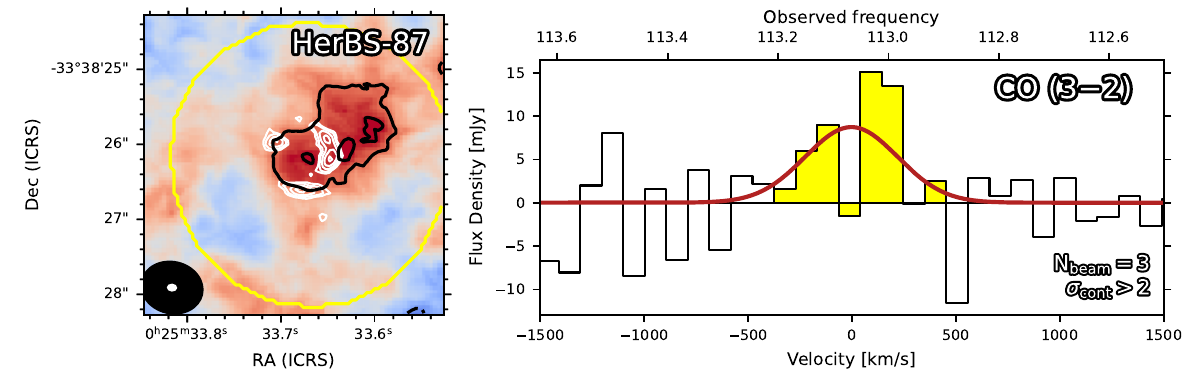}
\includegraphics[width=0.5\linewidth]{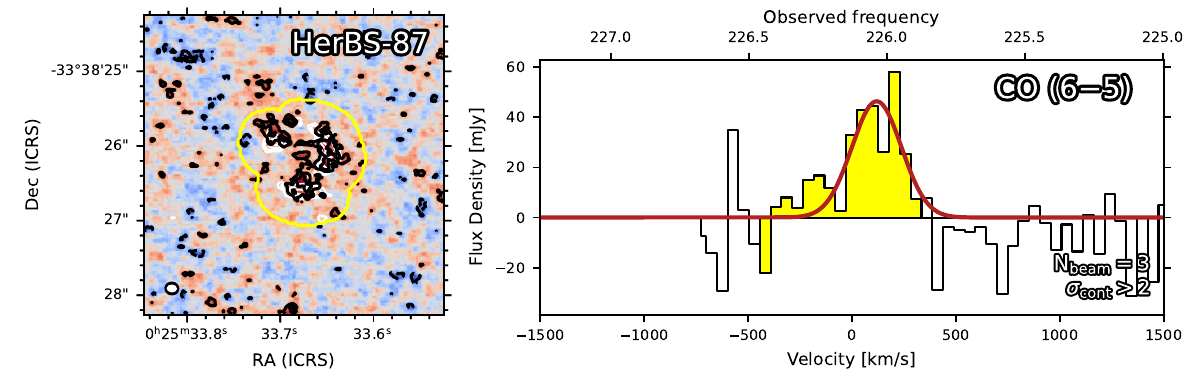}
\caption{The moment-0 (left) and line spectra (right) of HerBS-87 at $z = 2.059$}
\end{figure}\addtocounter{figure}{-1}
\begin{figure}
\includegraphics[width=0.5\linewidth]{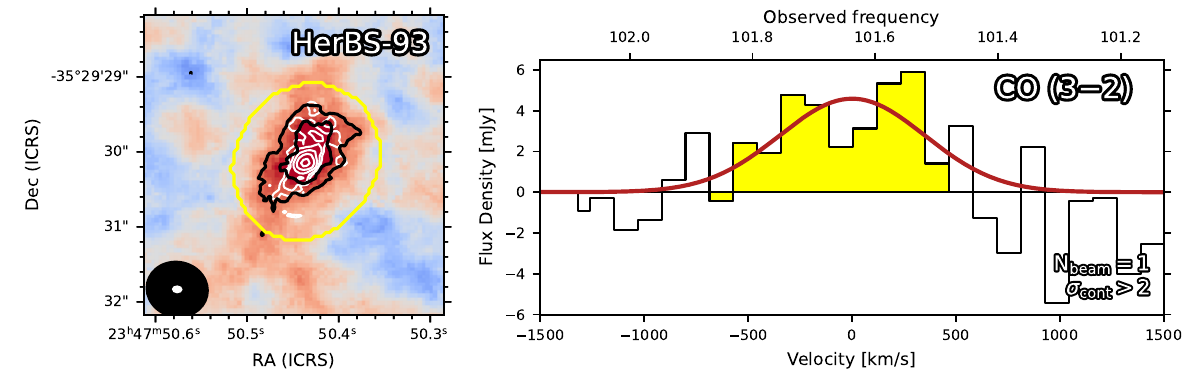}
\includegraphics[width=0.5\linewidth]{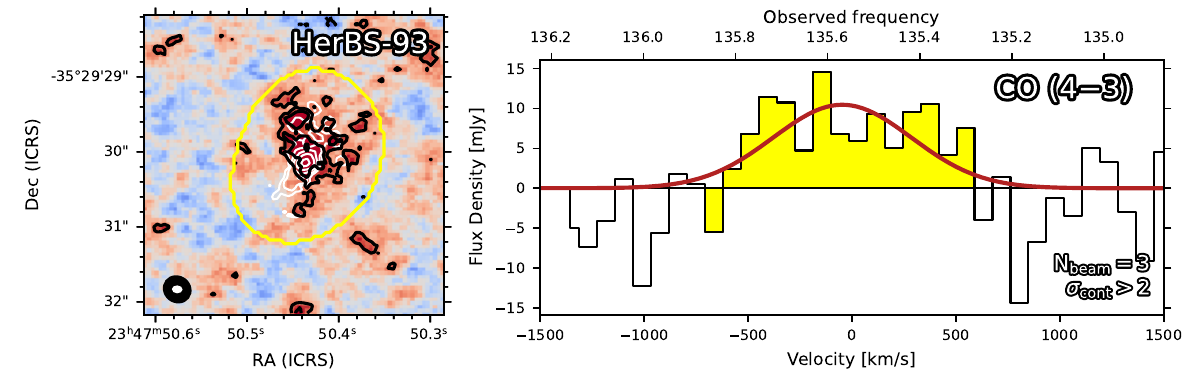}
\includegraphics[width=0.5\linewidth]{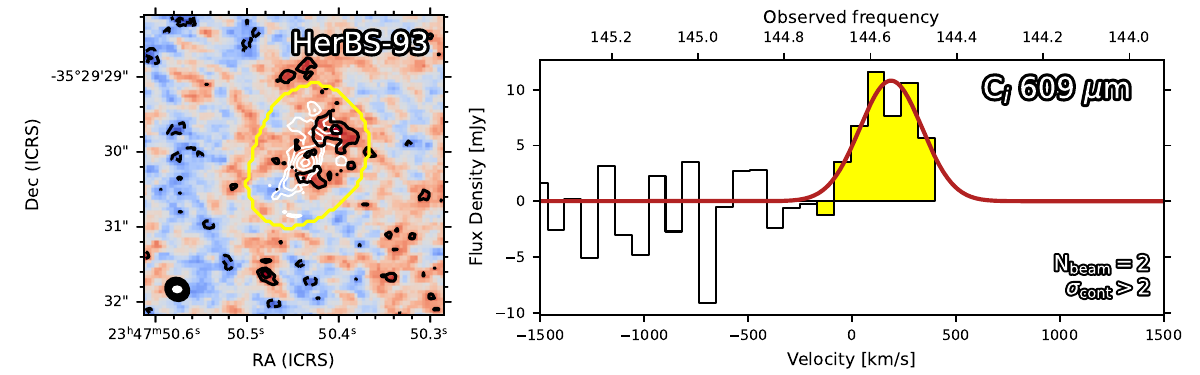}
\includegraphics[width=0.5\linewidth]{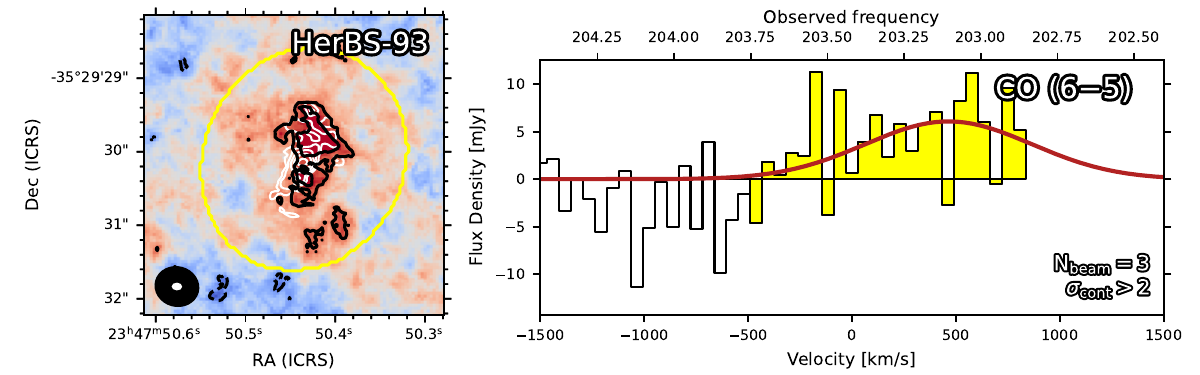}
\includegraphics[width=0.5\linewidth]{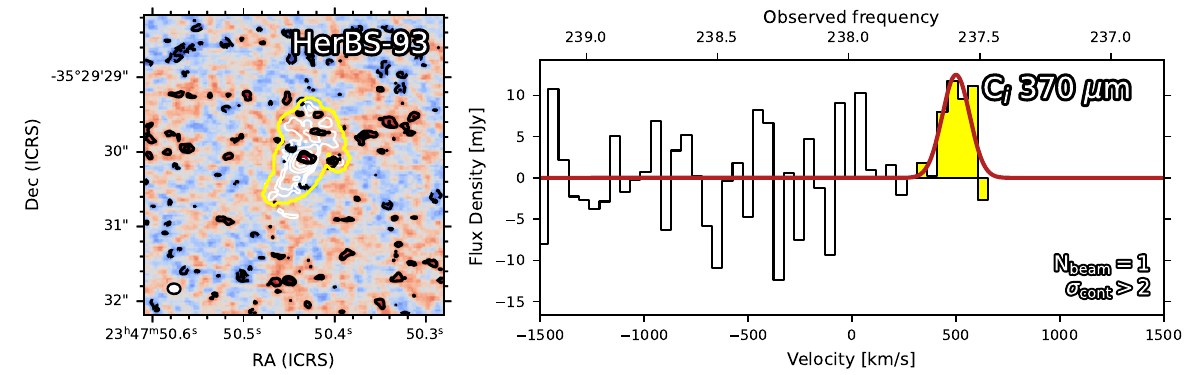}
\caption{The moment-0 (left) and line spectra (right) of HerBS-93 at $z = 2.402$}
\end{figure}\addtocounter{figure}{-1}
\begin{figure}
\includegraphics[width=0.5\linewidth]{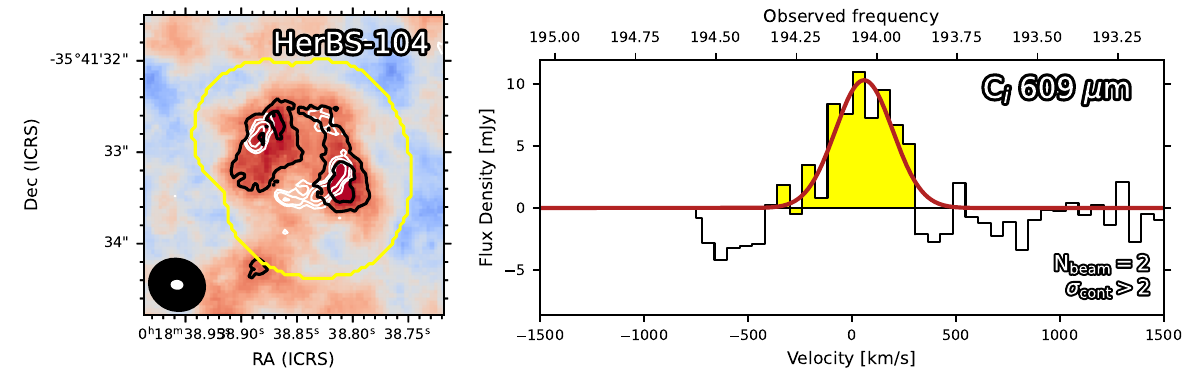}
\includegraphics[width=0.5\linewidth]{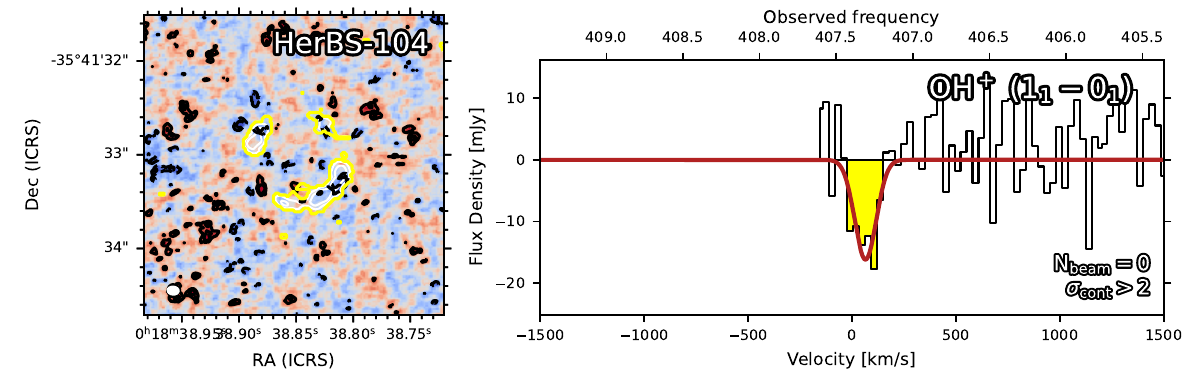}
\caption{The moment-0 (left) and line spectra (right) of HerBS-104 at $z = 1.536$}
\end{figure}\addtocounter{figure}{-1}
\begin{figure}
\includegraphics[width=0.5\linewidth]{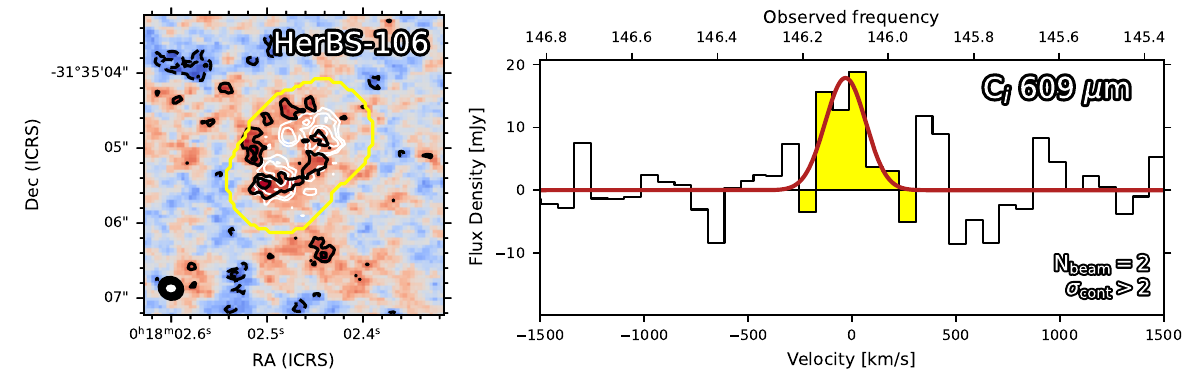}
\includegraphics[width=0.5\linewidth]{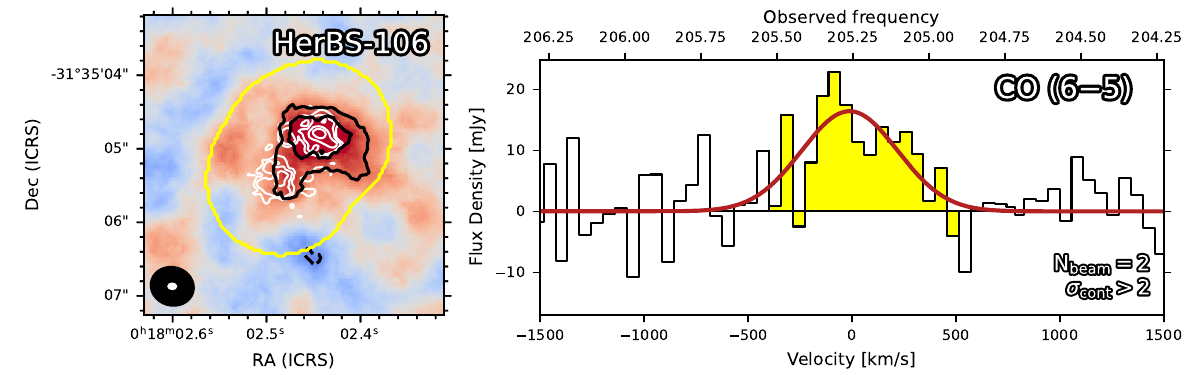}
\includegraphics[width=0.5\linewidth]{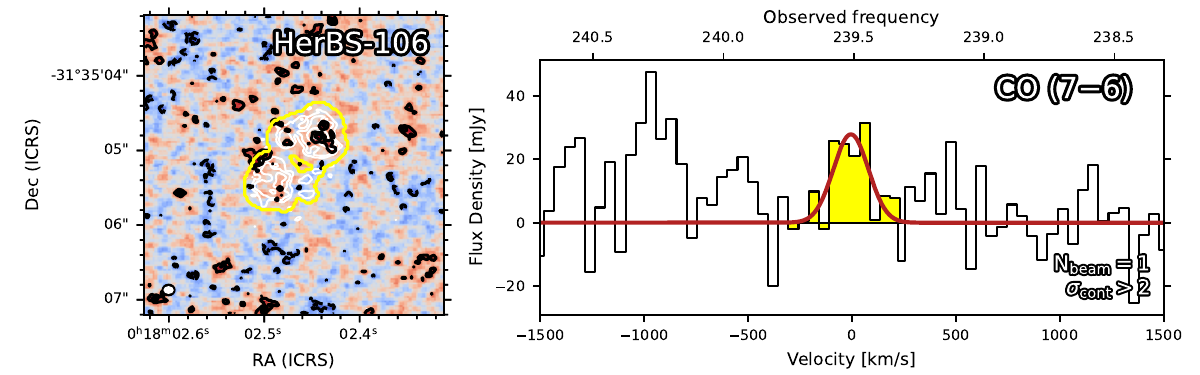}
\includegraphics[width=0.5\linewidth]{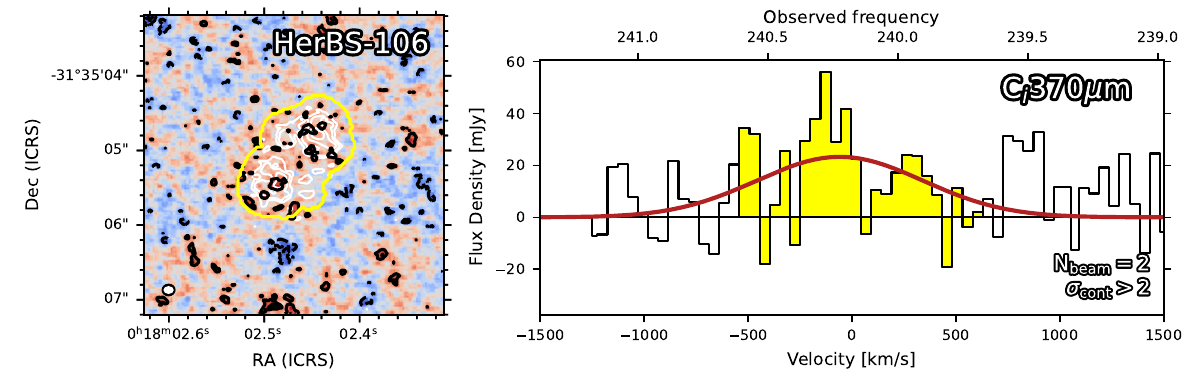}
\caption{The moment-0 (left) and line spectra (right) of HerBS-106 at $z = 2.369$}
\end{figure}\addtocounter{figure}{-1}
\begin{figure}
\includegraphics[width=0.5\linewidth]{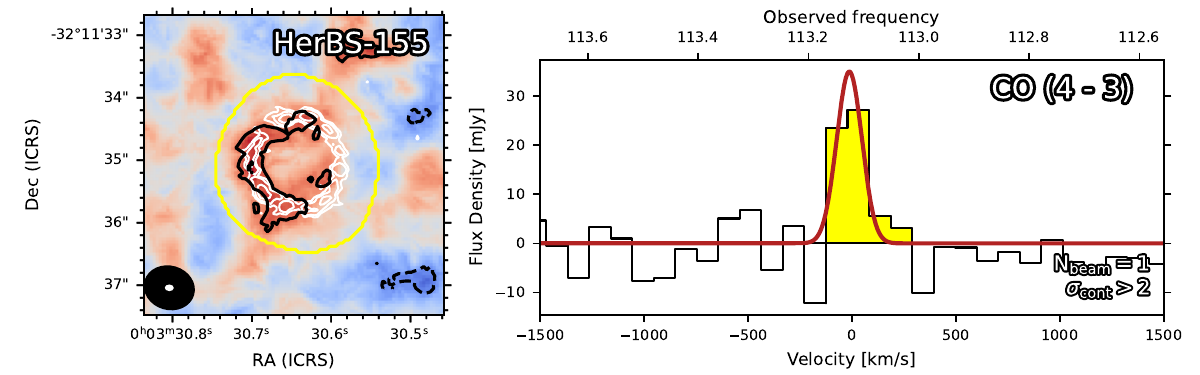}
\includegraphics[width=0.5\linewidth]{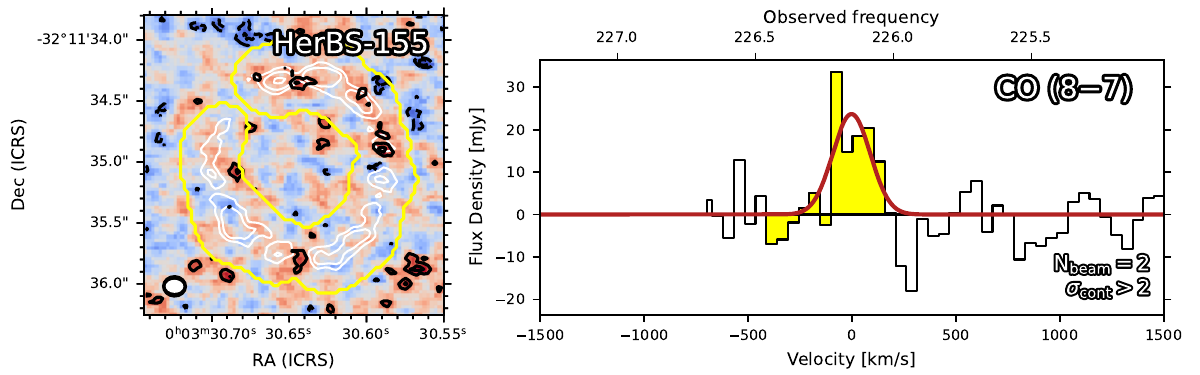}
\caption{The moment-0 (left) and line spectra (right) of HerBS-155 at $z = 3.077$}
\end{figure}\addtocounter{figure}{-1}
% \begin{figure}
% \includegraphics[width=0.5\linewidth]{linegraphs/HerBS-159A_OHp.pdf}
% \caption{The moment-0 (left) and line spectra (right) of HerBS-159A}
% \end{figure}\addtocounter{figure}{-1}
\begin{figure}
\includegraphics[width=0.5\linewidth]{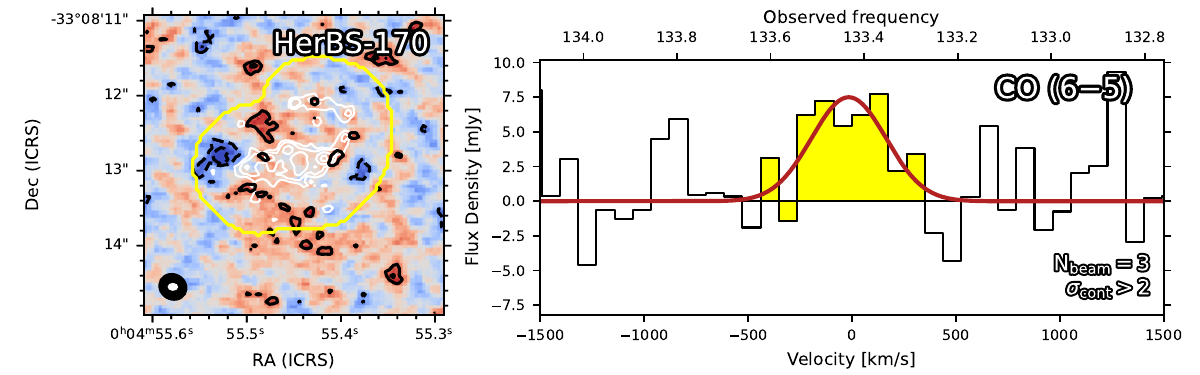}
\includegraphics[width=0.5\linewidth]{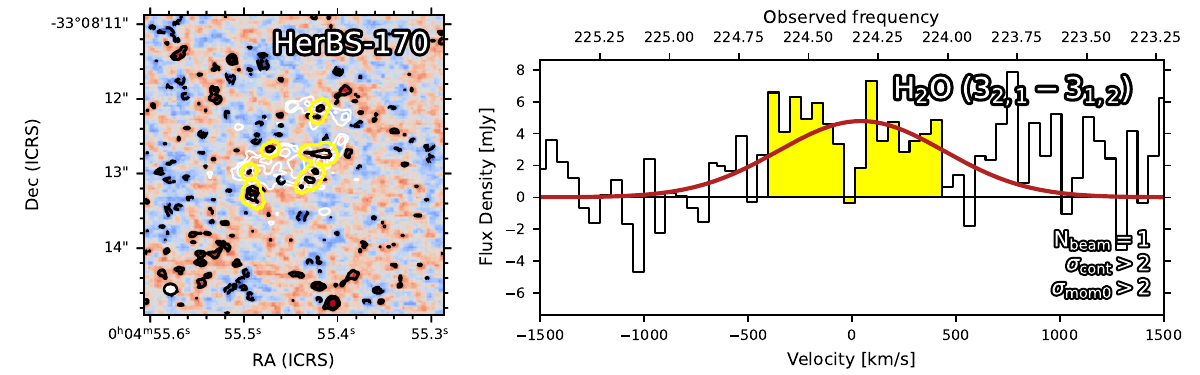}
\includegraphics[width=0.5\linewidth]{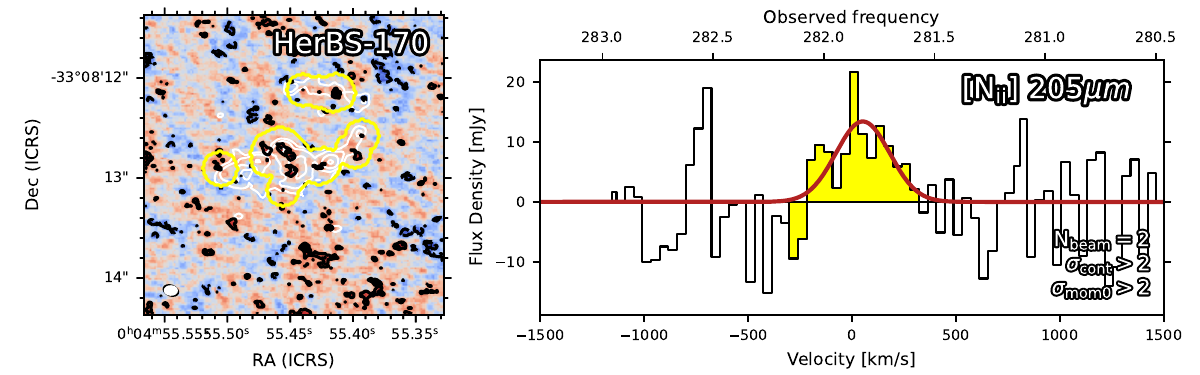}
\caption{The moment-0 (left) and line spectra (right) of HerBS-170 at $z = 4.182$}
\end{figure}\addtocounter{figure}{-1}
\begin{figure}
\includegraphics[width=0.5\linewidth]{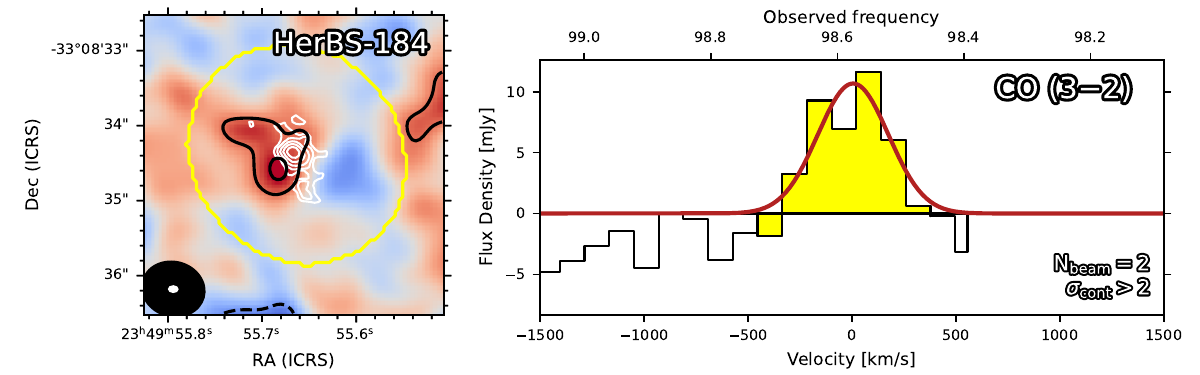}
\includegraphics[width=0.5\linewidth]{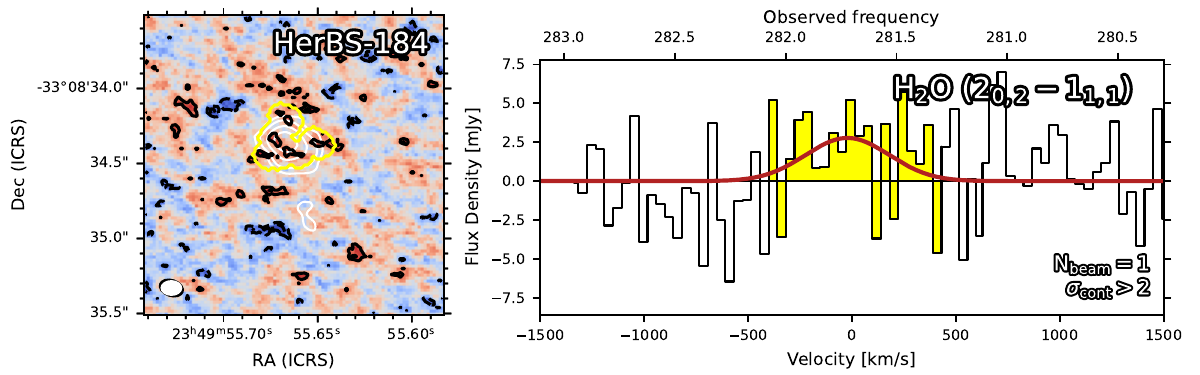}
\includegraphics[width=0.5\linewidth]{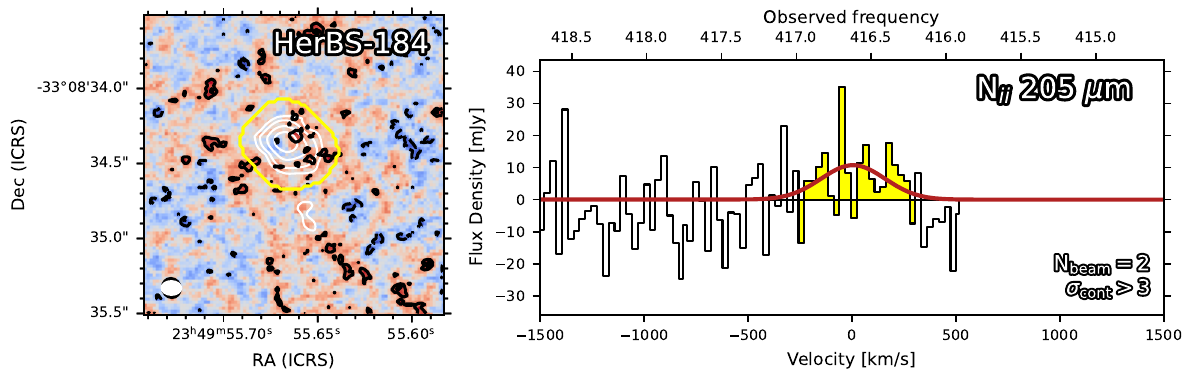}
\caption{The moment-0 (left) and line spectra (right) of HerBS-184 at $z = 2.507$}
\end{figure}
%%%%%%%%%%%%%%%%%%%%%%%%%%%%%%%%%%%%%%%%%%%%%%%%%%

\section{Infrared luminosities of ANGELS sources}
\label{sec:infraredLuminosities}
{\color{referee2}We provide the infrared luminosities for the ANGELS targets using three methodologies in Table~\ref{tab:infraredLuminosities}: 1) using the Eyelash galaxy template \citep[e.g.][]{ivison16},  2) using the {\it Herschel}-derived template from \cite{bakx18,Bakx2020Erratum}, and thirdly assuming a modified black-body with a $\beta_{\rm dust} = 2.0$, based on the method outlined in \cite{Bakx2021}. 
On average the luminosities of the first two methods are roughly 50~per cent higher than for a single-temperature fit, as much of the shorter-wavelength emission is missed in a single-temperature fit, which reflects mostly the colder dust \citep[see e.g.][]{Eales1999}. Meanwhile, the latter option also provides us an estimate of the dust masses for these targets, which are primarily at these colder dust temperatures.}

\begin{table}\color{referee2}
    \centering
    \caption{Infrared luminosities and dust masses for the ANGELS sources}
    \label{tab:infraredLuminosities}
    \begin{tabular}{llllccccccccc} \hline \hline
Source & $z$ & $r_{B7}$ & $\mu$ & $\mu L^{\rm eye}_{\rm IR}$ & $L^{\rm eye}_{\rm IR}$& $\mu L^{\rm B+18}_{\rm IR}$ & $L^{\rm B+18}_{\rm IR}$ & 
 $T_{\rm d}$ & $\mu L^{\rm fit}_{\rm IR}$ & $L^{\rm fit}_{\rm IR}$ & $\mu M_{d}$ & $M_{d}$ \\ 
 & & [0-1] & & [$10^{12}L_{\odot}$]& [$10^{12}L_{\odot}$]& [$10^{12}L_{\odot}$]& [$10^{12}L_{\odot}$] & [K] & [$10^{12}L_{\odot}$]& [$10^{12}L_{\odot}$] & [$10^{8}M_{\odot}$]& [$10^{8}M_{\odot}$]\\
 \hline
21  &   3.323 &   1     &   9     &  51.7 &    5.7    &  56.3 & 6.2 &     34.1              & 46.3  &   5.1 		& 109.4 & 12.1 \\
22  &   3.050 &   1     &  18.8   &  45.5 &    2.4    &  50.5 & 2.6 &     32.6              & 39.2  &   2.0 		& 121.7 & 6.4  \\
25  &   2.912 &   1     &   9.2   &  34.8 &    3.7    &  38.7 & 4.2 &     30.6              & 28.7  &   3.1 		& 130.9 & 14.2 \\
36  &   3.095 &   1     &   4.1   &  40.3 &    9.8    &  44.7 & 10.9 &    33.4              & 35.5  &   8.6 		& 95.4 & 23.2 \\
41  &   4.098 &   1     &   2.6   &  42.5 &    16.3   &  45.6 & 17.5 &    35.9              & 40.3  &   15.5    	& 70.3 & 27.0 \\
42A &   3.307 &   0.726 &   1     &  31.9 &    {\it 31.9}   &  35.4 & {\it 35.4} &    36.7  & 31.3  &   {\it 31.3}  & 47.7 & {\it 47.7} \\
42B &   3.314 &   0.274 &   1     &  12.0 &    {\it 12.0}   &  13.4 & {\it 13.4} &    36.8  & 11.8  &   {\it 11.8}  & 17.9 & {\it 17.9} \\
81A &   3.160 &   0.579 &   1     &  16.9 &    {\it 16.9}   &  18.6 & {\it 18.6} &    33.9  & 15.1  &   {\it 15.1}  & 40.8 & {\it 40.8} \\
81B &   2.584 &   0.421 &   1     &  8.5 &     {\it 8.5}    &   9.6 & {\it 9.6} &     29.2  & 6.8   &   {\it 6.8} 	& 37.2 & {\it 37.2} \\
86  &   2.564 &   1     &   1     &  17.8 &    {\it 17.8}   &  20.2 & {\it 20.2} &    28.1  & 13.7  &   {\it 13.7}  & 103.1 & {\it 103.1} \\
87  &   1.860 &   1     &   8.7   &  13.2 &    1.5    &  15.6 & 1.8 &     25.9              & 9.2   &   1.0 		& 113.7 & 13.0 \\
93  &   2.400 &   1     &   1     &  15.2 &    {\it 15.2}   &  17.5 & {\it 17.5} &    27.6  & 11.5  &   {\it 11.5}  & 97.5 & {\it 97.5} \\
104 &   1.540 &   1     &   6.3   &  10.7 &    1.7    &  12.6 & 2.0 &     24.6              & 6.6   &   1.0 		& 113.1 & 17.9 \\
106 &   2.369 &   1     &   2.3   &  20.9 &    9.0    &  24.2 & 10.5 &    31.7              & 17.9  &   7.8 		& 66.2 & 28.7 \\
155 &   3.077 &   0.860     &  28.3   &  20.2 &    0.7    &  22.2 & 0.8 &     30.6          & 16.6  &   0.6 		& 75.1 & 2.6  \\
159A&   2.236 &   0.690 &   1     &  10.1 &    {\it 10.1}   &  11.8 & {\it 11.8} &    28.5  & 7.8   &   {\it 7.8} 	& 54.6 & {\it 54.6} \\
159B&   2.236 &   0.310 &   1     &  4.5 &     {\it 4.5}    &   5.3 & {\it 5.3} &     28.5  & 3.5   &   {\it 3.5} 	& 24.5 & 24.5 \\
170 &   4.184 &   1     &   1     &  38.1 &    {\it 38.1}   &  41.5 & {\it 41.5} &    38.2  & 37.2  &   {\it 37.2}  & 45.0 & {\it 45.0} \\
184 &   2.507 &   1     &   1     &  18.6 &    {\it 18.6}   &  21.5 & {\it 21.5} &    30.6  & 15.4  &   {\it 15.4}  & 69.9 & {\it 69.9} \\ \hline
    \end{tabular}
        \raggedright \justify \vspace{-0.2cm}
\textbf{Notes:} 
Col. 1: Source name.
Col. 2: Spectroscopic redshift
Col. 3: Band 7 flux ratio between the component and other sources in the field.
Col. 4: {\color{referee3} Magnifications as derived in Section~\ref{sec:Lensing}}
Col. 5-6: Observed and magnification-corrected infrared luminosity derived from the Eyelash template \citep[e.g.][]{ivison16} using the $z_{\rm spec}$ fitted to the {\it Herschel} fluxes.
Col. 7-8: Observed and magnification-corrected infrared luminosity derived from the template in \citet{bakx18,Bakx2020Erratum} using the $z_{\rm spec}$ fitted to the {\it Herschel} fluxes.
Col. 9-13: Single-temperature modified black-body using the $z_{\rm spec}$ fitted to the {\it Herschel} fluxes, assuming $\beta_{\rm dust} = 2.0$, providing the dust temperature, observed and magnification-corrected infrared luminosities and observed and magnification-corrected dust masses using the methodology from \citet{Bakx2021}.
Italics indicate unchanged values in the case of an unlensed source.
\end{table}

\section{Comparison of BEARS and ANGELS fluxes}
{\color{referee}Figure~\ref{fig:compareBEARS_ANGELS} shows the velocity-integrated flux densties for the six galaxies with line measurements of the same CO transition in both BEARS and ANGELS, and compare this against the resolved observations extracted using the same technique as the BEARS study \citep{Urquhart2022}, where the aperture is increased in a curve-of-growth fashion until a signal-to-noise ratio of 5 is reached in the velocity-integrated flux density. The line fluxes of the resolved ANGELS observations are often higher than those extracted from the moderately-resolved BEARS observations, and could point to differences between the curve-of-growth approach from BEARS \citep{Urquhart2022}. Similarly, unresolved and resolved studies \citep{JvM1995, Czekala2021} have found different fluxes on the same object, such as the flux estimates from ALMA large program ALPINE when compared to the resolved CRISTAL study \citep{Posses2024}, or the flux estimates reported in resolved observations of bright AGNs \citep{Novak2019}. 
The origin of these issues is believed to arise from the deconvolution technique in \textsc{TCLEAN} that produces a combined image of background noise (in units of Jy~/~dirty~beam) and signal (assuming a Gaussian point-spread function in units of Jy~/~clean~beam). 
As a consequence, the fluxes from resolved maps might produce a different total flux estimate than those from unresolved observations.
}
\begin{figure}
    \centering
    \includegraphics[width=0.5\linewidth]{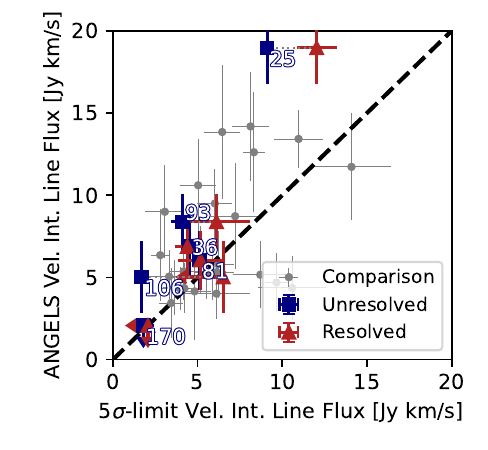}
    \caption{\color{referee}The velocity-integrated line fluxes of the BEARS (x-axis) and ANGELS (y-axis) flux extraction methods show discrepant flux densities across the six CO lines that are observed in both the unresolved ({\it blue squares}) and resolved ({\it red triangles} data.
    A comparison of the fields with only resolved data ({\it grey circles}) also show a bias towards higher fluxes when the ANGELS method is used for flux extraction, suggesting the flux difference between BEARS and ANGELS is both because of the use of resolved data (i.e., \citet{JvM1995,Czekala2021}) and because of a difference in methods of flux extraction. }
    \label{fig:compareBEARS_ANGELS}
\end{figure}

\section{RGB images from the dust continuum}
Figure~\ref{fig:RGBAngels} shows an RGB image using the Bands 6, 7 and 8 photometry of ANGELS across nineteen sources. These galaxies show a large diversity.
\begin{figure*}
    \centering
    \includegraphics[width=\textwidth]{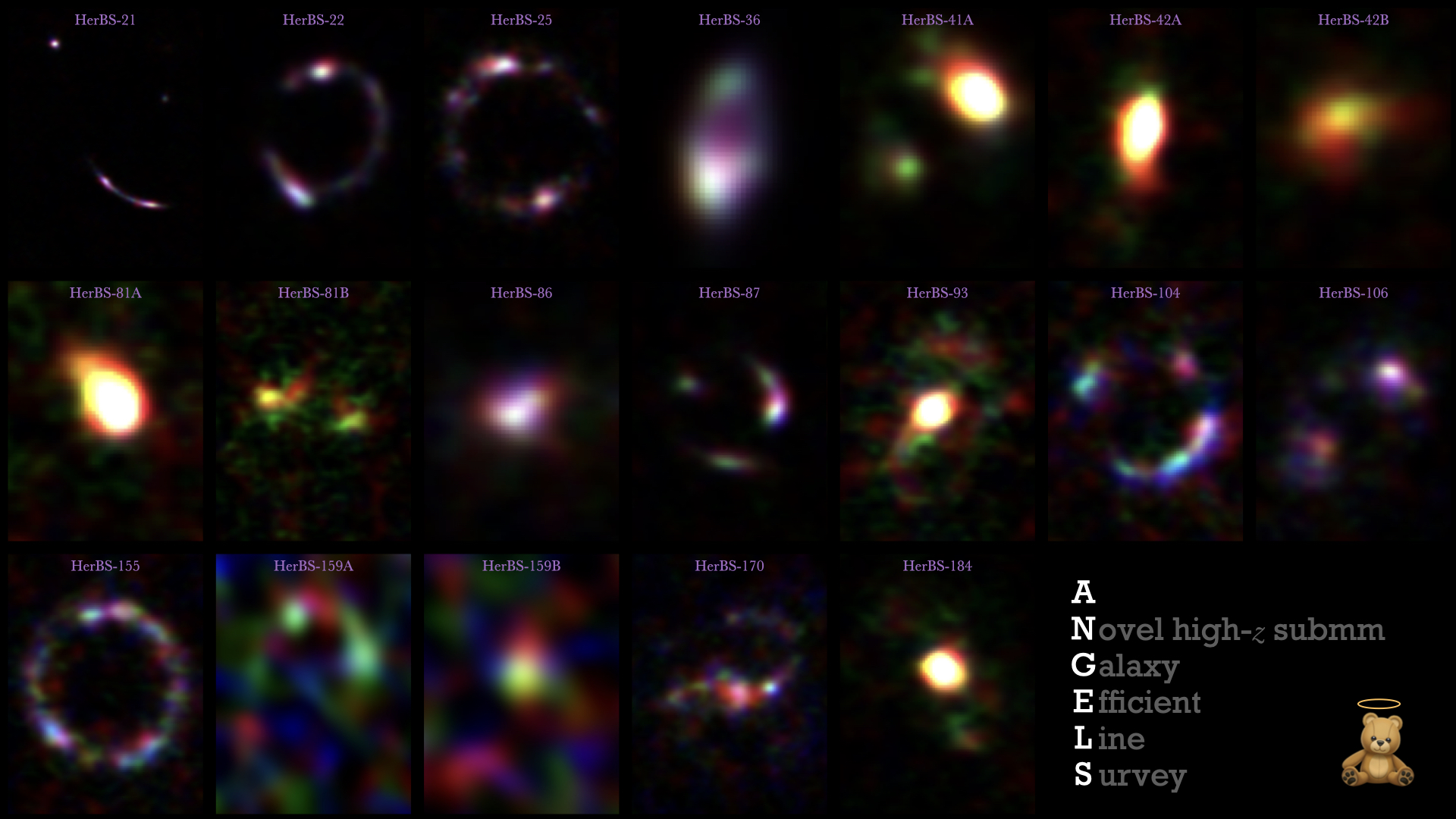}
    \caption{ \color{referee} A composite rgb image across all nineteen fields with Bands 6 (\it red), 7 ({\it green}) and 8 ({\it blue}) photometry shows the spread in morphologies across the ANGELS sources. }
    \label{fig:RGBAngels}
\end{figure*}

\section{Resolved comparison of dust to gas}
The data enables a spatially-resolved study of Schmidt-Kennicutt (see Figure~\ref{fig:ksresolved}). We derive this in a pixel-per-pixel comparison shown in Appendix Figure~\ref{fig:individualKS} for each of the sources with observed spectral lines.
\begin{figure*}
    \centering
    \includegraphics[width=\textwidth]{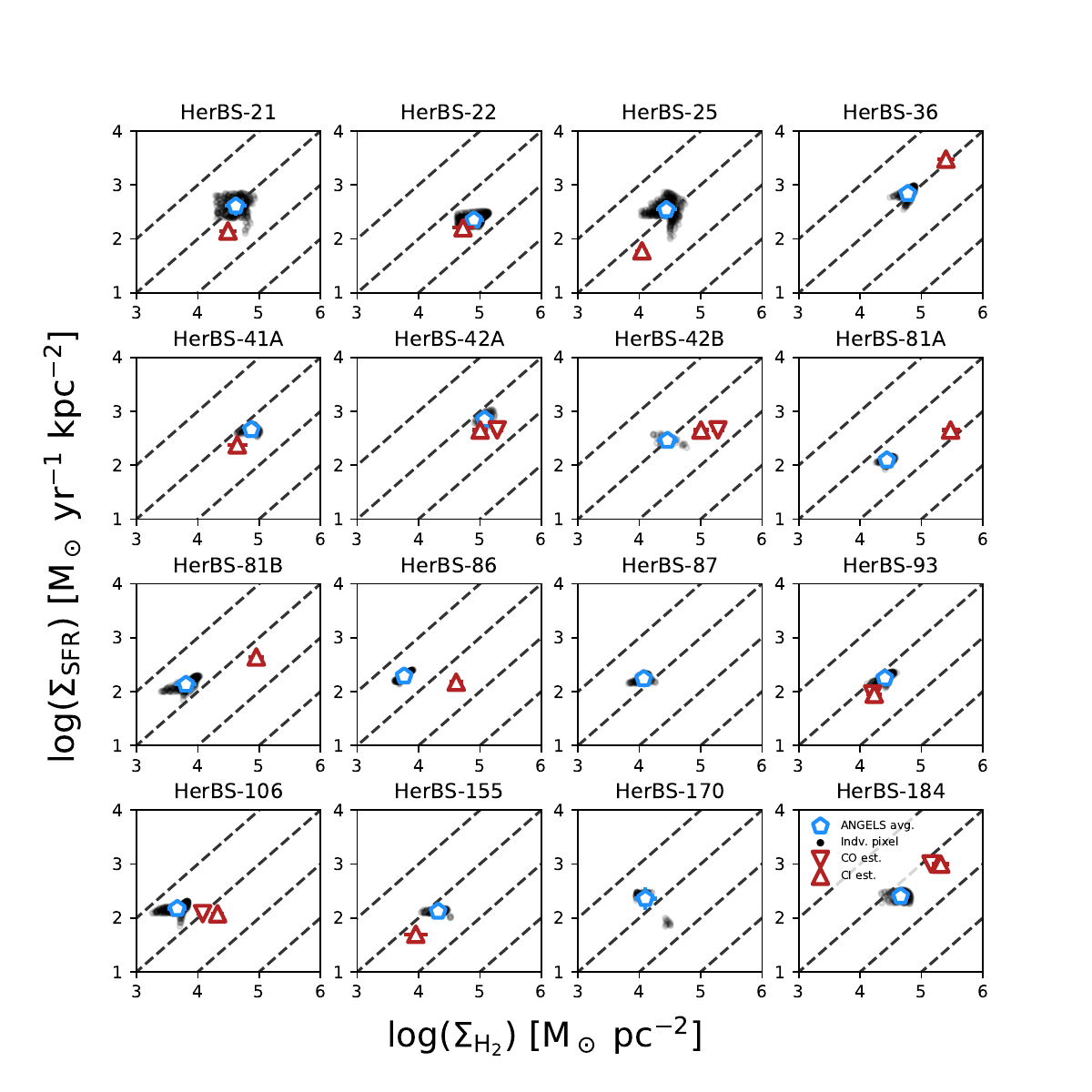}
    \caption{The pixel-per-pixel comparison of the Schmidt-Kennicutt relation for each source is shown for individual pixels ({\it black points}) and as the most likely value ({\it blue pentagon}) and associated errors. We show the constant depletion timescales as diagonal lines for direct comparison to Figure~\ref{fig:ksresolved}. The \textit{red upward and downward triangles} indicate the position of the source on the Schmidt-Kennicutt relation based on the unresolved observations in \citet{Hagimoto2023} for CO and \ci{}-derived dust masses, respectively.
    }
    \label{fig:individualKS}
\end{figure*}
\begin{figure*}
    \centering
    \includegraphics[width=\textwidth]{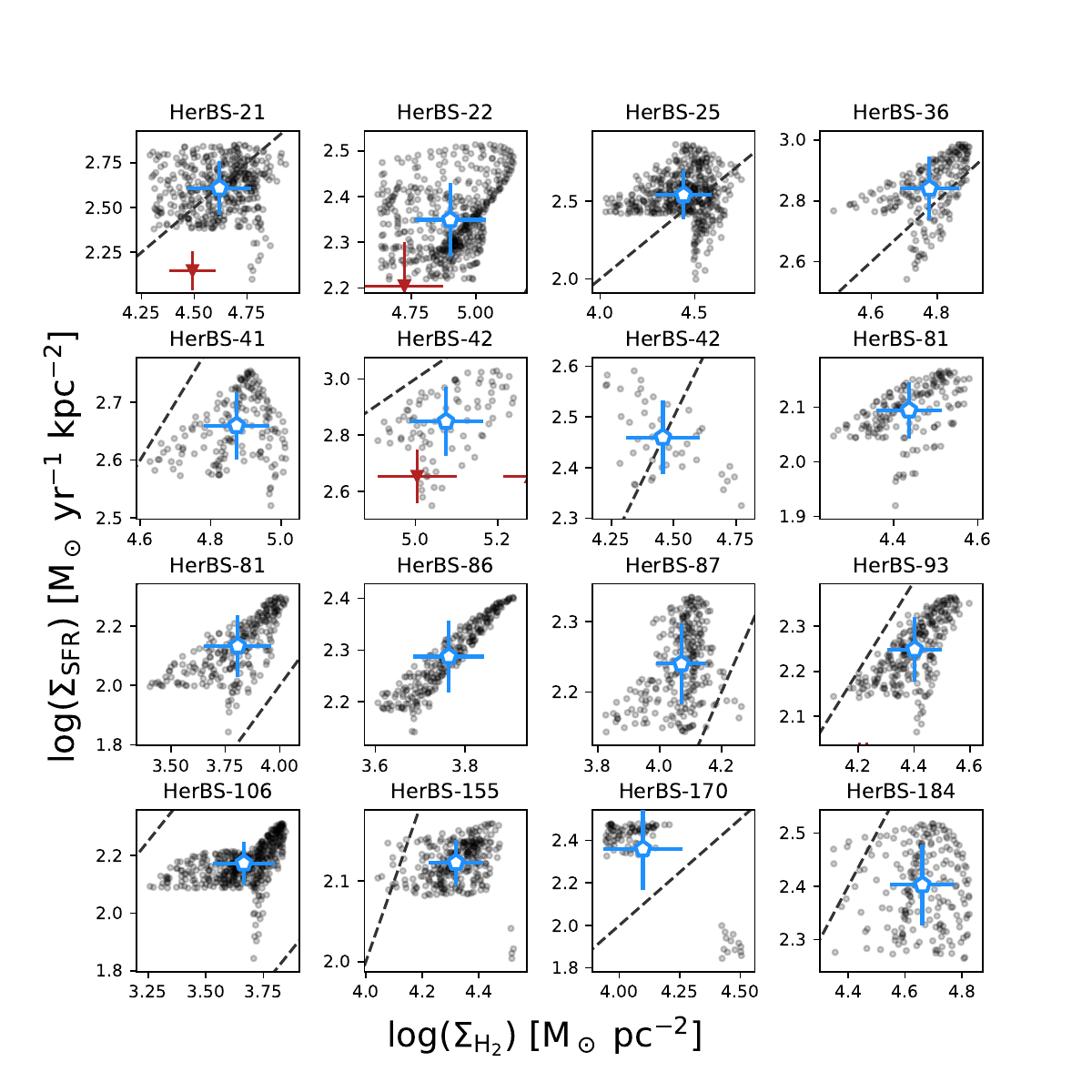}
    \caption{\color{referee2}A zoomed-in version of Fig.~\ref{fig:individualKS} {\color{referee3}focusing} on the pixel-per-pixel comparison of the Schmidt-Kennicutt relation {\it black points}) and its most likely value ({\it blue pentagon}). The constant depletion timescales are shown as diagonal lines for direct comparison to Figure~\ref{fig:ksresolved}. The variation of the gas depletion times across the sources is spread roughly 0.2 to 0.5~dex. in both surface star-formation rate and gas density. {\color{referee3} The sharp cut-offs seen most clearly for HerBS-25 and -106 originate from the }
    }
    \label{fig:individualKSzoomin}
\end{figure*}

% Don't change these lines
\bsp	% typesetting comment
\label{lastpage}
\end{document}